\documentclass[twocolumn]{aastex631}

\usepackage[caption=false]{subfig}
\usepackage[shortlabels]{enumitem}

\usepackage{rotating}
\usepackage{afterpage}

\shorttitle{MUSE Diagnostics: I. Coronal Heating}
\shortauthors{MUSE team}

\definecolor{mypink1}{rgb}{0.858, 0.188, 0.478}

\newcommand{\hinode}{{\em Hinode}}
\newcommand{\hinodee}{{\em Hinode/EIS}~}
\newcommand{\sohos}{{\em SOHO/SUMER}~}
\newcommand{\iris}{{\em IRIS}}
\newcommand{\muse}{{\em MUSE}}
\newcommand{\euvst}{{\em EUVST}}
\newcommand{\dkist}{{\em DKIST}}
\newcommand{\sst}{{\em SST}}
\newcommand{\hic}{{\em Hi-C}}
\newcommand{\psp}{{\em PSP}}
\newcommand{\eui}{{\em EUI}}
\newcommand{\est}{{\em EST}}
\newcommand{\gst}{{\em GST}}
\newcommand{\gregor}{{\em GREGOR}}
\newcommand{\spice}{{\em SPICE}}
\newcommand{\sdo}{{\em SDO}}
\newcommand{\punch}{{\em PUNCH}}
\newcommand{\aia}{{\em AIA}}
\newcommand{\solo}{{\em Solar Orbiter}}
\newcommand{\Bgol}{{\tt B\_en096014\_gol}}
\newcommand{\Bemer}{{\tt B\_en024031\_emer3.0}}
\newcommand{\Bnw}{{\tt B\_nw072100}}
\newcommand{\Bhion}{{\tt B\_en024048\_hion}}
\newcommand{\Bnpdns}{{\tt B\_npdns03}}

\newcommand{\Mplhe}{{\tt MURaM\_plE}}
\newcommand{\Pwaves}{{\tt PL\_waves}}
\newcommand{\Ptwist}{{\tt PL\_loop\_twist}}
\newcommand{\Pnano}{{\tt PL\_nanojets}}
\newcommand{\Cwaves}{{\tt Cip\_waves}}
\newcommand{\Ldc}{{\tt La\_DC}}
\newcommand{\Lac}{{\tt La\_AC}}
\newcommand{\Vawt}{{\tt VBA\_AWT}}
\newcommand{\Mac}{{\tt Mat-AC}}
\newcommand{\Rcone}{{\tt C1}}
\newcommand{\Reone}{{\tt E1}}
\newcommand{\Retwo}{{\tt E2}}
\newcommand{\Rhone}{{\tt H1}}
\newcommand{\heii}{\ion{He}{2}}
\newcommand{\heiiw}{\ion{He}{2}~304\AA}
\newcommand{\feix}{\ion{Fe}{9}}
\newcommand{\fexii}{\ion{Fe}{12}}
\newcommand{\fexv}{\ion{Fe}{15}}
\newcommand{\fexix}{\ion{Fe}{19}}
\newcommand{\fexxi}{\ion{Fe}{21}}
\newcommand{\feixw}{\ion{Fe}{9}~171\AA}
\newcommand{\fexiiw}{\ion{Fe}{12}~195\AA}
\newcommand{\fexvw}{\ion{Fe}{15}~284\AA}
\newcommand{\fexixw}{\ion{Fe}{19}~108\AA}
\newcommand{\fexxiw}{\ion{Fe}{21}~108\AA}

\newcommand{\alfven}{Alfv$\acute{\rm e}$n}

\newcommand{\longacknowledgment}{{\bf Acknowledgements} We would like to thank Helle Bakke, Stuart Bale,  Marc DeRosa, Lyndsay Fletcher, Lars Frogner, Boris Gudiksen, Sarah Jaeggli, Charles Kankelborg, Jorrit Leenaarts, Wei Liu, Nariaki Nitta, Hardi Peter, Jenna Samra, Toshifumi Shimizu, Sami Solanki, Harry Warren, Amy Winebarger, and Peter Young for their valuable contributions and discussions of the results.  We are grateful to Jim Lemen for his technical leadership in the MUSE investigation. We gratefully acknowledge support by NASA contract 80GSFC21C0011 (\muse\ Phase A). 

Some of this work was also supported by NASA contract NNG09FA40C (\iris), and NASA grants 19-HTMS19\_2-0025 ``Flux emergence and the structure, dynamics, and energetics of the solar atmosphere'', 80NSSC18K1285, and 80NSSC21K0737. P.A.\  acknowledges funding from the STFC Ernest Rutherford Fellowship (No. ST/R004285/2). K.K.\ acknowledges funding from a STFC grant (No. ST/T000384/1) and by a FWO (Fonds voor Wetenschappelijk Onderzoek – Vlaanderen) postdoctoral fellowship (1273221N). D.N.S.\ acknowledges funding from the Synergy Grant number 810218 (ERC-2018-SyG) of the European Research Council,
and the project PGC2018-095832-B-I00 of the the Spanish 
Ministry of Science, Innovation and Universities. S.D.\ is supported by a grant from the Swedish Civil Contingencies Agency (MSB) and the Knut and Alice Wallenberg foundation (2016.0019). I.D.M.\ has received support from the UK Science and Technology Facilities Council (Consolidated Grant ST/K000950/1), the European Union Horizon 2020 research and innovation programme (grant agreement No. 647214) and the Research Council of Norway through its Centres of Excellence scheme, project number 262622. T.V.D.\ was supported by the European Research Council (ERC) under the European Union's Horizon 2020 research and innovation programme (grant agreement No 724326) and the C1 grant TRACEspace of Internal Funds KU Leuven.

The simulations have been run on clusters from the Notur project, and the Pleiades cluster through the computing project s1061 and s8305 from the High End Computing (HEC) division of NASA.
Some of the results presented are based upon work supported by the National Center for Atmospheric Research, which is a major facility sponsored by the National Science Foundation under Cooperative Agreement No.\ 1852977. The computational resources and services used for simulations with the \Pwaves\ model were provided by the VSC (Flemish Supercomputer Center), funded by the Research Foundation Flanders (FWO) and the Flemish Government – department EWI. Numerical simulations with the \Cwaves\ model were carried out on Cray XC50 at the Center for Computational Astrophysics, NAOJ. To analyze the data we have used IDL and Python. Simulation \Mplhe\ was performed on resources provided by the Swedish National Infrastructure for Computing (SNIC) and the European Union’s Horizon 2020 research and innovation program under grant agreement No. 824135 (SOLARNET). For the coronal loop PLUTO 3D MHD simulations European PRACE is acknowledged for awarding access to the FERMI resource based in Italy at CINECA, through the project No. 2011050755 "The way to heating the solar corona: finely-resolved twisting of magnetic loops". The nanojets simulations used the DiRAC@Durham facility managed by the Institute for Computational Cosmology on behalf of the STFC DiRAC HPC Facility (www.dirac.ac.uk). The equipment was funded by BEIS capital funding via STFC capital grants ST\/P002293\/1, ST\/R002371\/1 and ST\/S002502\/1, Durham University and STFC operations grant ST\/R000832\/1. DiRAC is part of the UK National e-Infrastructure. 
The synthesis and analysis of the various numerical models have been performed with the aid of Google Cloud Platform (GCP, https://console.cloud.google.com, project  lunar-campaign-29341), allowing sharing of the synthetic data and models, developing common tools, and access to instances with various specifications and Graphics Processing Units (GPUs). This project has been granted by Google Cloud through the University of Oslo, Norway.
This research is also supported by the Research Council of Norway through its Centres of Excellence scheme, project number 262622, and through grants of computing time from the Programme for Supercomputing. 

\iris\ is a NASA small explorer mission developed and operated by LMSAL with mission operations executed at NASA Ames Research Center and major contributions to downlink communications funded by ESA and the Norwegian Space Centre.}

\begin{document}

\title{Probing the physics of the solar atmosphere with the Multi-slit Solar Explorer (MUSE): I. Coronal Heating}

\author[0000-0002-8370-952X]{Bart De Pontieu}
\email{bdp@lmsal.com}
\affil{Lockheed Martin Solar \& Astrophysics Laboratory,
3251 Hanover St, Palo Alto, CA 94304, USA}
\affil{Rosseland Centre for Solar Physics, University of Oslo, P.O. Box 1029 Blindern, N-0315 Oslo, Norway}
\affil{Institute of Theoretical Astrophysics, University of Oslo, P.O. Box 1029 Blindern, N-0315 Oslo, Norway}

\author[0000-0002-0405-0668]{Paola Testa}
\affil{Harvard-Smithsonian Center for Astrophysics, 60 Garden St, Cambridge, MA 02193, USA}

\author[0000-0002-0333-5717]{Juan Mart\'inez-Sykora}
\affil{Lockheed Martin Solar \& Astrophysics Laboratory,
3251 Hanover St, Palo Alto, CA 94304, USA}
\affil{Bay Area Environmental Research Institute, NASA Research Park, Moffett Field, CA 94035, USA.}
\affil{Rosseland Centre for Solar Physics, University of Oslo, P.O. Box 1029 Blindern, N-0315 Oslo, Norway}
\affil{Institute of Theoretical Astrophysics, University of Oslo, P.O. Box 1029 Blindern, N-0315 Oslo, Norway}

\author[0000-0003-1529-4681]{Patrick Antolin}
\affil{Department of Mathematics, Physics \& Electrical Engineering, Northumbria University, Newcastle Upon Tyne, NE1 8ST, UK}

\author[0000-0001-5507-1891]{Konstantinos Karampelas}
\affil{Department of Mathematics, Physics \& Electrical Engineering, Northumbria University, Newcastle Upon Tyne, NE1 8ST, UK}
\affil{Centre for mathematical Plasma Astrophysics (CmPA), KU Leuven, Celestijnenlaan 200B Bus 2400, 3001 Leuven, Belgium}

\author[0000-0003-0975-6659]{Viggo Hansteen}
\affil{Lockheed Martin Solar \& Astrophysics Laboratory,
3251 Hanover St, Palo Alto, CA 94304, USA}
\affil{Bay Area Environmental Research Institute, NASA Research Park, Moffett Field, CA 94035, USA.}
\affil{Rosseland Centre for Solar Physics, University of Oslo, P.O. Box 1029 Blindern, N-0315 Oslo, Norway}
\affil{Institute of Theoretical Astrophysics, University of Oslo, P.O. Box 1029 Blindern, N-0315 Oslo, Norway}

\author[0000-0001-5850-3119]{Matthias Rempel}
\affiliation{High Altitude Observatory, NCAR, P.O. Box 3000, Boulder, CO 80307, USA}

\author[0000-0003-2110-9753]{Mark C. M. Cheung}
\affil{Lockheed Martin Solar \& Astrophysics Laboratory,
3251 Hanover St, Palo Alto, CA 94304, USA}

\author[0000-0002-1820-4824]{Fabio Reale}
\affil{Dipartimento di Fisica e Chimica, Università di Palermo,  Piazza del Parlamento 1, 90134 Palermo, Italy}
\affiliation{INAF-Osservatorio Astronomico di Palermo, Piazza del Parlamento 1, 90134 Palermo, Italy}

\author[0000-0002-2344-3993]{Sanja Danilovic}
\affiliation{Institute for Solar Physics, Department of Astronomy, Stockholm University, AlbaNova University Centre, 106 91 Stockholm, Sweden}

\author[0000-0001-5274-515X]{Paolo Pagano}
\affil{Dipartimento di Fisica e Chimica, Università di Palermo,  Piazza del Parlamento 1, 90134 Palermo, Italy}
\affiliation{INAF-Osservatorio Astronomico di Palermo, Piazza del Parlamento 1, 90134 Palermo, Italy}

\author[0000-0002-4980-7126]{Vanessa Polito}
\affil{Bay Area Environmental Research Institute, NASA Research Park, Moffett Field, CA 94035, USA.}
\affil{Lockheed Martin Solar \& Astrophysics Laboratory,
3251 Hanover St, Palo Alto, CA 94304, USA}

\author[0000-0002-1452-9330]{Ineke De Moortel}
\affiliation{School of Mathematics and Statistics, University of St. Andrews, St. Andrews, Fife KY16 9SS, UK}
\affil{Rosseland Centre for Solar Physics, University of Oslo, P.O. Box 1029 Blindern, N-0315 Oslo, Norway}

\author[0000-0002-7788-6482]{Daniel N\'obrega-Siverio}
\affil{Instituto de Astrof\'isica de Canarias, E-38205 La Laguna, Tenerife, Spain}
\affil{Universidad de La Laguna, Dept. Astrof\'isica, E-38206 La Laguna, Tenerife, Spain}
\affil{Rosseland Centre for Solar Physics, University of Oslo, P.O. Box 1029 Blindern, N-0315 Oslo, Norway}
\affil{Institute of Theoretical Astrophysics, University of Oslo, P.O. Box 1029 Blindern, N-0315 Oslo, Norway}

\author[0000-0001-9628-4113]{Tom Van Doorsselaere}
\affiliation{Centre for mathematical Plasma Astrophysics (CmPA), KU Leuven, Celestijnenlaan 200B Bus 2400, 3001 Leuven, Belgium}

\author[0000-0002-9882-1020]{Antonino Petralia}
\affiliation{INAF-Osservatorio Astronomico di Palermo, Piazza del Parlamento 1, 90134 Palermo, Italy}

\author[0000-0003-0204-8385]{Mahboubeh Asgari-Targhi}
\affil{Harvard-Smithsonian Center for Astrophysics, 60 Garden St, Cambridge, MA 02193, USA}

\author{Paul Boerner}
\affil{Lockheed Martin Solar \& Astrophysics Laboratory,
3251 Hanover St, Palo Alto, CA 94304, USA}

\author[0000-0001-9218-3139]{Mats Carlsson}
\affil{Rosseland Centre for Solar Physics, University of Oslo, P.O. Box 1029 Blindern, N-0315 Oslo, Norway}
\affil{Institute of Theoretical Astrophysics, University of Oslo, P.O. Box 1029 Blindern, N-0315 Oslo, Norway}

\author[0000-0002-1253-8882]{Georgios Chintzoglou}
\affil{Lockheed Martin Solar \& Astrophysics Laboratory,
3251 Hanover St, Palo Alto, CA 94304, USA}
\affil{University Corporation for Atmospheric Research, Boulder, CO 80307-3000, USA}

\author[0000-0002-9288-6210]{Adrian Daw}
\affil{NASA Goddard Space Flight Center,Greenbelt, MD 20771, USA}

\author[0000-0001-7416-2895]{Ed DeLuca}
\affil{Harvard-Smithsonian Center for Astrophysics, 60 Garden St, Cambridge, MA 02193, USA}

\author[0000-0001-9638-3082]{Leon Golub}
\affil{Harvard-Smithsonian Center for Astrophysics, 60 Garden St, Cambridge, MA 02193, USA}

\author[0000-0002-1043-9944]{Takuma Matsumoto}
\affiliation{National Astronomical Observatory of Japan, 2-21-1 Osawa, Mitaka, Tokyo 181-8588, Japan}

\author[0000-0001-5503-0491]{Ignacio Ugarte-Urra}
\affil{Space Science Division, Naval Research Laboratory, Washington, DC 20375, USA}

\author{Scott McIntosh}
\affil{High Altitude Observatory, NCAR, P.O. Box 3000, Boulder, CO 80307, USA}

\author{the \muse\ team}

\begin{abstract}
  The Multi-slit Solar Explorer (\muse) is a proposed NASA MIDEX mission, currently in Phase A, composed of a multi-slit EUV spectrograph (in three narrow spectral bands centered around 171\AA, 284\AA, and 108\AA) and an EUV context imager (in two narrow passbands around 195\AA\ and 304\AA). \muse\ will provide unprecedented spectral and imaging diagnostics of the solar corona at high spatial ($\le 0.5$\arcsec), and temporal resolution (down to $\sim 0.5$~s) thanks to its innovative multi-slit design.   By obtaining spectra in 4 bright EUV lines (171\AA\ \feix,  284\AA\ \fexv,  108\AA\ \fexix-\fexxi) covering a wide range of transition region and coronal temperatures along 37 slits simultaneously, \muse\ will for the first time be able to ``freeze" (at a cadence as short as 10 seconds) with a spectroscopic raster the evolution of the dynamic coronal plasma over a wide range of scales: from the spatial scales on which energy is released ($\le 0.5$\arcsec) to the large-scale often active-region size ($\sim 170\arcsec \times 170\arcsec$) atmospheric response. We use advanced numerical modeling to showcase how \muse\ will constrain the properties of the solar atmosphere on the spatio-temporal scales ($\le$ 0.5\arcsec, $\le$ 20 seconds) and large field-of-view on which various state-of-the-art models of the physical processes that drive coronal heating, solar flares and coronal mass ejections (CMEs) make distinguishing and testable predictions. We describe how the synergy between \muse, the single-slit, high-resolution Solar-C \euvst\ spectrograph, and ground-based observatories (\dkist\ and others) can address how the solar atmosphere is energized, and the critical role \muse\ plays because of the multi-scale nature of the physical processes involved. In this first paper, we focus on how comparisons between \muse\ observations and theoretical models will significantly further our understanding of coronal heating mechanisms. An accompanying paper focuses on various aspects of solar activity such as solar flares and CMEs.  
\end{abstract}


\section{Introduction} \label{sec:intro}
The physical processes at work in the solar outer atmosphere, leading to the heating of coronal plasma to millions of degrees, to the acceleration of the solar wind, and to dynamic events such as flares and coronal mass ejections (CMEs) are still poorly understood. Although significant progress has been made recently, thanks to the ever increasing quality of solar observations as well as continuous advances in numerical modeling \citep[e.g.,][]{Reale2014,vanD2020,DePontieu2021}, the lack of spectroscopic measurements with sufficient spatial and temporal resolution and spatial coverage has hampered progress in understanding these phenomena, because they involve many spatial and temporal scales at once. 
The Multi-slit Solar Explorer (\muse, \citealt{BDP:MUSE,Cheung:SDC}), a mission proposed to NASA as a Medium-class Explorer and now in a Phase A study, aims to overcome these shortcomings of single-slit spectrometers by using an innovative approach using multiple slits and several different and narrow EUV spectral bands.
These novel observations of the solar corona, coupled with state-of-the-art numerical modeling, will provide unprecedented constraints on the physical mechanisms driving coronal heating and space weather events, and will allow \muse\ to address its science goals:
\begin{enumerate}
    \item Determine which mechanism(s) heat the corona and drive the solar wind,
    \item Understand the origin and evolution of the unstable solar atmosphere, and
    \item Investigate fundamental physical plasma processes.
\end{enumerate}

\begin{figure*}[!ht]
\centering
\includegraphics[width=0.95\textwidth]{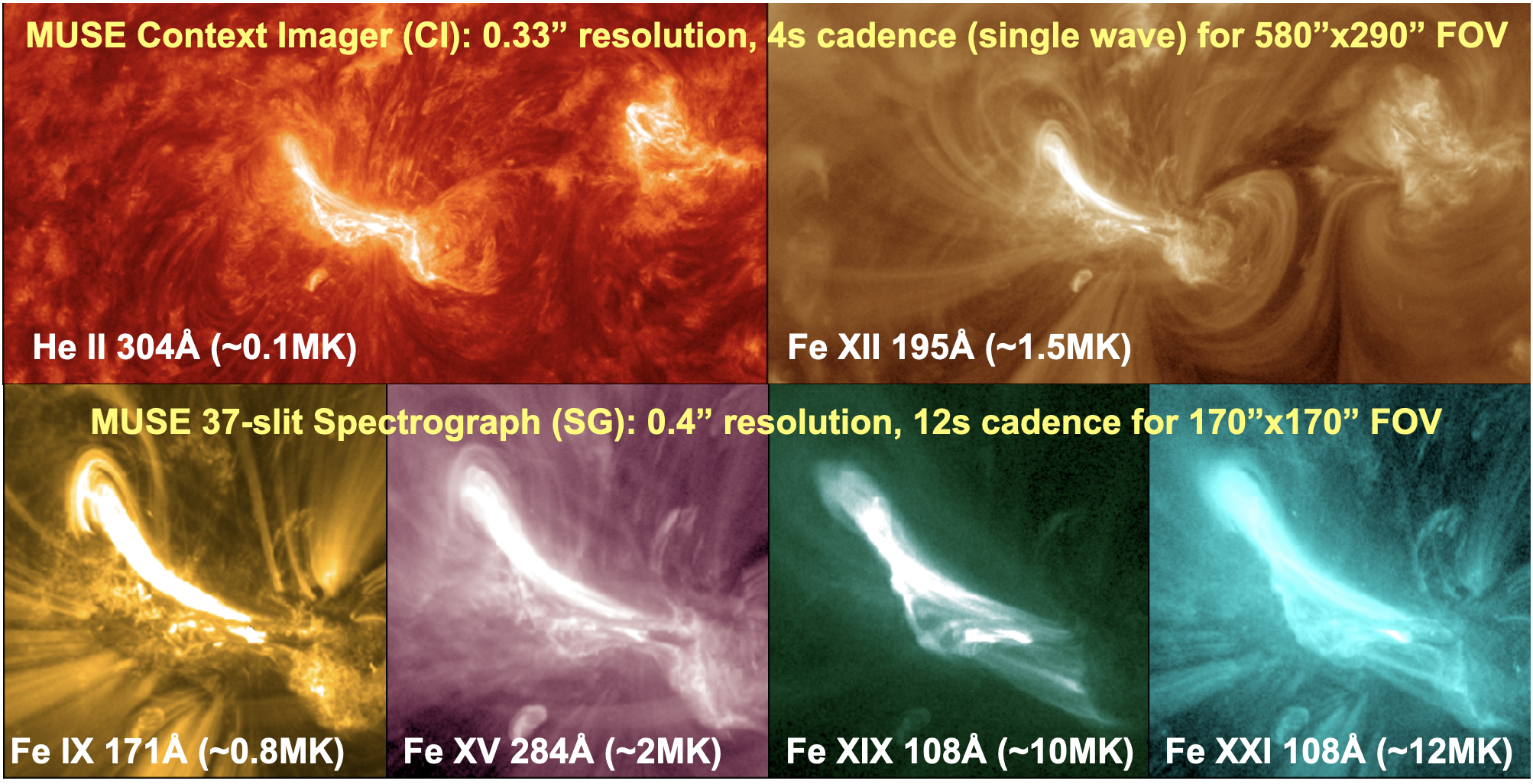}
\caption{Illustration of \muse\ EUV observations, including field-of-view and typical cadence, for the two instruments: the Context Imager (CI), and the 37-slit Spectrograph (SG).}\label{fig:MUSE_FOV}
\end{figure*}

\muse 's 37-slit extreme ultra-violet (EUV) spectrograph (\muse\ SG) operates at three wavelength bands (108, 171, 284 \AA), and will provide active region (AR) scale (field of view, FOV, is $170\arcsec \times 170\arcsec$) spectral rasters with a spatial resolution of 0.4\arcsec\ along the slits and at 0.4\arcsec\ spatial sampling across the slits, all at a cadence as fast as 12~s (see Fig.~\ref{fig:MUSE_FOV}).
Each parallel slit (along the y direction of the detector) produces its own two-dimensional spectral image on the detector, but offset from each other in the x-direction of the detector. By choosing isolated spectral lines, optimizing for the interslit spacing, and by using a compressed sensing method, it has been shown that the detector signal can be processed to retrieve physical properties (intensity, Doppler velocity, line broadening) simultaneously sampled by the 37 slits at the required accuracy (see \citealt{BDP:MUSE} and \citealt{Cheung:SDC} for a more detailed discussion). 

This innovative multi-slit design allows \muse\ to return AR-scale rasters at $\sim 30-100 \times$ the speed of existing EUV spectrographs, e.g., \sohos\ \citep{Wilhelm1995}, \hinodee\ \citep{Culhane2007}, and the upcoming Solar-C/\euvst\ \citep{Shimizu2019}. This allows \muse\ to effectively freeze the plasma dynamics in the corona and transition region (TR) over a field-of-view (FOV) that is the size of an active region, while delivering key spectroscopic information about the fundamental physical processes. In addition to the SG, the Context Imager (\muse\ CI) will provide 0.33\arcsec\ resolution narrowband images over a larger FOV (580\arcsec $\times$290\arcsec) in the 304 or 195 \AA~bands at 4~s cadence continuously. An even larger FOV of 580\arcsec $\times$580\arcsec can be obtained at 10~s cadence for a single passband. All of these observations can be supported indefinitely, enabled by the high data rate of \muse\ (21 Mbit~s$^{-1}$ continously). 

In this and a companion paper \citep{Cheung2021_muse_fl}, we highlight \muse's unique capabilities for addressing the mission's science goals. In addition, we consider the crucial contribution \muse\ would make towards realizing the science objectives of the Next Generation Solar Physics Mission (NGSPM). The following subsections provide a brief background on the NSGPM, the role of \muse\ in the Heliophysics Systems Observatory, and the current understanding of the processes driving coronal heating.

The remainder of the paper is structured as follows. 
In Section~\ref{sec:sim} we briefly describe the numerical models we use to predict \muse\ observables and diagnostics, further details are given in Appendix~\ref{app:sim}. In Section~\ref{sec:results}, we discuss the unique \muse\ contributions to addressing the NGSPM science objectives (listed in Table~\ref{table_so}), and present case studies highlighting the synergies with \euvst\ and  the 4m Daniel K. Inouye Telescope \citep[\dkist][]{Rimmele2020,Rast2021}.
In Section~\ref{sec:con} we summarize our results and draw our conclusions.

\begin{deluxetable*}{lccc}[!ht]
\tablecaption{\label{table_so} NGSPM Science Objectives and Corresponding Mission Science Objectives 
}
\tablehead{
\colhead{{\bf NGSPM Science Objectives \tablenotemark{a}}}  & \multicolumn{3}{l}{{\bf  Mission Science Objectives}} \\ \cline{2-4}
& \muse\ \tablenotemark{b} & \euvst\ \tablenotemark{c} & \dkist\ \tablenotemark{d}
}
\startdata
{\bf I. Formation mechanisms of the hot and dynamic outer solar atmosphere}  & & & \\ 
I.1 Understand the formation mechanism of chromospheric fine scale dynamic  & 1c & I-3-[1,2] & 5.1 \\
\hspace{1cm} structures and their influence in the corona  & & &  \\
I.2 Test the nanoflare heating hypothesis & 1b,3a & I-1-[1,2,3,4] &  5.2  \\
I.3 Test the wave-heating hypothesis & 1a,3b & I-2-[1,2,3] &  5.2 \\
I.4 Understand the role of magnetic flux emergence in the  heating of the & 1c &  &  5.5  \\
\hspace{1cm} chromosphere, transition region and corona &  &  &  \\
I.5 The sources and driving mechanisms of the  solar wind & 1a,1c & I-4-[1,2] & 5.2  \\
I.6 Formation mechanisms of solar prominences & 3b &  &  6.1 \\
\enddata
\tablenotetext{a}{https://hinode.nao.ac.jp/SOLAR-C/SOLAR-C/Documents/NGSPM\_report\_170731.pdf}
\tablenotetext{b}{\citet[][]{BDP:MUSE}}
\tablenotetext{c}{https://hinode.nao.ac.jp/SOLAR-C/SOLAR-C/Documents/2\_Concept\_study\_report\_part\_I.pdf}
\tablenotetext{d}{\dkist\ objective refers to Section number in \citet{Rast2021}.}
\end{deluxetable*}

\subsection{NGSPM}
\label{sec:ngspm}
The NGSPM is a mission concept developed by a panel of solar physics experts designated by JAXA, NASA, and ESA\footnote{https://hinode.nao.ac.jp/SOLAR-C/SOLAR-C/Documents/NGSPM\_report\_170731.pdf}. Following townhalls at international solar physics conferences and dozens of whitepaper submissions from the community, the Science Objectives Team (NGSPM SOT) developed NGSPM science objectives (SOs) based on several criteria, including the impact on solar physics and other disciplines and research fields,  the relevance to NASA/JAXA/ESA objectives, and the interest within the international solar physics community.
The NGSPM Science Objectives are generally focused on understanding how the interplay between plasma and magnetic fields causes heating in the solar atmosphere and leads to flares and coronal mass ejections. In Table~\ref{table_so} we list the NGSPM science objectives relevant to the coronal heating issue  addressed  in this  paper (i.e., "I. Formation mechanisms of  the hot and dynamic outer solar atmosphere"), together with the associated science objectives of the various NGSPM components. The companion paper \citep{Cheung2021_muse_fl} has an  analogous table listing the objectives relevant  to flares and eruptions (i.e., "II. Mechanisms of large-scale solar eruptions and foundations for prediction").

Flowing down from the SOs, the NGSPM report identifies a list of observational tasks, and the ``minimum set of instruments with which the NGSPM can address the greatest number of high-priority tasks consistent with the objectives of small length- and time-scale activity". The NGSPM report highlighted the need for simultaneous observations of the whole solar atmosphere from the photosphere to the hot corona. The suite of instruments identified by the NGSPM report as the most suitable to address the prioritized SOs are the following:
\begin{enumerate}
    \item 0.3\arcsec\ resolution coronal / transition region spectrograph,
    \item 0.2\arcsec-0.6\arcsec\ coronal imager, and
    \item 0.1\arcsec-0.3\arcsec\ resolution chromospheric/photospheric magnetograph/spectrograph
\end{enumerate}

The NGSPM report proceeds to recommend that the suite of instruments either be implemented on a single platform (e.g., the original Solar-C mission) or on multiple platforms as a ``distributed NGSPM mission". In this and a companion paper, we discuss the unique contributions of \muse, as both a high-resolution coronal/transition region spectrograph and coronal imager, to the NGSPM science goals, and how \muse, together with \euvst, and \dkist\ and other ground-based observatories (GBOs) more than matches the capabilities of a ``distributed NGSPM mission" to satisfactorily address the NGSPM SOs. At the time of writing,  \dkist\ is undergoing commissioning and \euvst\ has been selected for implementation by JAXA and NASA with a planned 2026 launch. \muse\ is currently in a Phase A study as a medium-class Heliophysics Explorer (MIDEX) mission, with a planned launch date of 2026 or 2028 (if selected).

\euvst\ is a traditional single-slit spectrograph covering various wavelength ranges in the EUV and FUV, enabling extensive thermal coverage from 20,000~K to 15~MK without significant gaps, allowing the tracing of plasma over a wide range of temperatures\footnote{https://solar-c.nao.ac.jp/en/}. It will obtain spectra along a single slit and achieve 0.4\arcsec\ resolution over the central 140\arcsec\ of the slit, and 0.8\arcsec\ resolution along the rest of the 280\arcsec\ long slit. \euvst\ is expected to achieve typical exposure times of order 1-5 seconds (depending on target), so that dense rasters over a FOV of $170\arcsec \times 170\arcsec$ at full resolution will take of order 7 to 35 minutes. \muse, on the other hand, will obtain similarly dense rasters (at selected temperatures) with a similar FOV but every 12s, i.e., of order 30-100$\times$ faster than \euvst. \euvst\ can obtain high raster cadences similar to those of \muse\ (e.g., 12s) but only over a very small field-of-view of order 4\arcsec\ $\times$ 140\arcsec, which is most often too small to capture loops, wave propagation, flares and CMEs. An advantage of the extensive line list of \euvst\ is that, over this small FOV (or a larger FOV at much lower cadence), it can trace the dynamic evolution, temperature, composition, and density of plasma from chromospheric to flaring temperatures. For sit-and-stare observations during flares, \euvst\ can achieve cadences of 200 ms. Similarly, during flares \muse\ can obtain sit-and-stare observations along half the length of its 37 slits at cadences of order 300 ms. \euvst\ will obtain slit-jaw images in  photospheric and chromospheric passbands (using a slitjaw imager like on \iris), and thus not obtain any coronal context, rendering interpretation of \euvst\ coronal and TR spectra very challenging, one of the reasons why the NGSPM report called for a coronal imager with similar resolution as the spectrograph. High-resolution coronal and transition region images from \muse\ will provide the necessary context. 

The latest generation of ground-based telescopes would not provide the extensive coronal diagnostics of \euvst\ or \muse\, nor the seeing-free, high-quality timeseries that space-based observatories can provide, but has a capability to provide spectropolarimetric and spectroscopic measurements of the photosphere and chromosphere, providing a wealth of information on the energetics and dynamics of magnetic field and plasma, at high ($< 0.2$\arcsec) spatial resolution and high cadence ($< 10$~s). Foremost among these observatories is \dkist, which can achieve an angular resolution as high as 0.03\arcsec.  \dkist\ is equipped with several instruments focused on spectropolarimetry, such as DL-NIRSP and ViSP. Such measurements, coupled with inversion approaches, provide information about the photospheric and chromospheric magnetic field. In addition, measurements of the magnetic field in the corona (off-limb) can be made with Cryo-NIRSP \citep{Kuhn2013} and COSMO \citep{Tomczyk2016} when the latter becomes available by 2026. High-resolution photospheric and chromospheric spectropolarimetry can also be obtained from other ground-based telescopes like the Swedish 1-m Solar Telescope \citep[SST, e.g., using CRISP and HeSP,][]{Scharmer2003}, the Goode Solar Telescope \citep[\gst, at Big Bear, e.g., using NIRIS and FISS]{Goode2013}, \gregor\ \citep{Schmidt2012}, and others. Further out into the future, the 4m European Solar Telescope \citep[\est,][]{Matthews2016} is expected to start around 2028, further enhancing the opportunity for coordinated observations of the magnetic field and lower atmospheric dynamics. 

\subsection{\muse\ {\rm x} \euvst\ synergies}
\label{sec:muse_euvst}

The complementary approach to spectroscopy (thermal coverage for \euvst, large FOV and high cadence for \muse) and imaging (photosphere and chromosphere for \euvst\, transition region and corona for \muse) means that \euvst\ and \muse\ will have major synergies. This synergy was recognized by the NGSPM report and will be further described in Section~\ref{sec:results}. The combination of \euvst\ and \muse\ observations will allow us to address the multi-scale and multi-thermal nature of critical physical processes in the solar atmosphere at the required high cadence. Energy is released and eruptions are triggered on small scales, often from driving processes in the lower atmosphere (which can be captured by \euvst\ and \dkist), and the energy release mechanism shows spatio-temporally coherent behavior ($\sim 10$~s and $\sim 400$~km) that \muse\ will resolve. The response of the atmosphere and magnetic field most often occurs on much larger spatial scales from jets (5-20~Mm) and active region loops (e.g., quiescent and flaring loops, 20-50~Mm) to active region-sized scales (e.g., CMEs, 50-150~Mm), which \muse’s extremely rapid imaging spectroscopy over an AR will be able to capture for the first time. 

\subsection{\muse\ in the Heliophysics Systems Observatory}

In this paper we focus on the unique contributions to NGSPM science objectives and the synergies with the missions highlighted in the NGSPM. However, there are also strong synergies with other missions that are part of the Heliophysics Systems Observatory, such as Parker Solar Probe \citep[\psp,][]{Velli2020}, \solo\ \citep{Mueller2020}, \punch\ \citep{DeForest2020}, Solar Dynamics Observatory \citep[\sdo,][]{Pesnell2012}, and the Interface Region Imaging Spectrograph \citep[\iris,][]{DePontieu2014}. 

MUSE’s high-resolution imaging and spectroscopy of the corona will fill a crucial gap in the capabilities of the Heliophysics System Observatory (HSO). The high-resolution spectroscopy of the corona will be a great complement to spectroscopy and imaging of the chromosphere and transition region with \iris, and the full-disk imaging at lower resolution by \sdo's Atmospheric Imaging Assembly \citep[\aia,][]{Lemen2012}. Similarly, \muse\ will be observing the source regions of the solar wind, measuring interchange reconnection, jets, and Alfv\'en waves, all of which are thought to feed into the wind, thus providing key information for in-situ measurements of the solar wind properties with \psp\ and \solo. \muse\ measurements of solar activity such as flares and coronal mass ejections will provide critical information of the upstream drivers of solar wind disturbances that are measured by \punch, \psp\ and \solo.

\muse\ also is very complementary to remote-sensing instruments onboard \solo: it will provide continuous high-resolution spectroscopy and imaging enabled by its very high sustained data rate which is more than 30$\times$ that of \iris, \euvst, and even higher multiplication factors for \solo.  In addition, coordinated observations of \muse\ and \solo\ offer the possibility of stereoscopic imaging or even spectroscopy during short time intervals when viewing angles from instruments like the Extreme Ultraviolet Imager \citep[\eui,][]{Rochus2020} or the Spectral Imaging of the Coronal Environment \citep[\spice,][]{Anderson2020} onboard \solo\ are favorable and when \solo\ is close to perihelion and obtains its highest resolution.

These exciting synergies will be described in more detail in a future publication.

\subsection{Outstanding challenges in understanding coronal heating}
\label{sec:muse_intro_coronal}

The detailed nature of the processes that power the corona and solar wind remains poorly constrained even though access to both new observations and numerical models has advanced the field rapidly the last ten years \citep{Carlsson2019,Reale2014,vanD2020,DePontieu2021}. We know that mechanical driving in the lower atmosphere transports sufficient energy flux to sustain the million-degree corona \citep{de-Pontieu:2007bd,Rast2021}, and that waves, currents and reconnection may carry or release substantial energy, but it remains unclear how important each is for the local energy balance, how this depends on the ambient environment (e.g., active regions, quiet Sun), and how the conversion of non-thermal to thermal energy works in detail. While previous missions like \hinode, \sdo, and \iris\ have shown tantalizing glimpses of how magneto-convective energy generated in the interior of the Sun drives solar activity and energizes the low solar atmosphere and corona, currently available observations lack coronal coverage and/or spatio-temporal resolution and are unable to arbitrate between competing theories of coronal heating \citep[e.g.,][]{HinodeReview2019,DePontieu2021}. These theories invoke processes like wave propagation, mode conversion, and dissipation through turbulence or resonant absorption, field-line braiding and magnetic reconnection, or non-thermal particle acceleration. As coronal heating models are beginning to be extrapolated to modeling the coronae of exoplanet host stars  \citep[][]{Alvarado-Gomez2016,Garraffo2017,Dong2018} to assess their habitability, it is timely to validate theories that work, and rule out those that do not.

\begin{deluxetable*}{llllcl}[!ht]
\tablecaption{\label{table_sims}Overview of Numerical Simulations}
\tablehead{
\colhead{Code}  &  \colhead{model} & \colhead{region} &\colhead{properties} & \colhead{NGSPM SO} & \colhead{Refs.\tablenotemark{a}}
}
\startdata
Bifrost  & \Bhion 
\tablenotemark{b} & quiet Sun network, braiding &3D MHD, non-eq.\ H & I.2-I.5 & [1] \\  
& \Bemer\ & network, braiding, flux emerg.\ &3D MHD & I.1,I.2,I.4 & [2] \\  
  & \Bgol\ \tablenotemark{b} & plage, spicules &  2.5D MHD, GOL & I.1-I.5 & [3,4] \\ 
  & \Bnw\ & network, braiding, flux emerg.\ &3D MHD & I.1,I.2,I.4 & [5] \\  
  & \Bnpdns\ & coronal hole, bright point  &  2D MHD & I.5 & [6] \\ \cline{1-6} 
MURaM   & \Mplhe\ &  plage, braiding, flux emergence &3D MHD & I.2-I.4 & [7,8] \\    \cline{1-6}
PLUTO  & \Ptwist\ & AR loops, braiding & 3D MHD & I.2 & [9] \\ 
  & \Pnano\ & loops interaction, braiding & 3D MHD & I.2 & [10]  \\ 
  & \Pwaves\ & AR loops, waves & 3D MHD, TWIKH & I.3 & [11]  \\ \cline{1-6}
CipMOCCT  & \Cwaves\ & AR loops, waves, impulsive driver & 3D MHD, TWIKH & I.3 & [12] \\ \cline{1-6}
Lare3d & \Ldc\ & AR loops, braiding & 3D MHD & I.2,I.3 & [13] \\ 
 & \Lac\ & AR loops, waves & 3D MHD & I.2,I.3 & [13] \\ \cline{1-6}
  & \Vawt\ & AR loops, waves & 3D RMHD, AWT & I.3 & [14] \\ \cline{1-6}
  & \Mac\ & QS loops, waves & 3D MHD & I.3 & [15] \\ \cline{1-6}
RADYN  & \Rcone, \Reone, \Retwo, \Rhone\ & AR loops, nanoflares & 1D HD, w/ NTE & I.2 & [16,17]\\ \cline{1-6} 
\enddata
\tablenotetext{}{GOL: Generalized Ohm's Law; non-eq. H: non-equilibrium Hydrogen ionization; TWIKH: Transverse Wave Induced Kelvin Helmholtz rolls; AWT: Alfv\'en wave turbulence; RMHD: reduced MHD; NTE: non-thermal electrons}
\tablenotetext{a}{References: [1] \cite{Carlsson:2016rt}; [2]
\cite{Hansteen2019}; [3] \cite{Martinez-Sykora:2017gol}; [4] \cite{Martinez-Sykora:2019hhegol}; [5] \cite{Hansteen_AGU2020}; [6] \cite{Nobrega-Siverio:2021inprep}; [7] \cite{Rempel:2017zl}; [8] \cite{Sanja_AGU}; [9] \cite{Reale2016a};  [10] 
\cite{Antolin2021}; [11] \cite{Karampelas2019b}; [12] \cite{Antolin_2019FrP.....7...85A}; [13] Howson  \& De Moortel (2021, in prep.);
[14] \cite{vanBallegooijen2017}; [15] \cite{Matsumoto2018}; [16] \cite{Polito2018}; [17] \cite{Testa2020a}.}
\tablenotetext{b}{Publicly available at http://sdc.uio.no/search/simulations.}
\end{deluxetable*}

Recent observations have confirmed the presence of many different types of waves throughout the corona that may carry a substantial amount of energy \citep{vanD2020}. Incompressible waves, e.g., Alfv$\acute{\rm e}$nic waves, are suspected to be important in heating the corona \citep{Matsumoto2016} and driving the solar wind \citep{McIntosh:2011fk}, but it remains unclear what role they play, as the limited spatio-temporal resolution of current observations leaves wave heating models poorly constrained \citep{Asgari-Targhi2014} with the exact wave energy content uncertain and direct observations of wave dissipation elusive \citep{McIntosh2012,Hahn2013,Antolin:2018fk}. Several competing numerical models of wave heating exist: (i) resonant absorption of kink mode waves and subsequent heating from wave dissipation through Kelvin-Helmholtz instability (KHI) vortices (first seen with \iris\ observations of prominence oscillations; \citealt{Okamoto2015,Antolin2015}), (ii) dissipation of \alfven\ waves, generated in the photosphere and propagating into the corona, through a turbulent cascade from interaction between counter-propagating waves (e.g., \citealt{vanBallegooijen2011}). However, current instrumentation cannot resolve the predicted signatures in the corona and cannot properly constrain the models. In both models the dissipation occurs on scales that are smaller than can be directly observed, but the predicted spatio-temporal intermittency, phase relations, amplitudes, and dominant periods are clearly different on scales of 0.5\arcsec and 20 seconds, as demonstrated in this paper. \muse\ will be able to determine whether such waves exist, and, if they do, discriminate between these models (Section~\ref{sec:waves}).

The dissipation of magnetic stresses in small-scale reconnection events (``nanoflares”) driven by braiding of field lines \citep{Parker:1988ys} is another major candidate process for coronal heating. Direct observations of this process have been rare: it is not clear whether the single case of organized, large-scale braided loops observed by \hic\ \citep{Cirtain:2013eu} and seen to lead to unwinding of the field and heating of the plasma is a common occurrence in the corona.
From advanced models of heating from braiding or reconnection, it is known that turbulent current sheets can spontaneously form down to scales well below the spatial scales accessible to remote sensing instruments. While \muse\ will not directly observe the smallest scales, turbulent reconnection simulations indicate that \muse\ has the spatiotemporal resolution to detect the dynamic response of the plasma to heating \citep{Hansteen:2015qv,Hansteen:2010uq,Pontin2017} and the resulting substructure of loops, and  to identify the presence and pervasiveness of turbulent current sheets in the solar corona. \muse\ observations will provide unprecedented observational constraints for braiding models. For example, recent models \citep{Hansteen:2015qv} predict that turbulent braiding can lead to outflow jets (v$\sim$100 km~s$^{-1}$). Such jets have now been detected in cool plasma with \iris\ \citep{Antolin2021}, but it is unclear how prevalent they are in the coronal volume.  Similarly, some models predict faint emission of transient high temperature plasma produced in response to heating events from braiding, for which observational evidence has been difficult to obtain, and it is not known how often such events occur \citep[e.g.,][]{Reale2009,Guarrasi2010,Testa2011,Miceli2012,Testa2012b,Brosius2014}. 
Braiding and resulting reconnection is also expected to provide clear signatures in the 1~MK upper TR of AR loops (so-called “moss”), which because of its small volume suffers less from superposition. Bidirectional reconnection outflows can cause line broadening, which \iris\ has provided tantalizing glimpses of in slow (30 min cadence, 60s exposure) rasters of moss \citep{Testa2016}, but firm evidence remains lacking because of observational limitations.  Observations of the upper TR moss also have revealed signatures of the deposition and thermalization  of non-thermal electrons generated by reconnection \citep{Testa2014,Testa2020a}, but at the currently observed resolution of the corona, it remains unclear how common such events are \citep{Graham2019}. 

A Poynting flux into the corona can be provided by both ``AC'' wave mechanisms, or ``DC'' braiding as convective motions push the sub-photospheric and photospheric field, but the emergence of a horizontal field into the outer atmosphere will also generate a Poynting flux and hence a possible source of coronal heating as newly emergent field interacts with the pre-existing ambient coronal field. Models of this interaction show that it will drive reconnection events where the reconnecting field lines can meet at large angles, giving rise to bursty brightenings, large flow velocities up to the \alfven\ speed, various types of waves, and large non-thermal line widths \citep[e.g.][]{Hansteen2019}. These events will be followed by the draining of mass brought up by the emerging field, visible in the chromospheric lines but also lines formed at coronal temperatures such as \feix. The clearest examples of these interactions are found in the vicinity of newly forming ARs, but the role of flux emergence in powering the AR corona in general is unclear. Similarly, recent studies of the H$\beta$ line at the SST have shown that flux emergence with associated (though weak) Ellerman bomb like reconnection events may be ubiquitous, even in the quiet Sun \citep{Joshi2020}. In the latter case it may be the field generated by the near surface local dynamo that is the source of the emerging field. Fairly weak fields do not easily emerge into the chromosphere and it is currently not clear how much energy such events deposit in the chromosphere \citep[e.g.,][]{Gosic2021} and corona.

Another source of high-angle reconnection could be cancellation events, where photospheric fields of opposite polarity are driven together by photospheric motions resulting in reconnection strong enough to generate nanoflares and thus coronal heating \citep{Priest2018,Chitta2018}.

Recent advanced models encompassing the whole solar atmosphere, including important processes such as ion-neutral coupling, also predict heating to coronal temperatures associated with chromospheric spicules \citep{Martinez-Sykora:2018gf,DePontieu2017}. Such heating has been suggested based on imaging observations \citep{DePontieu2017,De-Pontieu:2009fk,De-Pontieu:2011lr} but contested using lower resolution spectra \citep{Madjarska:2011fk}. The heating associated with such jets is predicted to be caused by a complex sequence of events including dissipation of electrical currents \citep{DePontieu2017} and Alfv\'enic waves \citep{Antolin2018}. Current observations are not adequate to settle this issue.

Some of these processes may also play a role in providing energy to the solar wind. For example, spicules and other jets (e.g., coronal jets) occur at the roots of the fast solar wind. Higher resolution observations are key to address how such jets are formed, test models for jet generation \citep{DePontieu2017,Sterling2015}, and their role in the solar wind \citep{Cirtain2007,Wang2020}. It is known that an extra acceleration beyond the critical point is needed to achieve the high velocities measured \citep{Hansteen2012} in the fast solar wind at 1~AU. The source of this additional acceleration has long been thought to lie in Alfv\'en waves, but the properties, propagation, and dissipation of these waves have remained elusive. Alfv\'enic waves have been seen at the roots of the corona along spicules and coronal jets at high amplitudes \citep{de-Pontieu:2007bd,Cirtain2007} but coronal spectroscopic observations that could elucidate the propagation, damping, and dissipation of such waves have been scarce or at low-resolution \citep{Tomczyk:2007vn, Hahn2013}. High-resolution observations are required to provide insight into this important energy source at the roots of the solar wind.

Such observations will also be able to constrain the driver and intermittency of “AR outflows” that feed into the slow solar wind \citep{Bryans:2010lr}; and the role of opening of closed loops through reconnection in driving the slow 
wind \citep{Tu2005}.  

\begin{figure*}[!ht]
    \centering
    \includegraphics[width=1\textwidth]{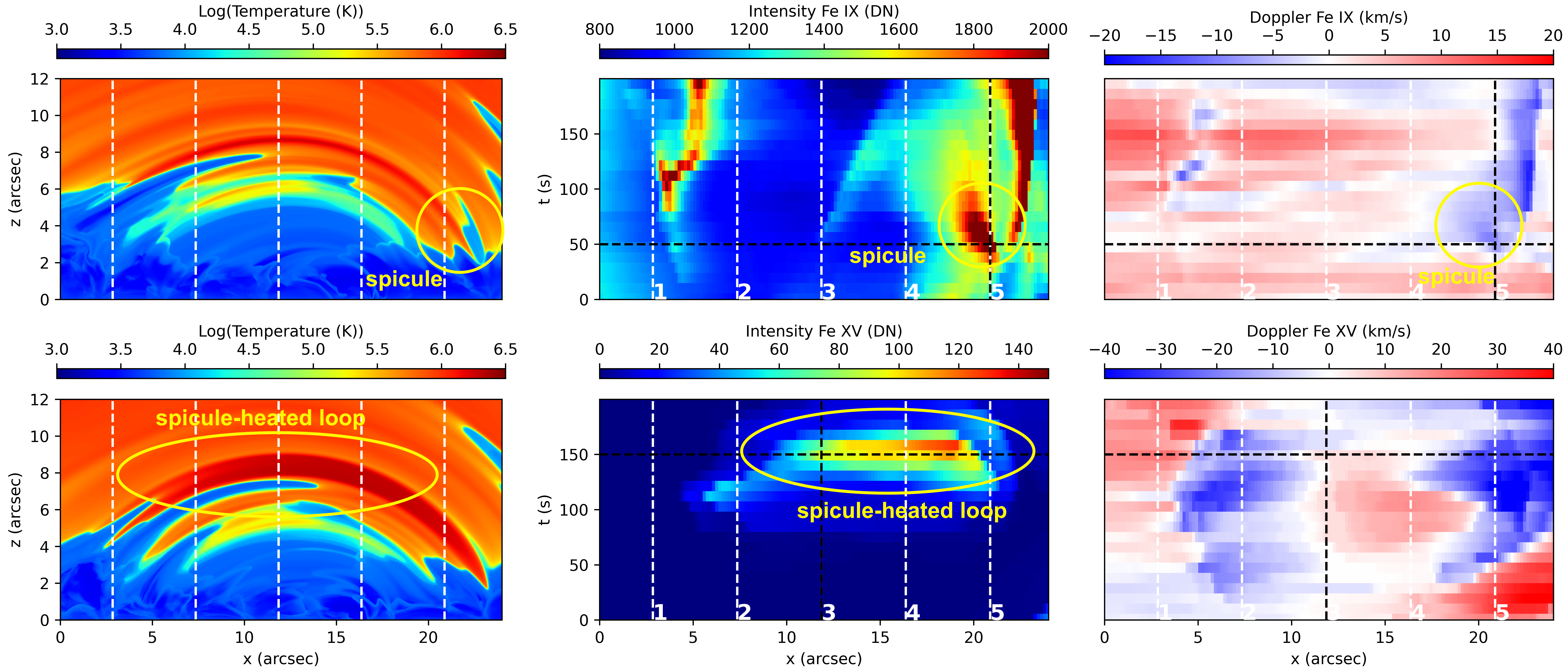}
	\caption{\label{fig:spicmaps} 
    Advanced models predict that coronal heating associated with chromospheric spicules is fundamentally a multi-scale process that cannot be tracked without \muse's multi-slit spectroscopy at high cadence and large field-of-view. Here we show the temperature structure of model \Bgol\ (see Appendix~\ref{app:sim} for details) and synthetic \muse\ \feixw\ and \fexvw\ intensity (middle panel) and Doppler shift (right panel) for two times 100~s apart. The initial spicule impact is detectable as a small-scale blueshifted brightening in \feix, while dissipation of spicule-related currents and waves occurs quickly thereafter, forming a loop (through evaporation, as evident in the  blue-shifted \fexv\ emission at the footpoints, in the bottom right panel) at large distances, offset from the spicule location. The \muse\ intensities are calculated by assuming a time integration of 10 and 20s for \feix\ and \fexv. Count rates are calculated as described in Appendix~\ref{app:synthesis}.
	}
\end{figure*}

\section{Numerical simulations}\label{sec:sim}

Advanced numerical modeling is of crucial importance for the interpretation of solar observations. In this paper we show how comparisons between \muse\ observables and predictions from models will provide stringent tests of state-of-the-art models, and will allow to distinguish between competing models.
The \muse\ science team includes modeling experts with access to several advanced numerical models, able to simulate a broad range of solar features (from spicules, to warm and hot loops, to coronal jets) and with a variety of physical mechanisms leading to atmospheric heating (from magnetic braiding to \alfven\ waves to flux emergence). 
We exploit this large set of numerical models to highlight novel \muse\ diagnostics that will lead to significant breakthroughs in our understanding of the processes leading to the formation and energization of the solar corona. 
Here we focus on the simulations used to model the non-flaring solar corona, while additional models of flares and coronal mass ejections (CMEs) are used in the companion paper \citep{Cheung2021_muse_fl}.

The range of models we use includes self-consistent 3D and 2.5D radiative MHD models, such as the ones using the Bifrost \citep{Gudiksen:2011qy} and MURaM codes \citep{Rempel:2014sf,Rempel:2017zl}, as well as 3D MHD models of loops (e.g., \citealt{Reale2016a,Karampelas2019a}), and 1D HD models computed using the RADYN code (e.g., \citealt{Carlsson:1992kl,Allred2005,Testa2014,Polito2018,Testa2020a}). Some of the latter models, although idealized, are more flexible and well suited to study some specific physical processes not easily modeled or isolated in the more complex atmospheric models. 

\begin{figure*}[!ht]
    \centering
    \includegraphics[width=1\textwidth]{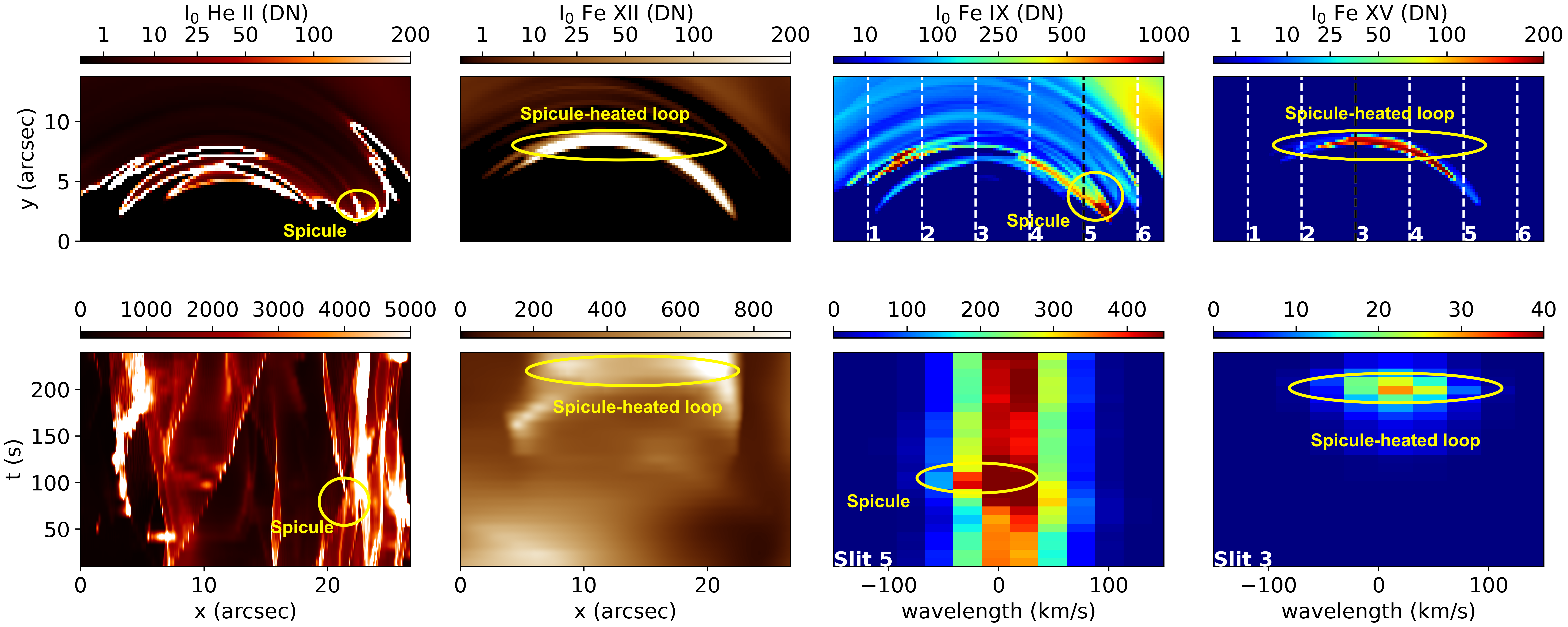}
	\caption{
	\label{fig:spicmom} 
	\muse\ synthetic observables for a 2.5D Bifrost model (\Bgol; see Table~\ref{table_sims}, Appendix~\ref{app:sim}, and Figure~\ref{fig:spicmaps}) producing spicules and associated coronal heating. 
	Top row shows side views (i.e., as seen at limb) of the intensity at t=240s , from left to right, in: the two \muse\ CI passbands (\heiiw, and \fexiiw), the \muse\ SG lines (\feixw, and \fexvw).
	Second row shows the top view (i.e., as seen on disk) of the space-time intensity maps in the two \muse\ CI passbands (\heiiw, and \fexiiw) in the left two panels. The right panels show the temporal evolution of the \muse\ SG \feix\ and \fexv\ spectra along slit 5 (at the footpoint, where the spicule occurs) and slit 3 (near loop top) respectively (see Figure~\ref{fig:spicmaps}.
    Time integration is 10~s for all \feix\ and for the limb view of \fexv , and 20s for  \fexv\ in the bottom-right panel.
    Count rates are calculated as described in Appendix~\ref{app:synthesis}. The associated animation shows the time evolution of the temperature (top left) and velocity (bottom left) of the simulation, including the side view of the intensities shown in the top row of this figure (right four panels). Animations of this figure can be found online.
}
\end{figure*}

In Table~\ref{table_sims} we provide a concise overview of the models used here to synthesize \muse\ observables and showcase the diagnostic potential of \muse. 
A description of the properties of the numerical models, and codes, including reference to previous papers that have analyzed them in detail, are provided in Appendix~\ref{app:sim}.

\section{Unique MUSE contributions to NGSPM science objectives}\label{sec:results}

\subsection{Impact of chromospheric fine scale dynamic structures on the formation of the corona}~\label{sec:spic}

The most common chromospheric fine-scale dynamic structures that protrude into the corona are chromospheric spicules. Some theoretical models predict that significant heating to coronal temperatures occurs in association with spicules, e.g., through reconnection and/or the dissipation of electrical currents \citep{Martinez-Sykora:2017sci} or transverse waves (e.g., from resonant absorption and resulting Kelvin-Helmholtz instabilities, \citealt{Antolin:2018fk}). All of these processes are predicted to occur on short time-scales of the order of 20~s, with the heating occurring over a wide range of spatial scales, typically offset by up to tens of arcseconds from the injection of spicular material (Figures~\ref{fig:spicmaps} and \ref{fig:spicmom} and the associated animation).  

These predictions highlight the desire for a comprehensive NGSPM approach where each instrument provides important observations to address whether spicules play a significant role in the heating of the corona. \dkist\ (or other GBOs) will provide measurements of the magnetic field in the photosphere (ViSP scans covering a few arcsec at $\sim 1$~min cadence) and chromosphere (DL-NIRSP scans of $30 \times 30$ arcsec at $\sim 30$~s cadence) to understand the magnetic field conditions that lead to spicule formation, while DL-NIRSP and VTF  scans through chromospheric lines will provide timing and location of spicular upflows. Such measurements will also provide insight into the magnetic and shock waves expected during spicule formation. 

The thermal evolution of the spicular plasma from chromospheric to coronal temperatures can be studied using rapid \euvst\ rasters covering 2\arcsec\ $\times 140$\arcsec\ at 10~s cadence. While these rasters will seamlessly cover all temperatures from 20,000 to several million degrees at high cadence, their small FOV will make it difficult to track the impact of spicules on the transition region and corona without context imaging at TR (e.g., \heiiw\, Fig.~\ref{fig:spicmom} and associated animation) and coronal temperatures. In addition, such small rasters can only track the spicular signal close to the loop footpoints, missing the dissipation of currents and \alfven\ waves that leads to heating along the associated coronal loop \citep{Martinez-Sykora:2017sci,De-Pontieu:2017pcd}. Models predict this to occur quickly after spicular injection, but up to tens of arcseconds away from the footpoints as electrical currents and waves propagate away from the spicule and are dissipated. In Fig.~\ref{fig:spicmaps} (top row) and the animation associated with Fig.~\ref{fig:spicmom}, the spicular upflows are visible as short-lived brightenings and associated strong upflows in \feixw. The coronal loops that form in response to spicules are relatively short (30\arcsec) in this simulation because of the imposed magnetic field geometry. However, in the solar atmosphere such loops will typically be much longer (20 to 100 Mm). \euvst\ context images only cover photospheric and chromospheric temperatures and thus cannot capture the full coronal response. Rapid \muse\ rasters covering $170\arcsec \times 170\arcsec$ at $\sim20$~s cadence, combined with TR and coronal images, will capture the coronal loops (Fig.~\ref{fig:spicmaps}, bottom row) associated with the spicular injection \citep{De-Pontieu:2017pcd} at the required high resolution.

\begin{figure*}[h!]
    \centering
    \includegraphics[width=0.99\textwidth]{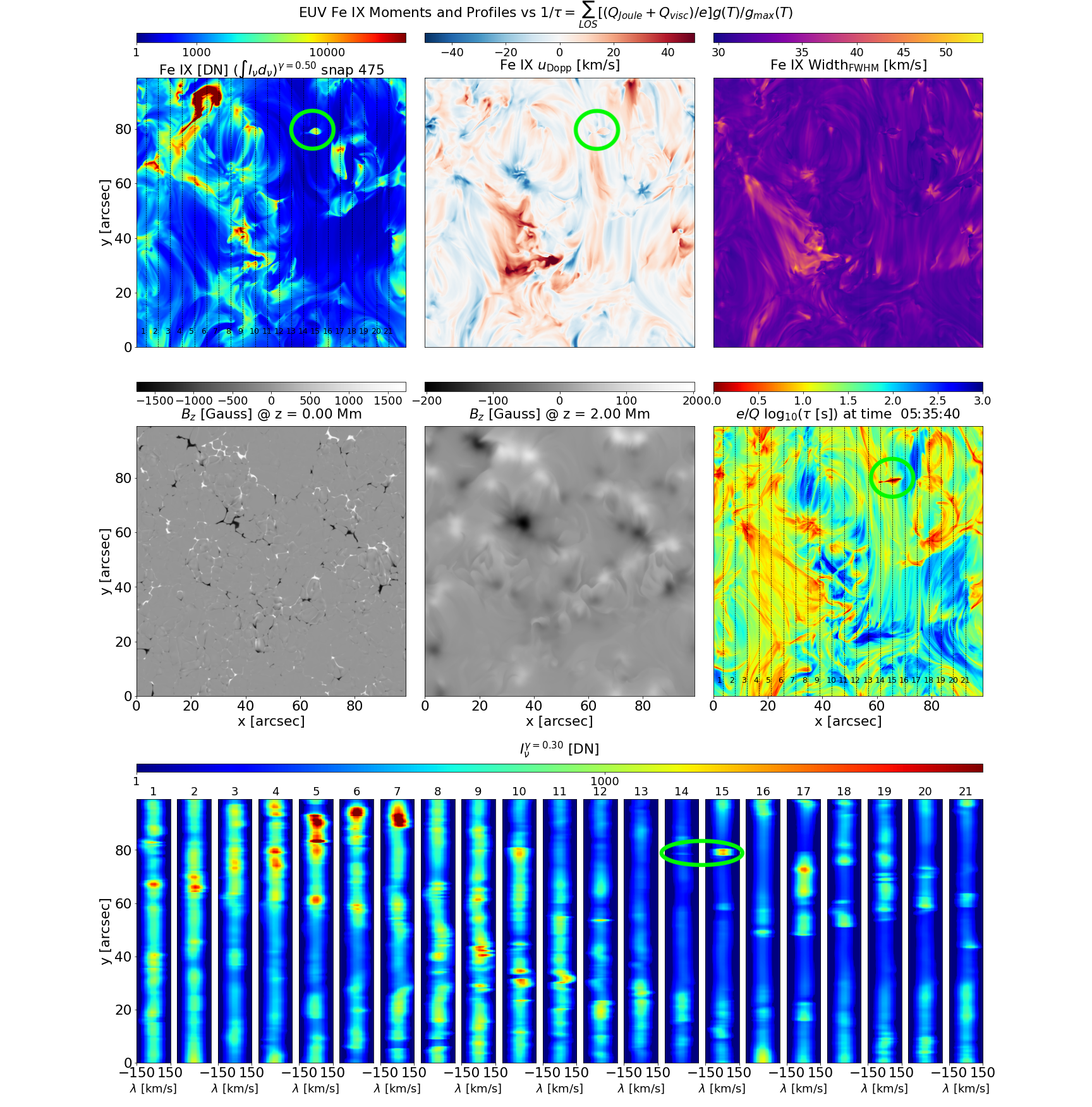}
	\caption{\label{fig:viggo1} 
    \muse\ synthetic observables, from Bifrost 3D MHD model \Bnw\ (see Table~\ref{table_sims} and Appendix~\ref{app:sim}), diagnosing heating through braiding. The upper three panels show the intensity, line shift, line width of the \fexvw\ line, while the second row of panels show the photospheric field, the field 2~Mm above the photosphere, and a measure of the heating rate in the temperature range where this line is formed: $(Q_{\rm Joule}+Q_{\rm visc})/e\times G(T)/G_{max}(T)$, filtered over the normalized $G(T)$ for the line in question and summed over the line of sight. The  upper left and lower right panels also show the potential location of 21 of the 37 \muse\ slits for a ``sit and stare'' observation: The bottom row of panels show the line profiles of the \fexv\ line at these locations. 
	Braiding events occur throughout the simulated region as the magnetic field is driven by photospheric motions.  Particularly strong heating is found in a small region near the top of the domain where $x\sim [20,40]$ and $y\sim [85,95]$ and along a diagonal that stretches from $(x,y)\approx (15\arcsec,60\arcsec)$ to $(x,y)\approx (40\arcsec,50\arcsec)$ which outlines a long current sheet that shows continual episodic heating events. Note that the \fexv\ line width responds strongly to these heating events all along the current sheet. Nano-jets caused by reconnection resulting from braiding can also be found throughout this simulation. An example is shown in the green circles in the two left panels of the top row, and the right panel of the middle row.
	The assumed exposure time is 1.5~s. Count rates are calculated as described in Appendix~\ref{app:synthesis}. Animations of this figure and the \fexv\ spectral band can be found online.
	}
\end{figure*}

\muse\ \feixw\ spectra at the loop footpoints (3rd panel of bottom row of Fig.~\ref{fig:spicmom}, and 3rd column of Fig.~\ref{fig:spicmaps}) will show strong blueward excursions when spicules are initially launched, while a statistical analysis of Doppler shifts and broadening can be used to determine whether there is a significant coronal response along the loops (e.g., evaporative upflows, bottom left panel of Fig.~\ref{fig:spicmaps}, right column of Fig.~\ref{fig:spicmaps} and the animation associated with Fig.~\ref{fig:spicmom}). In addition, if there is a coronal loop forming, the spatio-temporal evolution of intensity, Doppler shift, and broadening can be used to determine whether dissipation of magnetic waves (as deduced from phase relationships between the plane-of-the-sky, POS, and line-of-sight, LOS, oscillations; \citealt{Antolin:2018fk}), shock waves (as deduced from propagation speed, \citealt{Skogsrud2016}) or electrical current dissipation (as deduced from propagation speed,  \citealt{Martinez-Sykora:2017sci}) is the dominant mechanism for the coronal heating. \muse\ images in \heiiw\ will show the TR counterparts of chromospheric spicules (left column, Fig.~\ref{fig:spicmom}), while the \fexiiw\ images will show both the formation of the hot loop through evaporation and the subsequent cooling, as shown towards the end of the time range of Fig.~\ref{fig:spicmom} (second column). 

By providing spectroscopy over the whole length of coronal loops, including both footpoints, \muse\ will be able to capture the full history of loops, including preceding spicular activity and resulting current or wave dissipation, as well as subsequent cooling. In this fashion, \muse\ will  provide the missing link to firmly establish whether any associated heating along coronal loops can statistically be tied to spicular injection, and determine its importance in the global energy balance of the corona.

\subsection{The nanoflare heating hypothesis}\label{sec:nano}

Nanoflares are small-scale impulsive events in which energy ($10^{24}$-$10^{26}$~erg) is released as a result of magnetic reconnection caused by braiding. Models for nanoflares predict clear observables that can be used to not only test the model accuracy, but also to constrain the properties of the energy release. The rapid reconfiguration involved in braiding events and resulting plasma response are expected to occur on short timescales of the order of 20~s and a wide range of spatial scales, from 0.5\arcsec\ to several tens of arcseconds as reconnection leads to strong plasma outflows in the form of nanojets \citep{Antolin2021}, loops are heated and filled with plasma, and conjugate loop footpoints respond to the energy release \citep{Testa2014,Testa2020a}. 

\begin{figure*}[!ht]
    \centering
    \includegraphics[width=0.99\textwidth]{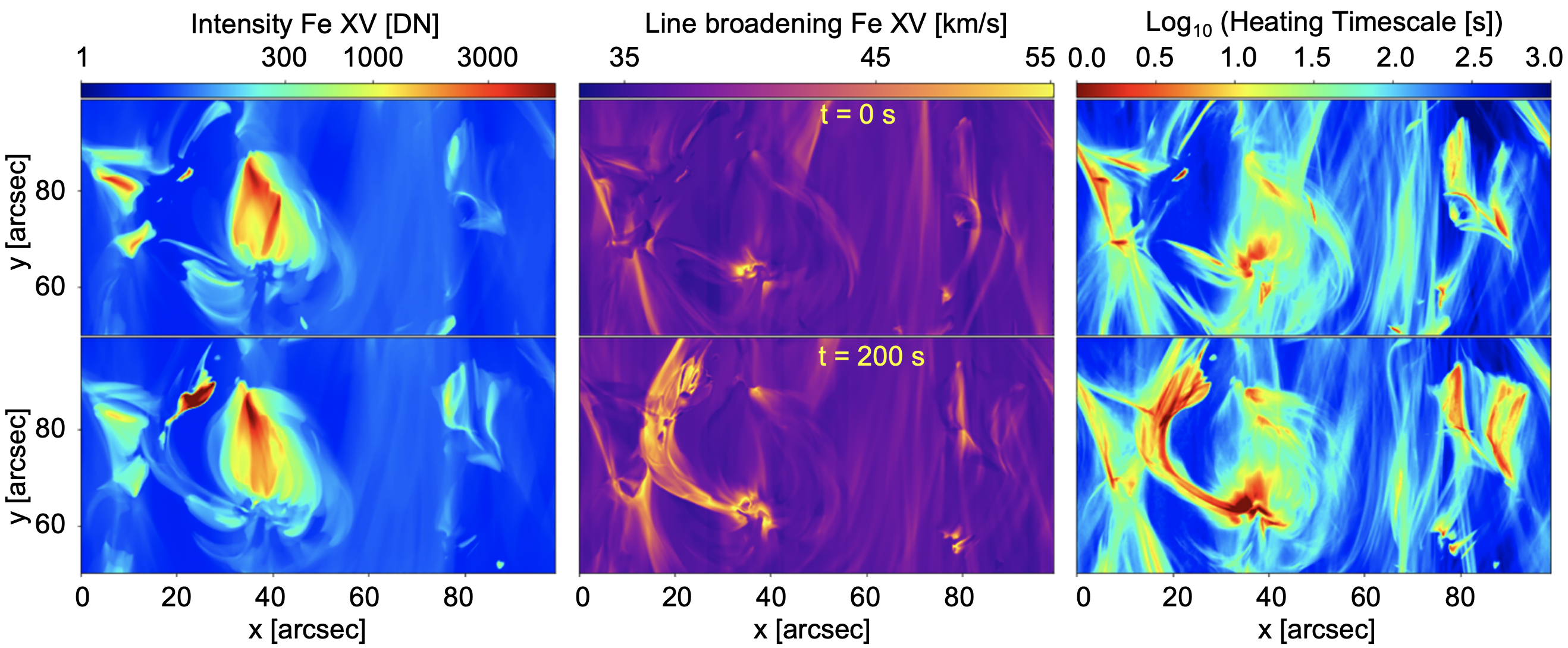}
	\caption{\label{fig:viggo2_csr}  Models predict that coronal heating associated with braiding and flux emergence is highly dynamic and shows spatial coherence on a wide range of scales (from 0.5-40\arcsec) leaving tell-tale signatures that cannot be tracked without \muse’s multi-slit spectroscopy and imaging at high cadence and large FOV. Here we show an example from model \Bnw, in which current dissipation leads to drastic changes in heating rate (right) which, in turn, causes rapid brightenings (left) and turbulent motions (middle) that can be captured with \muse\ e.g., in the \fexvw\ line shown here.
 	The assumed exposure time is 1.5~s. Count rates are calculated as described in Appendix~\ref{app:synthesis}.}
\end{figure*}

Both theoretical predictions from models and recent observations indicate that a combined NGSPM approach is desired to determine the properties of these events and their importance for the coronal energy balance. Measurements of the evolution of the magnetic field configuration (e.g., using field extrapolations based on \dkist\ ViSP or DL-NIRSP measurements) are important to understand the overall topology. This will help determine to what extent the events are caused by large-angle reconnection (e.g., resulting from flux emergence,  see Section~\ref{sec:flux_emergence}) or more gentle braiding events.

\subsubsection{Spatio-temporal correlations related to braiding}
In numerical models heated by braiding (see Figure~\ref{fig:viggo1}, and, e.g., \citealt{Hansteen:2010uq,Hansteen:2015qv}) reconnection occurs between field lines that are nearly parallel to each other as they are jostled by granular and supergranular motions in the photosphere and below where the field lines are tied. Current sheets form and are dissipated, which leads to heating of plasma.

\begin{figure*}[!ht]
    \centering
    \includegraphics[width=0.95\textwidth]{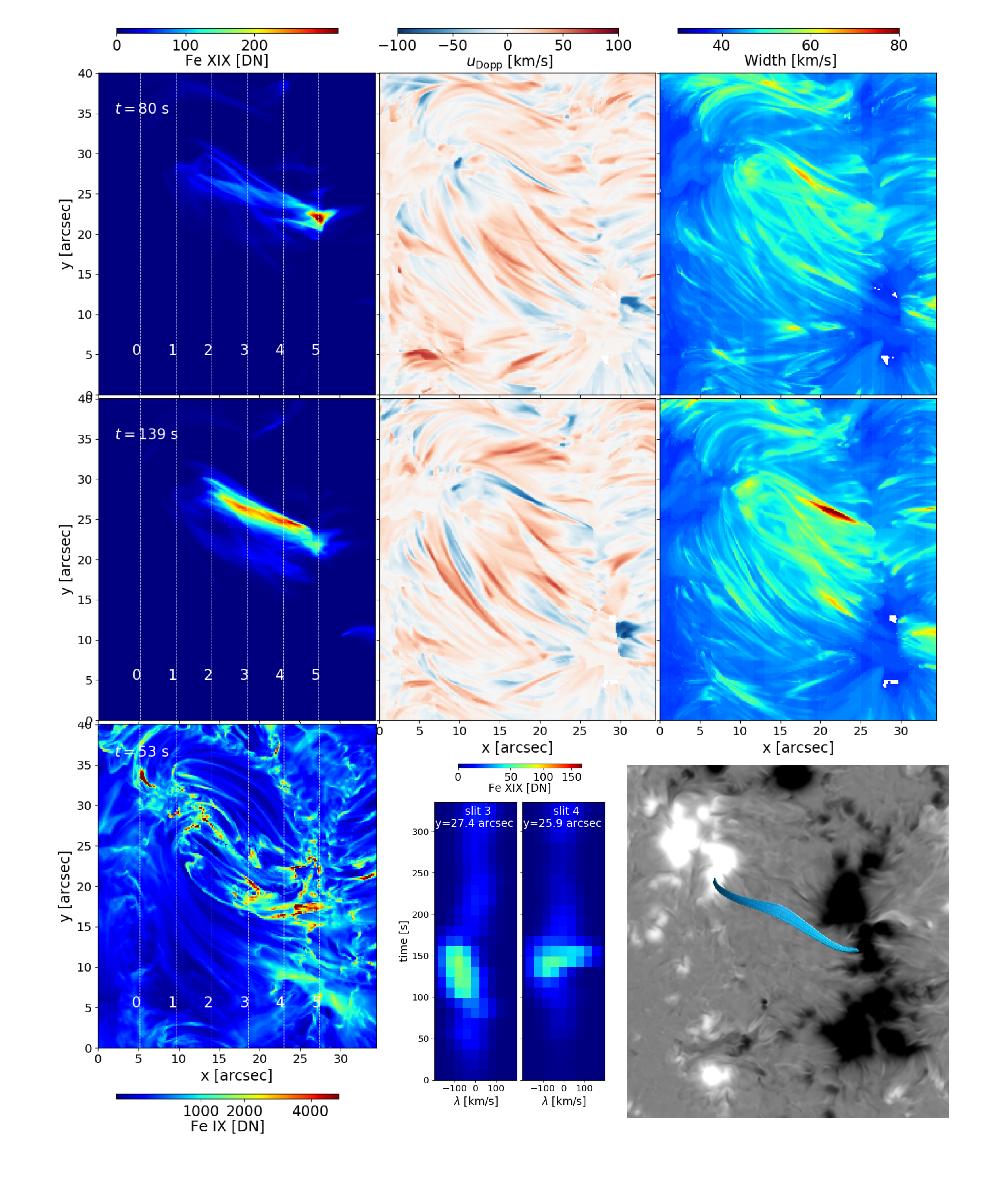}
	\caption{\label{fig:sanja1} \muse\ emission for AR core hot loops  in non-flaring conditions (from model \Mplhe; see Table~\ref{table_sims}, and Appendix~\ref{app:sim}). Multiple slit configurations can simultaneously sample the footpoint brightenings in TR lines caused by the short-lived heating events (see also Figure~\ref{fig:radyn} for the corresponding 1D RADYN simulations) and the flows and heating in the coronal portion of the loops.  
	Synthetic \muse\ observations of \fexixw\ at two different  times (top two rows show \fexix\ moments), reveal rapidly evolving hot loops, and the \feixw\ intensity map (bottom left panel) shows brightenings at the loop footpoints at the beginning of the heating event. We also provide two examples of \muse\ \fexix\ spectra as a function of time (bottom middle panels) showing the dynamic nature of the \fexix\ emission. The bottom right panel displays the vertical magnetic field in the chromosphere (black and white plate) with overplotted magnetic field lines of the hot loop shown in blue. 
	We assume an exposure time of 6~s for \fexix\ and 1~s for \feix. Count rates are calculated as described in Appendix~\ref{app:synthesis}. 
	}
\end{figure*}

To trace the full thermal evolution of plasma in coronal loops or loop footpoints, \euvst\ sit-and-stare observations or rasters, both at high cadence ($\sim 10$-20~s), can be used. Given the cadence requirement, such rasters will necessarily be limited in FOV to $< 5$\arcsec$  \times 140$\arcsec, capturing only a single loop footpoint region or small parts of a loop, not both footpoints at the same time, nor the evolution of the full loop.  \euvst\ data of this type, when fortuitously targeted, can help discriminate the evaporative response (to heating events) from plasma cooling into a passband, determine plasma densities, or study the highly variable emission in very hot coronal lines ($>5$-$10$~MK) that is expected to occur in loops with low-frequency nanoflares (see e.g., \citealt{Reale2019a,Reale2019b,Testa2020a,Testa2020b}, and Figure~\ref{fig:sanja1}). However, the small FOV of the high-cadence \euvst\ rasters and lack of coronal context will in many cases render interpretation ambiguous or impossible without \muse’s simultaneous high-resolution (0.5\arcsec) and high-cadence ($\sim 10$~s) coronal images and spectra over the full length of the coronal loop. For example, high-cadence \muse\ CI 195\AA\ images will reveal whether events caught under the \euvst\ slit are the result of loops that are visibly braided (as seen with \hic; \citealt{Cirtain:2013eu}) or whether the field is relaxed as braiding continuously occurs \citep[as some models predict, e.g.,][]{Hansteen:2015qv}, which remains an open question.

\begin{figure*}[!ht]
    \centering
    \includegraphics[width=0.99\textwidth]{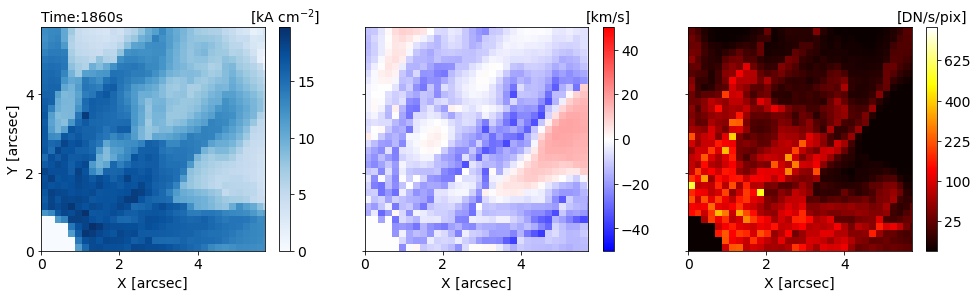}
	\caption{\label{fig:fabio} Diagnostics of heating by braiding in a 3D MHD model of twisted loops. Horizontal cross-section at one footpoint of a twisted coronal loop (model \Ptwist; see Table.~\ref{table_sims}, and Appendix~\ref{app:sim}). We show the current density from the model (left), as well as the synthetic \muse\ \feixw\  Doppler velocity (middle), and intensity (right), assuming a 1~s integration time. The line emission is integrated over a layer 2000~km thick above the transition region. Count rates are calculated as described in Appendix~\ref{app:synthesis}.}
	\label{fig:rea16a}
\end{figure*}

Such braiding models predict that the current sheets are highly filamentary but show collective behavior on timescales of the order of 20~s from the sub-granular ($\sim 0.5$\arcsec) to the supergranular scale ($\sim 10$-40\arcsec). As shown in Fig.~\ref{fig:viggo1}, heating and resulting flows associated with such current dissipation (right panel) are predicted to cause strong spatially correlated changes in the 0th (e.g., strong heating), 1st (e.g., rapid outflows, nanojets, plasmoids) and 2nd (e.g., from turbulence or plasmoids related to reconnection) moment of coronal spectral lines across scales from 0.5 to 40\arcsec. Given the rapid evolution and large spatial scales, \muse\ observations are key to test this prediction of braiding models. This is illustrated in Fig.~\ref{fig:viggo2_csr} which shows that the heating in these models rapidly changes on timescales of $\sim$60~s and spatial scales of at least 40\arcsec. In addition, the current dissipation and associated heating leads to strong brightenings and broadening of the \muse\ spectral lines (\fexvw\ in this example).

More generally, large FOV, high-resolution \muse\ CI images are needed to determine whether the events under the \euvst\ slit are local in nature or in response to events outside of the \euvst\ raster. For example, \muse\ images and spectra can capture the spatial distribution and temporal evolution of heating events and resulting flows and turbulence all along loops on timescales of 10-20~s and thereby determine whether these events are caused locally by flux cancellation (e.g., as deduced from magnetograms from GBOs) and whether such cancellation is a major driver for coronal heating, as recently proposed \citep{Chitta2018,Chitta2019,Chitta2020}. Coronal imaging, at the same resolution as \euvst\ or the \muse\ spectra, and at high cadence, is key for this purpose.

\begin{figure*}[!ht]
     \centering
    \includegraphics[width=0.99\textwidth]{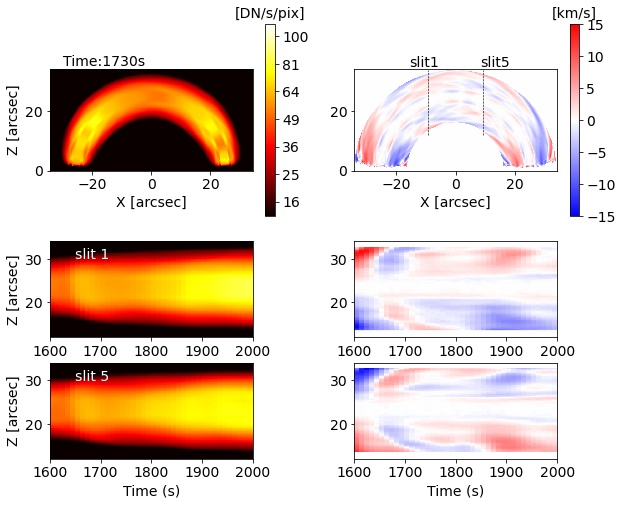}
	\caption{\label{fig:fabio2} 
	The Doppler velocity patterns across and along the loop for a twisted coronal loop model (\Ptwist; see Table~\ref{table_sims}, and Appendix~\ref{app:sim}) change rapidly and need high-cadence spectroscopy over a large FOV to be detected.
	Top panels show the side view of \muse\ \fexvw\ intensity and Doppler velocity maps. The bottom panels show the  temporal evolution of the \muse\ \fexv\ intensity (left)  and Doppler shift (right) across two slits $\sim 20$\arcsec\ apart. As in Figure~\ref{fig:fabio} we show the \muse\ observables assuming $0.167$\arcsec\ pixels. Count rates are calculated as described in Appendix~\ref{app:synthesis}.
	}
	\label{fig:rea16b}
\end{figure*}

Some braiding models suggest that twisting of loops can lead to heating (e.g., \citealt{Reale2016a}). In Figures~\ref{fig:fabio} and ~\ref{fig:fabio2} we show two more examples of how \muse\ spectral observations of the plasma emission and dynamics over the full length of coronal loops, at high spatial and temporal resolution, constrain the heating properties. We use a 3D MHD model of coronal loops heated by reconnection caused by the twisting that is triggered by the photospheric plasma flow patterns (model \Ptwist; see Section~\ref{sec:sim} and Appendix~\ref{app:sim}).
One prediction of such models is that short-lived ($\sim 20$-60~s) currents in the lower atmosphere (visible as heating patterns in chromospheric images from, e.g., \dkist) should correlate well with heating and turbulence at the upper TR footpoints of coronal loops (“moss”), which can be determined from \muse\ spectroscopic observations. In Figure~\ref{fig:fabio} the maps of current density from the model are compared to the Doppler map (component along $z$) and the intensity in the \muse\ \feixw\ line, at the loop footpoint. At the time shown, the current density is above the threshold for dissipation in most of the layer, and shows fine-scale structuring due to the irregular twist pattern. The related heating has triggered the evaporation of dense plasma from the chromosphere upwards to fill the coronal part of the loop. This is clearly tracked by the blue patterns in the Doppler map and by the bright structures in the intensity maps, which closely resemble the current patterns. Clearly, \muse\ \feixw\ observations of upper TR moss will be able to determine whether such correlations between intensity and Doppler shifts occur, and whether they are associated with heating in the chromosphere (e.g., through comparison with \dkist\ measurements of plasma temperature in the upper chromosphere).

\begin{figure*}[!ht]
    \centering
    \includegraphics[width=0.95\textwidth]{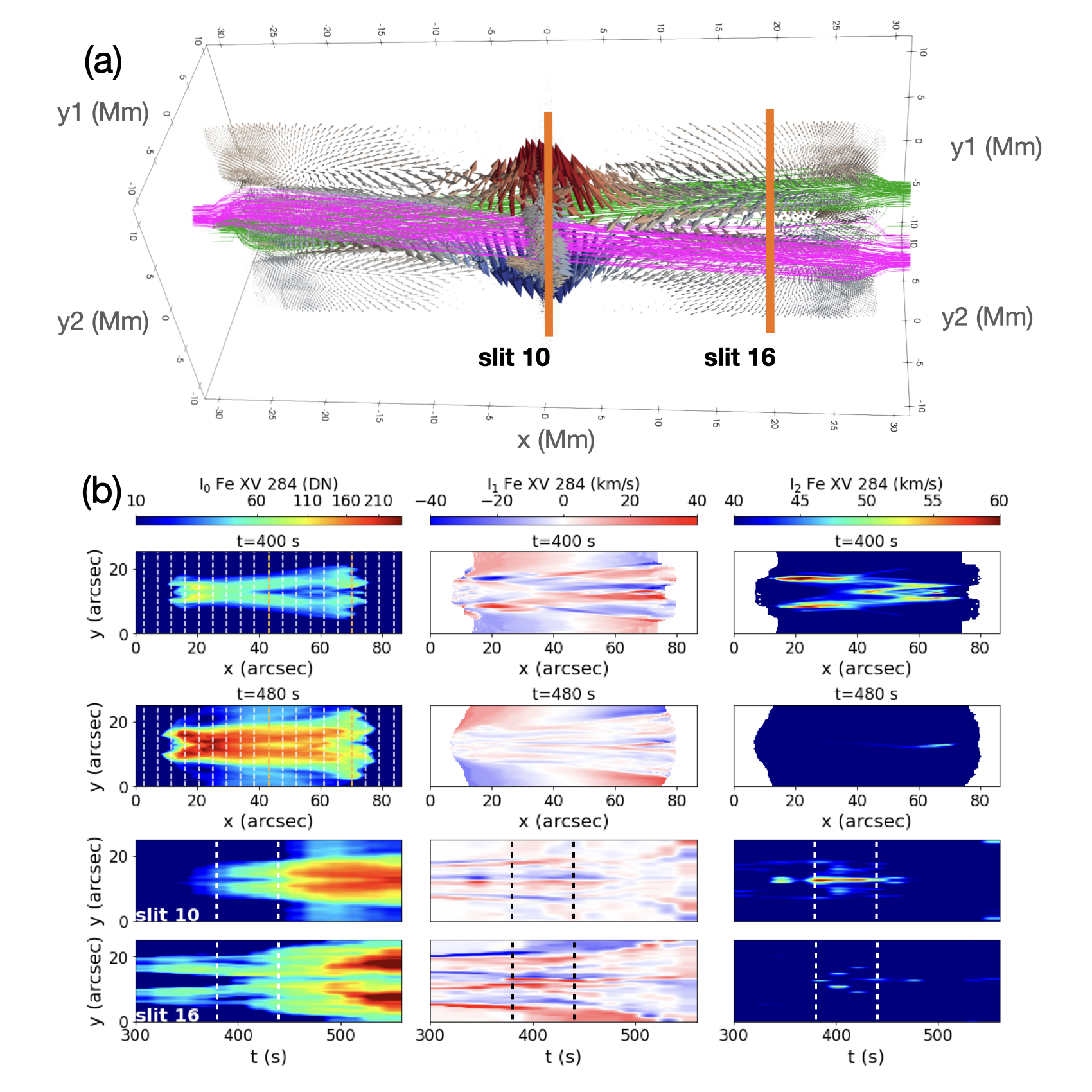}
	\caption{\label{fig:paolo} \muse\ observables for nanojets resulting from magnetic reconnection, as predicted by model \Pnano\ which simulates interacting AR loops (see Table~\ref{table_sims}  and Appendix~\ref{app:sim}). The setup and geometry of the model is shown in the panel (a) which shows the velocity field at $t=400$ $s$ in the 3D MHD simulation (blue and red arrows, with values ranging from $-200$ [blue] to $200$~km~s$^{-1}$ [red]) and some magnetic field lines (green and magenta) representing the two tilted flux tubes. Orange lines identify the \muse\ slits configuration we analyze.
	In the first two rows of (b) we show the moments of the \fexvw\ line (at $t=400$~s and at $t=480$~s) with the 3D MHD simulation rotated by $45^{\circ}$ (in the $y_1 y_2$ plane), similar to panel (a). The bottom two rows of (b) show the temporal evolution of the \fexv\ moments along the two selected slits.
	We assume an  exposure time  of 8s, and a pixel size 0.167\arcsec. The vertical dashed lines mark the time window during which the jets occur. Count rates are calculated as described in Appendix~\ref{app:synthesis}.}
\end{figure*}

The side view of this loop (see Figure~\ref{fig:rea16b}) illustrates diagnostic signatures associated with the twisting of the loop, which is characterized by Doppler shift patterns that rapidly change over short time scales ($< 1$~min) along and across the full extent ($\sim$ 90\arcsec) of the loop. The intensity images (left column) are not uniform along the loop because of substantial transient flows coming up from the footpoints (right column). The Doppler map (upper right panel) shows alternating (across the loop diameter) blue and red patterns because the flows follow the twisted field lines and a cross-loop speed component appears (therefore explaining the lack of  significant velocities along the loop central axis). The Doppler patterns are not uniform because of loop fine structuring, driven by the irregular heating across the footpoints. The plasma velocities associated with these twisting motions change rapidly on time scales of order 60s along the full length of the loop, as illustrated in the bottom two rows, which show the temporal evolution of intensity across the loop at two different slit locations. These rapid changes are a key diagnostic of this type of driver and any resulting heating, and can only be captured with \muse\ spectral rasters.

Recently, the Extreme Ultraviolet Imager on board Solar Orbiter has found ubiquitous small-scale brightenings in the quiet corona 174~{\AA} imaging passband (covering similar temperatures as the \muse\~\feixw\ passband but without spectroscopy) near disk center. These brightenings have been named ``campfires'' \citep{Berghmans2021} and may represent events caused by braiding. \muse\ spectroscopic measurements of intensity, Doppler shifts, and line broadening over its full field of view are needed to confirm that such campfires indeed are generated through braiding caused by the motions of well separated footpoints. Such measurements will provide strict constraints on forward models and determine whether they have a  spatio-temporal filling factor that is sufficient to produce energy sufficient to heat the quiet sun and network dominated corona. 

\subsubsection{Strong flows and nanojets}

Some braiding models  suggest that reconnection resulting from braiding can lead to strong flows on small spatial and temporal scales. Such small-scale ($< 1$\arcsec) high-velocity (100~km~s$^{-1}$) events have been spectroscopically detected with \iris\ \citep{Antolin2021},  but only at TR temperatures in cooling loops (coronal rain). However, this has not ben detected in coronal loops of a typical active region yet, because of the poor spatio-temporal resolution of current coronal observations. It is also unclear to what extent the observed jets were uniquely formed because of the peculiar magnetic field environment or topology. 

Magnetic field measurements at high spatial and temporal resolution, e.g., from \dkist, can help elucidate the role of topological changes in generating nanojets. However, the main challenge is to determine, through high-resolution coronal and TR spectroscopy and imaging, how common these types of events are, and whether they are as common as predicted by braiding models.

\begin{figure*}[!ht]
    \centering
    \includegraphics[width=0.9\textwidth]{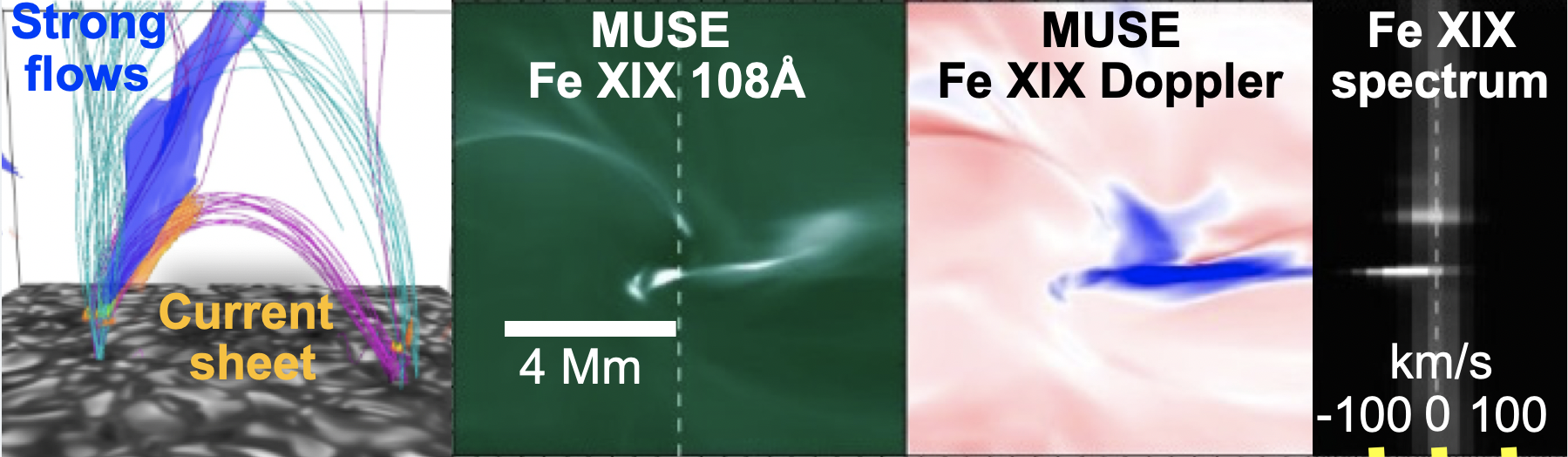}
	\caption{\label{fig:braiding_prop} 
	\muse\ will provide observations of flows and heating resulting from reconnection of braided magnetic field, constraining radiative MHD models and determining the role of braiding in coronal heating. We show \muse\ \fexixw\ synthetic observables from a 3D MHD Bifrost (model \Bemer\, in particular spatial maps of \fexix\ intensity and Doppler shift (middle panels, and the \muse\ \fexix\ spectrum along a slit (right panel). In the left panel we show the magnetic field configuration from the model. The model spans a region that covers $24\times 24\times 17$~Mm. For this relatively weak event, the maximum intensity is of order 10 DN for exposure times of 30s, while the Doppler range is [-50,50]~km~s$^{-1}$. The count rates are calculated as described in Appendix~\ref{app:synthesis}.}
\end{figure*}
\muse\ will provide key diagnostics of nanojets, as illustrated in Figure~\ref{fig:paolo}, where we show \muse\ synthetic observables from a 3D MHD model where magnetic reconnection is caused by the interaction of AR loops (\Pnano; see Section~\ref{sec:sim}, Table~\ref{table_sims}, and Appendix~\ref{app:sim}), and produces reconnection-driven "nanojets" ejected from the reconnection site.
As an example we show \muse\ \fexvw\ spectra along two slits, one located at the center of the domain, where the jet velocities are the strongest, and another one located $\sim 30$\arcsec\ away from the center, where the dynamics of the slower braiding phase are observed before the jet can be detected.
The plasma line-of-sight velocities show distinctive signatures of the ongoing braiding of the field lines with the two loops tilting in opposite directions. 
Note that the chosen line-of-sight leads to a significant overlap between the two jet components directed in opposite directions, which in turn leads to Doppler velocities smaller than the actual jet speed, and an increase in the line width, clearly observable in \fexvw, at the location of the jets (slit 10).
The temporal evolution of the observed emission along the two slits clearly shows the heating of the loops, as revealed by an increase of the hotter \fexv\ emission (and decrease of the cooler \feix\ emission, not shown here). The \muse\ emission at the time and location of the nanojets is characterized by (a) two spatially resolved peaks in \fexv, that correspond to two opposite and neighboring Doppler velocity enhancements; and (b) a sudden and significant increase in line width.

High-speed flows resulting from reconnection caused by braiding are also predicted in Bifrost models of an active region, as illustrated by Fig.~\ref{fig:braiding_prop}. In this case, a current sheet is formed (left panel) and the associated reconnection leads to heating and strong flows that are visible in the \fexixw\ line, in part helped by the lack of other 10 MK plasma along the line-of-sight. This particular case occurs in an active region with low coronal densities so the predicted count rates are low ($\sim 10$ DN in 30s). However, density variations of order a factor of 3 can be expected at these temperatures, which would increase the count rate to $\sim$ 100~DN.

It is thus clear that while nanojets caused by braiding events are relatively small-scale with a spatial extent of several arcseconds, the resulting heating of the loop in which the nanojet quickly deposits energy occurs over spatial scales of tens of arcseconds along the loop and 0.5\arcsec\ across the loop, and can only be captured with \muse.  Furthermore, these events are predicted to occur all along loops (as shown by \iris\ observations), thereby making multi-slit observations essential for capturing their spatio-temporal distribution. \muse\ observations of the spatio-temporal distribution of nanojets will also help determine whether the reconnection driving these nanojets occurs as a result of an MHD avalanche, excessive braiding, and the role of instabilities (e.g., kink or KHI) \citep{Antolin2021, Sukarmadji_2021_inprep}.

\subsubsection{Hot loops and footpoint diagnostics of non-thermal particles} \label{sec:hotloops}

Simultaneous high-cadence \muse\ spectra over a large FOV that capture the full extent of loops are also critical to fully diagnose the properties of coronal nanoflares and non-thermal electrons from the energy deposition at loop footpoints \citep{Testa2014}. \euvst\ measurements at a single footpoint can be useful to help constrain those properties by using forward modeling, but they cannot fully constrain all free parameters in such models \citep{Testa2020a}. In addition, it is difficult to interpret the footpoint variability without context imaging of the hot corona. It is the combination of \euvst\ rasters with \muse\ imaging and spectra at both conjugate footpoints and along the whole loop (including the very hot $\sim 10$~MK plasma generated by the nanoflares) that provide the strict constraints. These NGSPM measurements will address the occurrence frequency of such events, their importance for the coronal energy budget, and the properties of the non-thermal particles accelerated during nanoflares. In addition, the \muse\ observations can determine whether the heating occurs in the coronal volume or closer to the footpoints (e.g., from local reconnection) by measuring time differences between variations of spectral line properties at conjugate footpoints.

\begin{figure*}[!ht]
    \centering
    \includegraphics[width=0.99\textwidth]{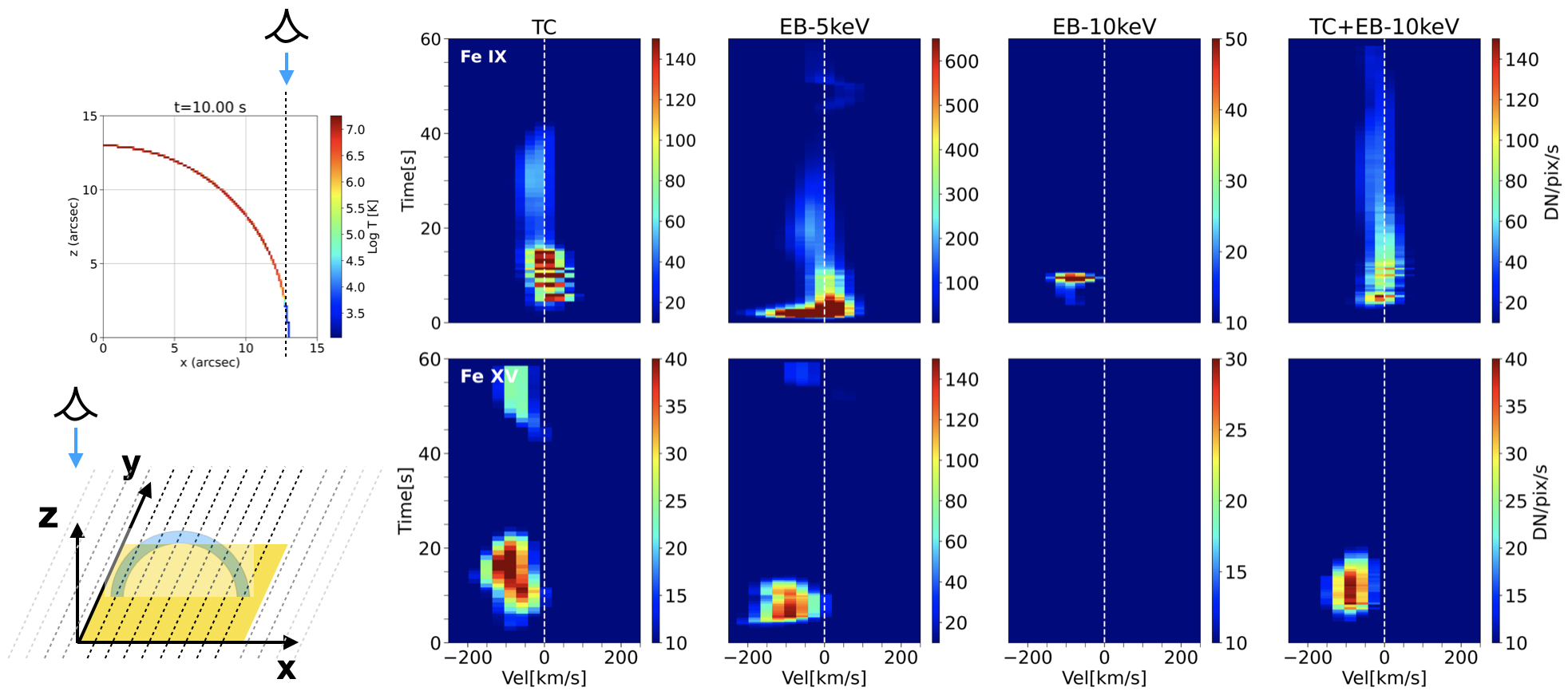}
	\caption{\label{fig:radyn} RADYN simulations of nanoflare heated loops predict that \muse\ footpoint brightenings provide key diagnostics of the heating mechanism(s). The  left panels show the geometry: we mapped the 1D loop simulations to a 2D semicircular loop (lying in the xz plane; the top left panel shows the simulated plasma temperature along half-loop at t=10~s for model \Rhone), and we are assuming the line-of-sight coincides with the z direction (i.e., the \muse\ slits are in the xy plane, see sketch in bottom left panel). \muse\ will allow to observe the atmospheric response to impulsive heating of the loop plasma in different loop locations simultaneously, and uniquely constrain the properties of the heating events (see also Fig.~\ref{fig:sanja1}). Here we show the temporal evolution of synthetic \muse\ spectra at the loop footpoint ($x \sim 13$\arcsec, i.e., on the right-hand side in top left panel) in \feixw\ (top row) and \fexvw\ (bottom row), assuming 1~s integration times, for four different assumptions about the heating properties (in all cases the heating duration is 10s, and the heating flux is $1.2 \times 10^9$~erg~cm$^{-2}$~s$^{-1}$): thermal conduction (TC), non-thermal electrons (NTE) with low energy cut-off $E_C =5$~keV, NTE with $E_C =10$~keV, and a hybrid model with TC and NTE with  $E_C =10$~keV (from left to right; they correspond to models \Rcone, \Reone, \Retwo, \Rhone\ of Table~\ref{table_sims} respectively). Count rates are calculated as described in Appendix~\ref{app:synthesis}. 
	}
\end{figure*}

The hot emission of coronal loops in the active region core, in non-flaring conditions, is typically observed to be transient (e.g., \citealt{Testa2012b,UgarteUrra2014,UgarteUrra2019,Testa2020b}). Short-lived and localized brightenings (underscoring the need for high spatial and temporal resolution) have been observed, with \hic\ \citep{Testa2013}, \iris\ \citep{Testa2014,Testa2020a}, and \aia\ \citep{Graham2019}, at the footpoints of these AR core loops in their initial heating phases.  
\iris\ spectral observations of these footpoint brightenings have provided new diagnostics of particle acceleration in nanoflares and NTE properties \citep{Testa2014,Testa2020a}.
\muse\ will significantly improve on the existing single-slit  diagnostics provided e.g., by \iris, in several ways. For instance, the multi-slit nature of \muse\ observations will provide for the first time simultaneous spectral observations at both footpoints, which will allow to constrain the location of the heating and, together with high resolution \dkist\ magnetic field data, will allow to explore the scenario of magnetic reconnection in the corona (e.g., in braiding, flux emergence) vs.\ a flux cancellation scenario \citep{Syntelis2019}. The multi-slit observations will also constrain much more tightly the properties of the heating by observing (1) the evolution and dynamics of the plasma over the whole loop length (from the TR to the corona), at high temporal cadence, in different lines, and (2) many more of these footpoint brightenings, including brightenings in the same overall event (see e.g., \citealt{Testa2020a}) which will constrain the spatial and temporal distribution of the heating. 

3D MHD simulations of the solar atmosphere can produce hot transient loops, similar to the ones observed in AR cores, as we show here for instance for model \Mplhe\ (see Section~\ref{sec:sim} and Appendix~\ref{app:sim}). Note that the hot loops in this simulation have projected lengths of $\sim 15$~Mm, on the lower end of the observed range (e.g., \citealt{Testa2020a} observe projected loop lengths in a range of $\sim 7-70$~Mm). \muse\ synthetic observables from this model (Figure~\ref{fig:sanja1}) show how \muse\ multi-slit spectral observations will simultaneously sample the footpoints brightenings observed at the beginning of the heating event (e.g., in \feix, bottom left panel), and the rapid evolution of the hotter plasma in the coronal portion of the loop (see \fexix\ intensity maps and spectra), and associated plasma flows (see e.g., maps of \fexix\ broadening, and \fexix\ spectra). 

However, it is challenging to include heating by NTE in 3D MHD simulations (but see \citealt{Bakke2018,Frogner2020} for first attempts, although so far applied only to cooler coronal conditions). Therefore, here we use RADYN 1D loop models to explore the novel \muse\ diagnostics of nanoflare heating (see Section~\ref{sec:sim}, Table~\ref{table_sims}, and Appendix~\ref{app:sim}). 
In Figure~\ref{fig:radyn} we show examples of how the heating properties and mechanism of energy transport (e.g., thermal conduction vs.\ NTE) can be diagnosed with \muse, and shows interesting differences with respect to \iris\ diagnostics \citep{Testa2014,Polito2018,Testa2020a}. For instance, while in \iris\ low-TR spectra (\ion{Si}{4}) the case of direct heating (and energy transport by thermal conduction TC) was virtually indistinguishable from heating by NTE with low values for the low-energy cutoff $E_c$ ($\sim 5$~keV), in \muse\ the latter produces blueshifted and very broadened \feixw\ emission while the former (TC) causes redshifted and narrow emission. 
The \muse\ CI will provide necessary complementary imaging revealing both the coronal morphology and the  spatial and temporal variability of the footpoints at high resolution.
As discussed above \euvst\ will provide important additional constraints (including density) by observing in a broad temperature range at high spatial resolution in the smaller areas covered by its slit, but will miss most events and not capture the associated coronal loop.

\begin{figure*}[!ht]
    \centering
    \includegraphics[width=1.0\textwidth]{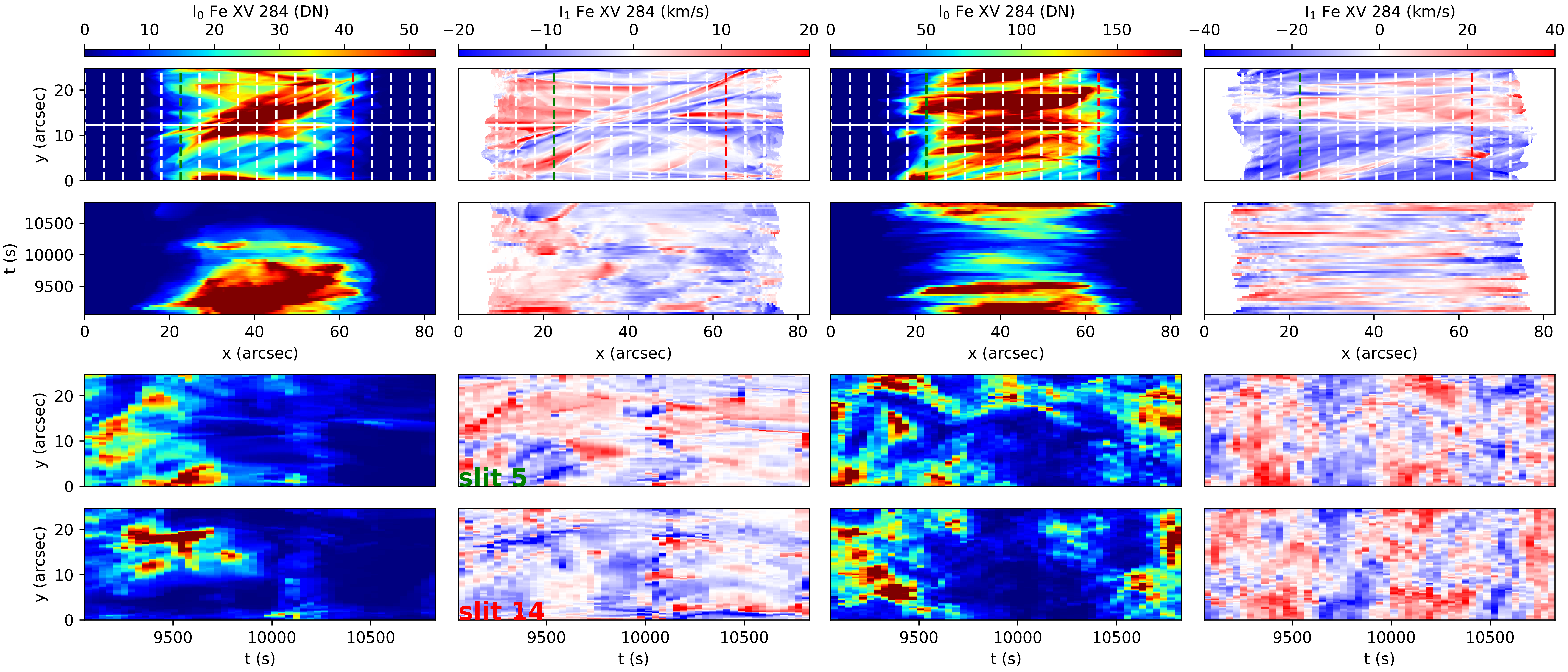}
    \caption{Comparisons between \muse\ observables predicted by idealized numerical loop models in which loop footpoints are shuffled by either drivers on DC-like (model \Ldc; see Section~\ref{sec:sim} and Appendix~\ref{app:sim}) or AC-like (model \Lac) timescales \citep[see ][ for a detailed explanation of the driving timescales]{Howson2020}  -- i.e., mimicking heating by braiding or waves respectively --  show clear differences. 
    We show intensity and Doppler velocity for the DC (left two columns) and AC (right two columns) model in the \muse\ \fexvw\ line integrated over 1.5s. The top row shows a snapshot at t= 9054~s where x represents the field aligned coordinate. The LOS has been taken to be perpendicular to the magnetic field. The second row shows a time-distance image along the central y-coordinate (horizontal white line in top row). The 3rd and 4th rows show time-distance images in the cross-field direction along slit 5 (vertical dashed green line in top row) and slit 14 (vertical dashed red line in top row).
    Correlated oscillatory signals at separate locations along loops are predominant in AC models (right column) and rare in DC models (second column). Count rates are calculated as described in Appendix~\ref{app:synthesis}.
    }
    \label{fig:ineke1}
\end{figure*}

\subsection{The wave-heating hypothesis}
\label{sec:waves}

Various types of waves (e.g., shock waves and \alfven\ waves) are generated in the lower atmosphere through the interaction between magnetic flux concentrations and convective motions, or reconnection. The propagation and eventual dissipation of these waves plays an important role in the dynamics and energetics of the chromospheric footpoints of coronal loops. The subsequent rapid propagation ($\sim 1000$~km~s$^{-1}$) along coronal loops ($\sim 100$~Mm) and the eventual dissipation of transverse MHD waves in particular has been proposed as a heating mechanism for the corona \citep{vanD2020}. A combined NGSPM approach is desired to address whether or where wave heating dominates the local energy balance, from the photosphere all the way into the corona, and for various types of regions on the Sun. 

Such an approach can track waves from the photosphere through the plasma $\beta =1$ layer in the chromosphere \citep{Bogdan2003}, e.g., along spicules or other chromospheric features, using DL-NIRSP or ViSP spectropolarimetric data, VTF scans or other GBO observations. Rapid ($\sim 20$~s cadence) \euvst\ rasters centered at loop footpoints (with a FOV of 4\arcsec\ $\times 140$\arcsec) can be used to determine which fraction of the wave energy flux (from density and velocity measurements) is reflected by the steep gradients of the transition region or transmitted into the low corona \citep[e.g.,][]{Matsumoto2018}. Correlations between heating signatures and decreasing line broadening along relatively short or fortuitously slit-aligned spicules as they undergo rapid temperature changes can elucidate signatures of wave-related heating. However, because of the high propagation speed, large transverse motions, and curved coronal loop morphology, \euvst's small FOV of high cadence rasters and lack of high-resolution coronal imaging will prevent it from properly capturing the subsequent propagation and signatures of dissipation along coronal loops.

High cadence \muse\ images ($\sim 5$~s) and multi-slit spectral scans ($\sim 12$~s) or sit-and-stare (along 37 slits with a cadence of 2~s) over the whole length of the loop are key to distinguish between DC (braiding) and AC (wave) models of heating.  For example, comparisons between \muse\  observables predicted by idealized numerical loop models in which loop footpoints are shuffled by either drivers on DC or AC timescales, i.e., either faster or slower than the loop crossing time for an Alfv\'enic  wave, illustrate that both mechanisms (braiding versus waves) can, in principle, heat plasma to coronal temperatures. However, observables at the high spatial resolution, high cadence, and large FOV of \muse\ show clear differences, as illustrated in Fig.~\ref{fig:ineke1}: correlated oscillatory signals at separate locations along loops are predominant in AC models (right column) and rare in DC models (2nd column).  

In addition, these types of measurements are key to reveal the propagation and fate of these waves and their role in heating the loops (if any). 
They provide key constraints for, and allow to discriminate between, the various more advanced numerical models that have recently been developed to describe the propagation and dissipation of waves, and subsequent heating of plasma.

\begin{figure*}[!ht]
    \centering
    \includegraphics[width=0.99\textwidth]{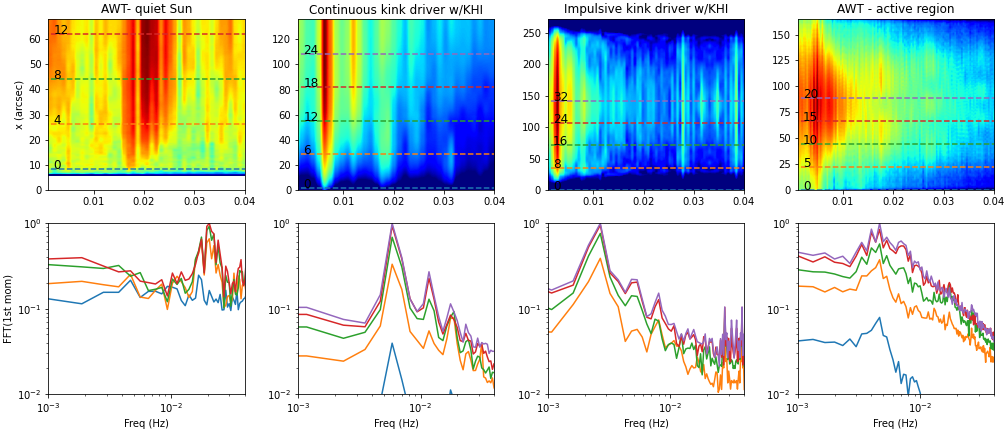} 
	\caption{\label{fig:matsufft} \muse\ can simultaneously observe oscillatory (e.g., velocity) signals all along loops. Comparison of the observed FFT power spectrum (top row, with selected loop locations shown in bottom row) with model predictions provides a rigorous test for various wave dissipation models. The top row shows maps of the FFT power spectrum (of Doppler shift vs.\ time at a specific location) for each location along the loop. The bottom row shows the FFT power spectrum at specific slit positions marked on the top row. From left to right, the loop models shown are: \Mac\ (Alfv\'en wave turbulence -AWT- model in a quiet Sun coronal loop), \Pwaves\ (continuous AC driver with KHI formation), \Cwaves\ (impulsive AC driver with KHI formation), and \Vawt\ (AWT model in an AR coronal loop).}
\end{figure*}

\subsubsection{Dissipation mechanisms}

For example, model predictions of FFT wave power spectra obtained simultaneously along a full loop differ sharply between various mechanisms, requiring \muse\ multi-slit measurements. This is illustrated in Fig.~\ref{fig:matsufft} which shows: (i) a model of wave mode coupling, propagation and dissipation in a coronal loop in quiet Sun conditions, in which the dissipation mostly occurs through Alfv\'en wave turbulence (\citealt{Matsumoto2018}; model \Mac), (ii) a model in which a loop is continuously driven with a transverse driver which couples to an azimuthal mode through resonant absorption, followed by the development of a KHI instability and subsequent dissipation (\citealt{Karampelas2018...9K}; model \Pwaves), (iii) a model in which a loop is impulsively driven with a single pulse of a kink mode, followed by resonant absorption and dissipation through KHI (\citealt{Antolin_2017ApJ...836..219A}; model \Cwaves), and (iv) a model of Alfv\'en wave turbulence (AWT) in an active region coronal loop assuming RMHD (\citealt{Asgari-Targhi2014}; model \Vawt). As shown in this figure, simultaneous measurements of oscillatory wave power (e.g., from Doppler shifts) along the length of coronal loops can discriminate between these various models with some showing specific frequencies, others a broad-band spectrum, and all showing different dependence of the wave power along the loop length. For example, in the quiet Sun AWT simulation low frequencies are less likely to reach the corona because of reflections below the transition region. The continuously driven kink model shows harmonics of the primary driving frequency, in contrast to the impulsively driven kink model. The active region AWT model (\Vawt) shows a power law distribution with the slope varying depending on the location along the loop.
\muse\ observations that are simultaneous along 37 slits and that cross a loop along its full length are key to provide these types of constraints.

\begin{figure*}[!ht]
    \centering
    \includegraphics[width=0.99\textwidth]{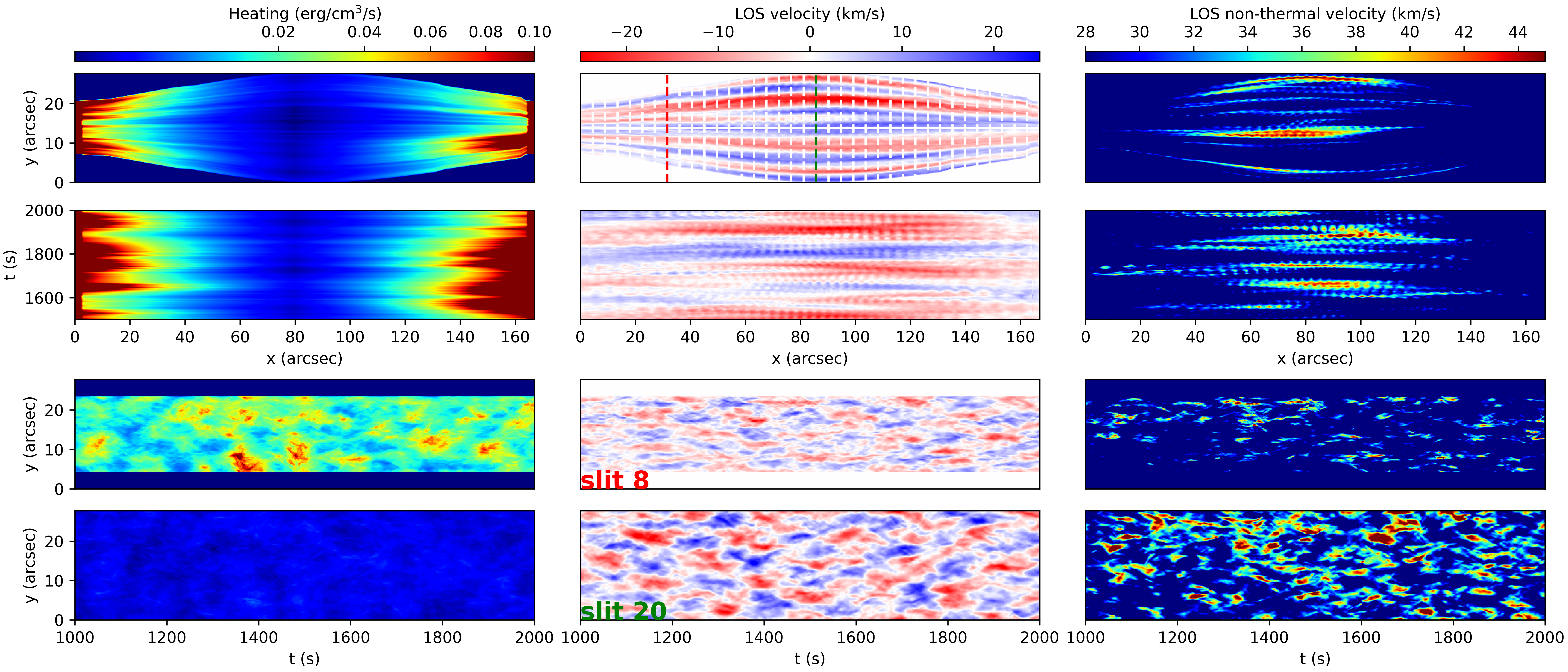}
	\caption{\label{fig:mah1} Predicted spatial and temporal properties of heating rate, Doppler shift, and non-thermal line broadening from a reduced MHD model based on \alfven\ wave turbulence (\Vawt; see Table~\ref{table_sims} and Appendix~\ref{app:sim}).
	In the top row we show the side view (the loop footpoints are at the two extremes of the x range and the loop top at the center, i.e., $x \sim 80$\arcsec) of the heating rate (left column), line-of-sight velocity, and non-thermal broadening. The second row shows a time-distance image along the central y-coordinate of the top row. The 3rd and 4th rows show the time-distance images in the cross-field direction, along slit 8 and 20, marked in the top middle panel by a red and blue vertical line respectively.  
	}
\end{figure*}

A more detailed study shows that the predicted spatial and temporal patterns of the intensity, Doppler shift, and broadening can also provide critical constraints and allow discrimination between these various wave dissipation mechanisms. For example, the random oscillatory signals (Fig.~\ref{fig:mah1}) predicted by the \alfven\ wave turbulence model \Vawt\ are qualitatively and quantitatively very different from those predicted by resonant absorption models (Fig.~\ref{fig:patrick2}). In the latter models, loops are driven by kink waves which couple to azimuthal motions that lead to the formation of the Kelvin-Helmholtz instability and spatially extended TWIKH (Transverse Wave Induced KH) rolls. Wave dissipation then occurs through turbulence established via, e.g., the Kelvin-Helmholtz instability (KHI) and phase mixing, which, as models predict, has unique observable consequences, including spatio-temporal correlations between variations in intensity, Doppler shift, and broadening  \citep{Antolin:2018fk,Karampelas2019a,Karampelas2019b,Guo2019}.

\begin{figure}[!ht]
\includegraphics[width=0.45\textwidth]{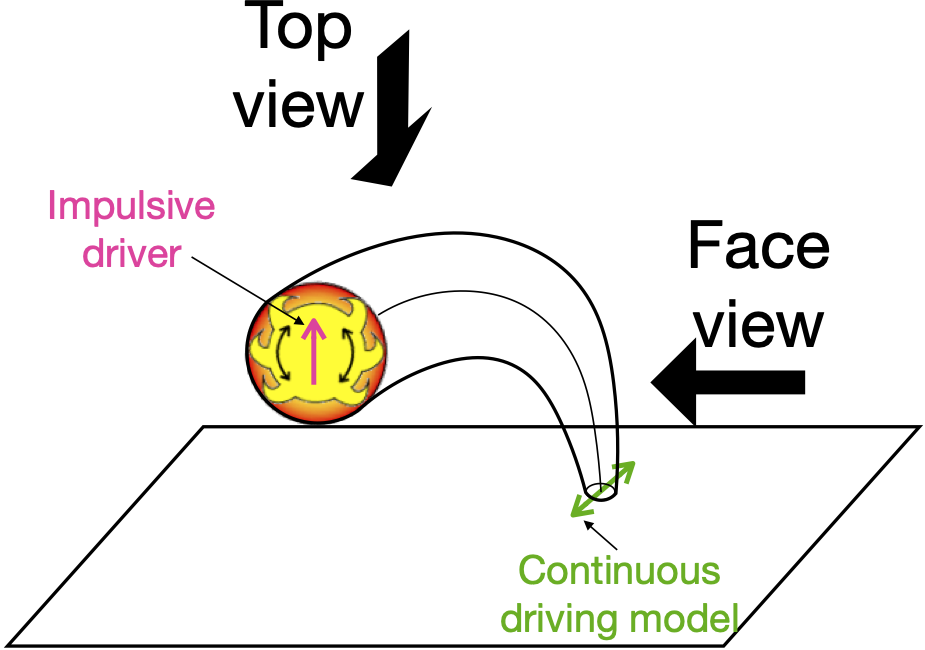}
\caption{\label{fig:LOSint} Sketch to illustrate the geometry of straightened loop models (e.g., \Pwaves, \Cwaves) mapped to a curved, semicircular, loop, and different lines of sight: the {\bf {\it face view}} is perpendicular to the loop plane (essentially coinciding with the typical side view of straightened loop models; see Fig.~\ref{fig:patrick2}); the {\bf {\it top view}} implies significant integration of the loop length when close to the loop footpoints, and allows to observe more of the field aligned flows  (see Figs.~\ref{fig:patrick0}, \ref{fig:patrick1}, \ref{fig:kostas}).}
\end{figure}

\begin{figure*}[!ht]
    \centering
    \includegraphics[width=0.99\textwidth]{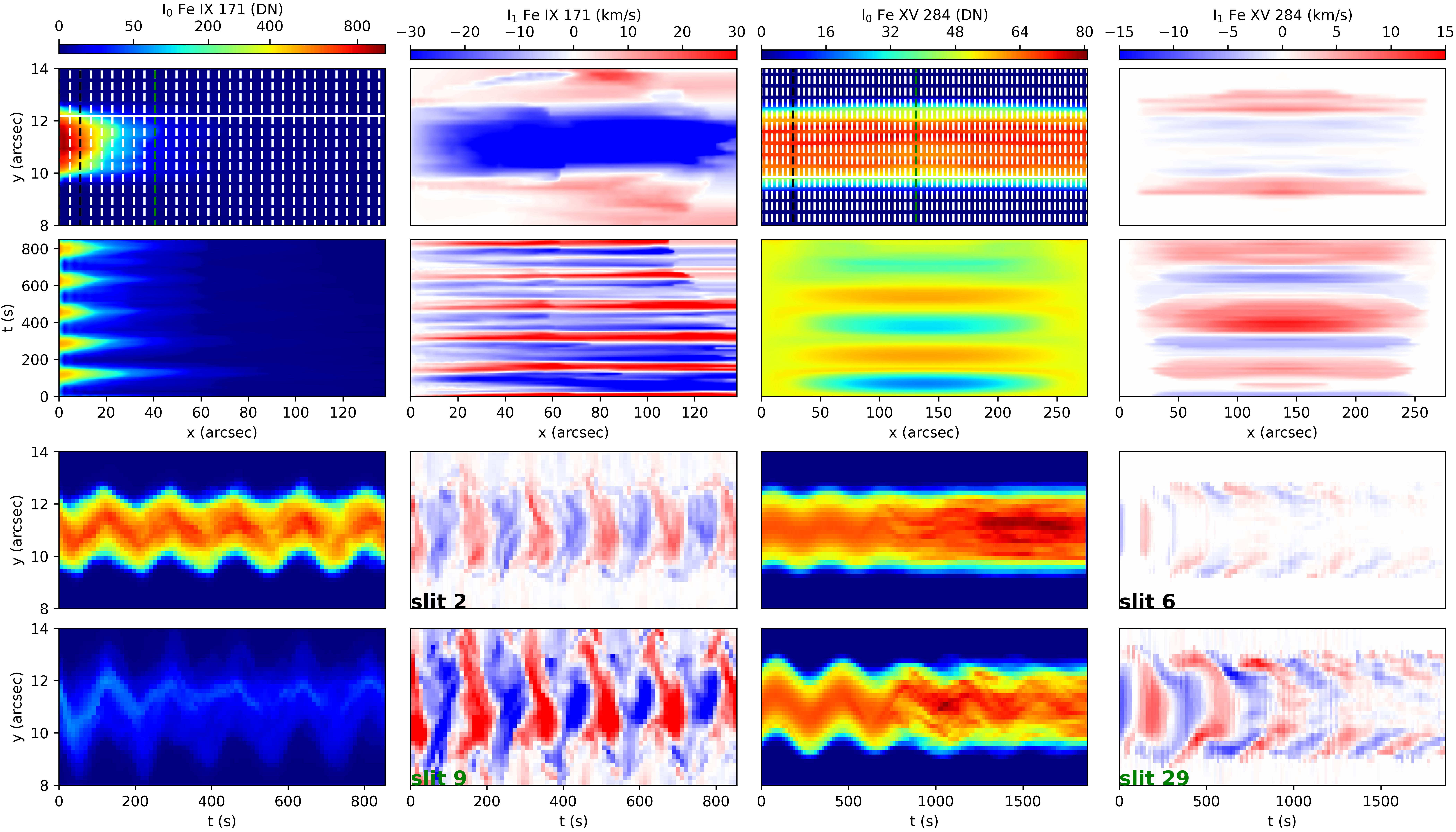}
	\caption{\label{fig:patrick2}
	The comparison of \muse\ observables predicted from two models of oscillating loops, shows clear detectable differences between the continuously driven (\Pwaves) and the impulsively driven (\Cwaves) cases, providing clear diagnostics for distinction.
	We show the 0th (1st and 3rd columns) and 1st moment (2nd and 4th columns) for \muse\ \feixw\ (model \Pwaves, left two columns) and \fexvw\ (model \Cwaves, right two columns). The x coordinate is along the loops, and we show half the loop for model \Pwaves\ (footpoint located on the right end of the x-axis), and the full loop for model \Cwaves\ (footpoints located at the two extremes of the x-axis). The LOS corresponds to the face view (see Fig.~\ref{fig:LOSint}) but is at a 45$\degr$ inclination with respect to the kink oscillation direction. The integration times are 1.5~s and 5~s for the \Pwaves\ and \Cwaves\ models, respectively.
	The top panels show the loops at a time when the KHI is already well developed, while the second row of panels track the evolution of the oscillation along the boundary of each loop. Note here that due to gravitational stratification and the mixing of plasma due to the KHI, only the loop footpoint is visible for model \Pwaves\ \citep[also see][]{Karampelas2019b}. 
	The two bottom rows show time-distance maps for the 0th and 1st moments, at two different heights (and slits, counting from left to right) along the loops. Note the increase in amplitude with height for the 1st moment, with the herringbone patterns and small-scale structure best visible towards the apex. Note also the damping of the oscillatory motion in the 0th moment and decrease/increase in the 1st moment at the loop core/boundary due to resonant absorption, only visible in the impulsively driven model. Count rates are calculated as described in Appendix~\ref{app:synthesis}. An animation of this figure can be found online.  
	}
\end{figure*}

\begin{figure*}[!ht]
    \centering
    \includegraphics[width=0.99\textwidth]{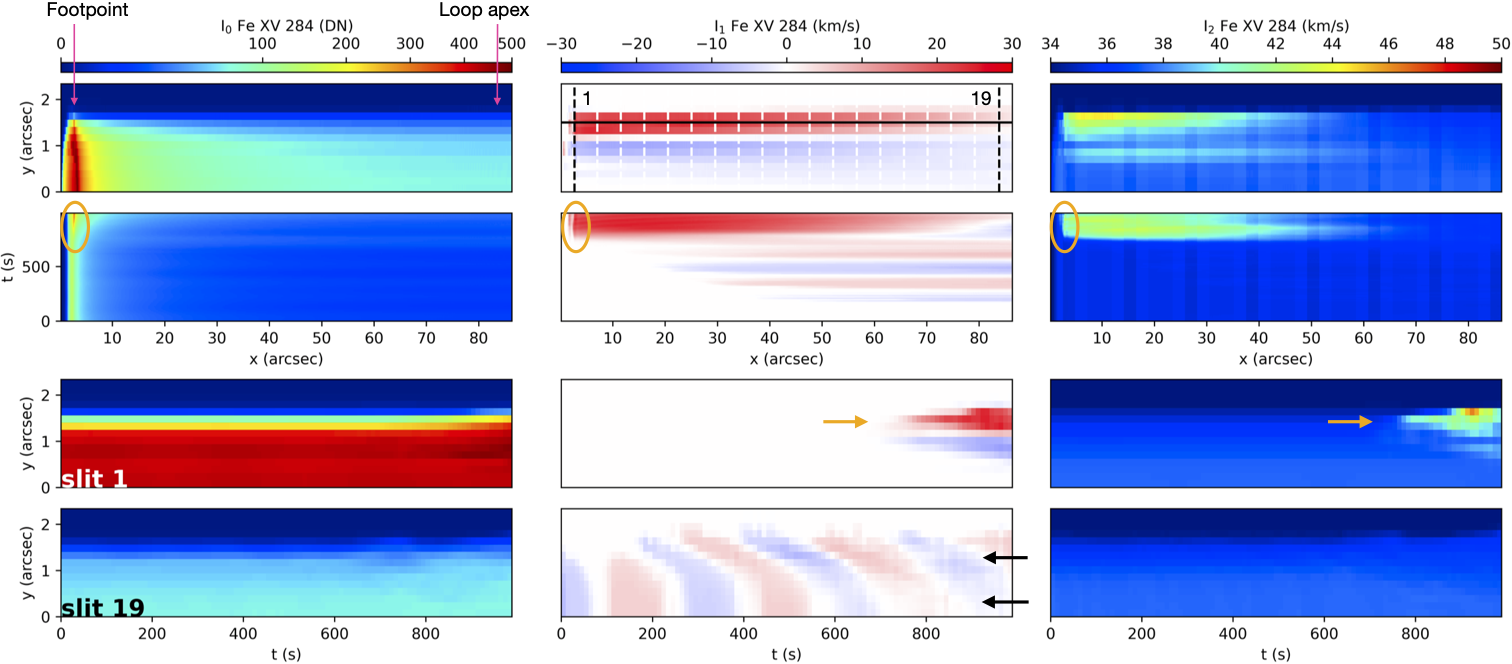}
	\caption{\label{fig:patrick0} Long term characteristics of \muse\ synthetic observables (top view; see Figure~\ref{fig:LOSint}) for a simulation modeling a coronal loop undergoing an impulsively excited and vertically polarized kink oscillation (\Cwaves; see Table~\ref{table_sims}, and Appendix~\ref{app:sim}). The loop coordinate axis is transformed onto a curved trajectory to mimic a semi-circular coronal loop (as in the RADYN models of Figure~\ref{fig:radyn}). Here we show the \fexvw\ synthesis for half of the loop with a time integration of 5~s, with x=0 (and slit 2) near the left footpoint and larger x (higher slit number) corresponding to rays crossing the loop higher up in the corona, up to the loop apex. 
	The top row shows a snapshot at t=950~s, when the KHI is fully developed across the loop. 
	The second row shows the time-distance maps of the moments, at the boundary of the loop (marked by the horizontal black line in top middle panel). Note the distinct change in the 1st and 2nd moments at $t\approx700~$s, indicating the onset of the KHI. An increase in the intensity at the footpoints	is marked with an orange circle. 
	The two bottom rows show time-distance maps of the \fexv\ moments, for two slits at different heights along the loops: slit 1 is at the footpoint, while slit 19 is at the apex (half of the loop). The black arrows mark the kink mode attenuation and accompanied amplitude increase of the Alfv\'en waves at the boundary due to resonant absorption. The orange	arrows mark the KHI onset. Count rates are calculated as described in Appendix~\ref{app:synthesis}. 
}
\end{figure*}

\begin{figure*}[!ht]
    \centering
    \includegraphics[width=0.99\textwidth]{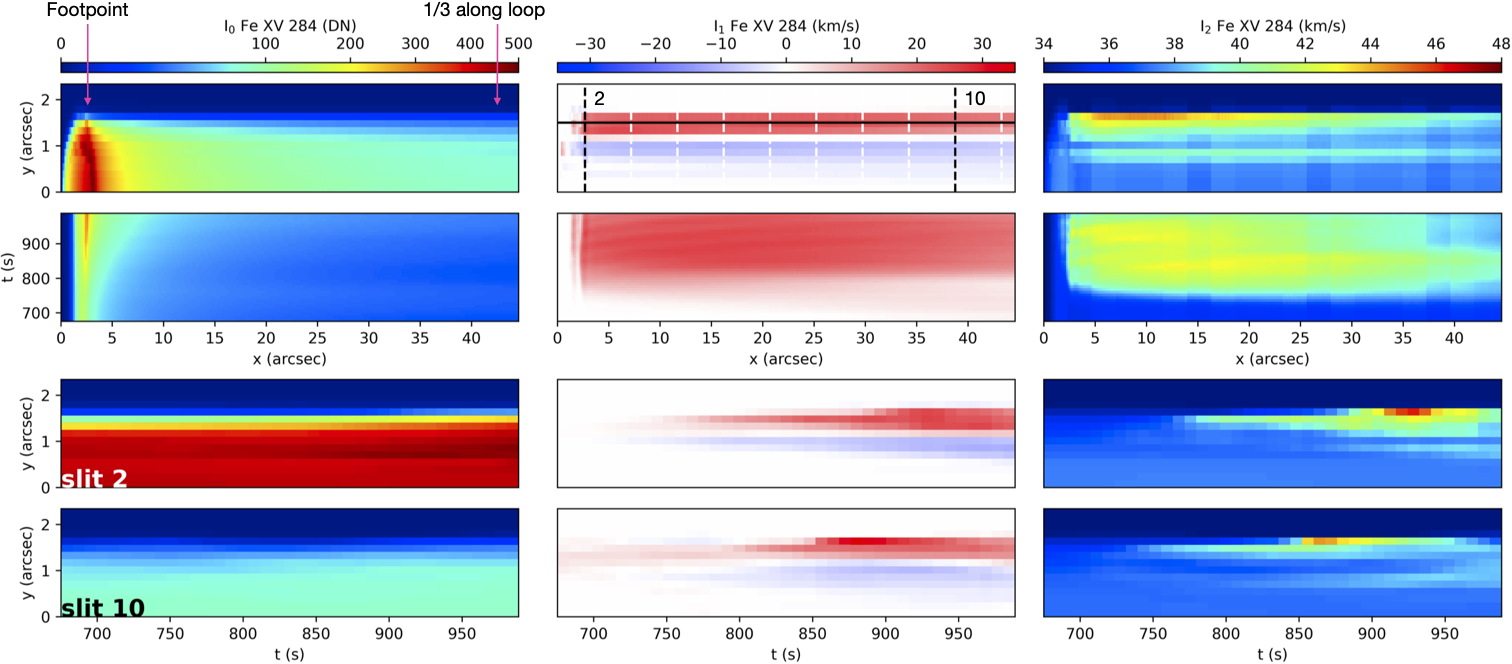}
	\caption{\label{fig:patrick1} Short term characteristics of \muse\ synthetic observables (top view; see Figure~\ref{fig:LOSint}) for a simulation modeling a coronal loop undergoing an impulsively excited and vertically polarized kink oscillation (\Cwaves; see sec.~\ref{sec:sim} and Appendix~\ref{app:sim}). The panel configuration, distance and time axis, LOS view and exposure time are the same as in Fig.~\ref{fig:patrick0} for the curved loop. Here we show the \fexvw\ synthesis for the first 1/3 of the curved loop focusing on a 5~min interval during which the KHI develops.
	The top row shows a snapshot at t=950~s, at the end of this stage. 
	The KHI leads to steady upflows (blueshifts) / downflows (redshifts) with increased line widths at the loop boundary all along the loop (slits 2 and 10). Count rates are calculated as described in Appendix~\ref{app:synthesis}. 
}
\end{figure*}

\begin{figure*}[!ht]
    \centering
    \includegraphics[width=0.9\textwidth]{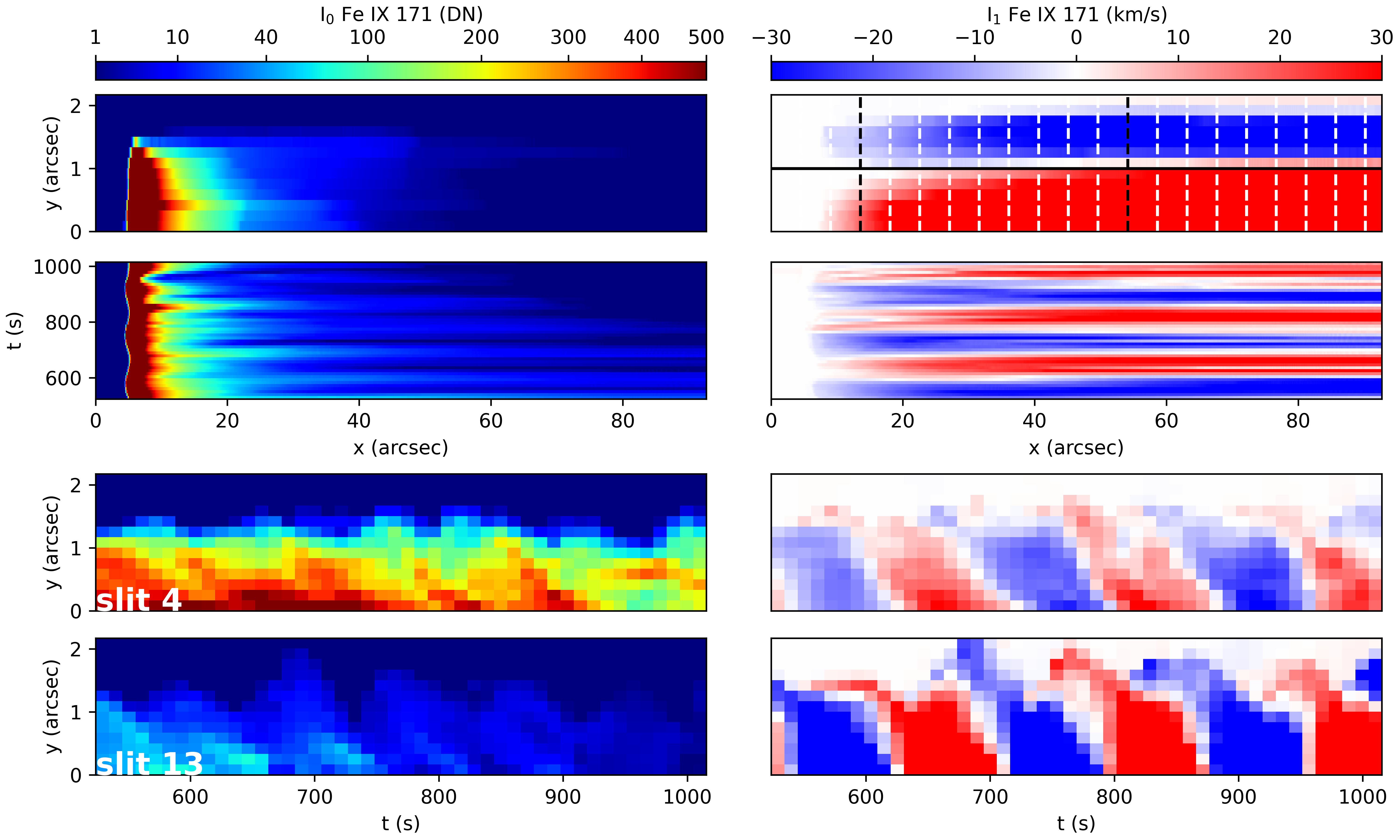}
    \caption{\label{fig:kostas} 
    \muse\ synthetic observables of intensity (left) and Doppler shift (right) for a top view (see Figure~\ref{fig:LOSint}) for model \Pwaves\ (see sec.~\ref{sec:sim} and  Appendix~\ref{app:sim}), similar to model \Cwaves\ of Figure~\ref{fig:patrick1} but with a continuously driven transverse oscillation. Here we show the \feixw\ synthesis for $2/3$ of the loop, with x=0 (and low slit numbers) near the footpoint and larger x (higher slit numbers) towards the apex. The exposure time is 1.5~s. The top row shows the loop at time t=1015~s, when the KHI has developed over the entire loop cross-section. 
    The second row shows the time-distance maps of the moments, close to the loop boundary of the loop (marked by the horizontal black line in top right panel).
	The two bottom rows show time-distance maps of the \feix\ moments, for two slits at different heights along the loops (marked by the vertical black dashed lines in the top right panel): slit 4 is close to the footpoint, while slit 13 is higher in the corona. Count rates are calculated as described in Appendix~\ref{app:synthesis}. 
    }
\end{figure*}

This is shown in a set of figures that illustrate two models of a loop: (i) one that is initially driven by a continuous kink mode driver (model \Pwaves; Figure~\ref{fig:patrick2} and~\ref{fig:kostas}), (ii) and the other driven by an impulsive kink mode driver (model \Cwaves; Figures~\ref{fig:patrick2}, \ref{fig:patrick0}, and~\ref{fig:patrick1}),  both leading to KHI and subsequent turbulence. The geometry and viewing angles are illustrated in Fig.~\ref{fig:LOSint}. 

The continuously driven model simulates an active region coronal loop, during its cooling phase, undergoing a decay-less transverse oscillation (e.g., \citealt{Anfinogentov2013}). In this model, the continuously driven transverse oscillation leads to the development of the Kelvin-Helmholtz instability and of spatially extended TWIKH (Transverse Wave Induced KH) rolls. These TWIKH rolls disrupt the initially monolithic loop profile, leading to a turbulent cross-section with a wide temperature range. Inside this loop cross-section, both energy cascade and extensive plasma mixing take place, giving rise to heating effects from energy dissipation and
mixing of plasma with different temperature (see also \citealt{Karampelas2019a}), and leading predominantly to an apparent heating effect. This has an effect on the synthesized loop emission, and can lead to a gradual appearance or disappearance from view for parts of the loop. The evolution of the 0th, 1st, and 2nd moments can lead to observational evidence for the development of the KHI in oscillating coronal loops, while observations at different passbands can constrain heating \citep{Antolin_2016ApJ...830L..22A}. 

The impulsively driven model is very similar in general set-up, except that the driver is now a single impulse leading to a kink mode (red arrow in Fig.~\ref{fig:LOSint}) and that we have a coronal loop hotter and denser than its environment. In this model, the global kink mode oscillation rapidly damps due to resonant absorption, and most of the wave power is transferred to the loop boundaries, where it couples with localized azimuthal Alfv\'en waves \citep{Goossens_2009AA...503..213G}. A reduction/increase in the Doppler velocity along the loop axis/loop boundary is therefore expected, contrary to the continuously driven kink model for which the continuous input of energy counterbalances the damping. The KHI sets in after a couple of periods, energized by resonant absorption \citep{Antolin_2019FrP.....7...85A}, and is characterized by its stranded structure in all moments and appearance of high-frequency perturbations, only detectable at high resolution \citep{Antolin_2017ApJ...836..219A,Antolin:2018fk}. The TWIKH rolls appear first at the apex and the compressive, localized twists propagate downwards as azimuthal Alfv\'en waves. The magnetic field is reshuffled and sound waves are generated, leading to localized, high-speed steady upflows/downflows at the loop core/boundary. Such upflows/downflows are mostly absent in the continuous kink driver model, due to the constantly changing boundary conditions at the loop's resonant frequency.

Figure~\ref{fig:patrick2} shows the two resonant absorption models when a loop is viewed from the side ({\em face view} of Fig.~\ref{fig:LOSint}).
The top panels of Figure~\ref{fig:patrick2} show the loops at a time when the KHI is already well developed. Note here that due to gravitational stratification and the mixing of plasma due to the KHI, only the loop footpoint is visible for model \Pwaves\ \citep[also see][]{Karampelas2019b}. The panels in the second row track the evolution of the oscillation along the boundary of each loop. The effects of resonant absorption, phase mixing and the KHI are shown in the panels of the 1st moment, with the herringbone pattern and the stronger values found higher up along each loop. 

It is clear from Figure~\ref{fig:patrick2} that the phase relationship between the intensity and Doppler shift is fundamentally different from that predicted by the Alfv\'en wave turbulence model (Fig.~\ref{fig:mah1}). Transverse MHD waves subject to resonant absorption typically lead to strongly periodic oscillatory Doppler shift signals (Fig.~\ref{fig:patrick2}), in contrast to random signals in the \alfven\ wave turbulence model (\Vawt). Clear wedge-like features are seen in the Doppler shift timeseries (bottom two rows), as well as strand-like features in the intensity (left and 3rd column, bottom two rows). 

There are also clear differences between the predicted signals of intensity and Doppler shift of spectral lines in both resonant absorption models. This is illustrated in the face view of both models (Fig.~\ref{fig:patrick2}), but also in the top view of Figs.~\ref{fig:patrick0}, \ref{fig:patrick1}, and~\ref{fig:kostas}.
For example, in Fig.~\ref{fig:patrick2} we can observe qualitative differences between the driven oscillation of model \Pwaves, with the strong Doppler shifts found across the whole loop cross-section, and the impulsive oscillation of model \Cwaves, where the Doppler shifts are more prominent near the loop edge as resonant absorption and the KHI develop. These differences can be detected at \muse’s high spatial resolution. 

In addition, simultaneous measurements at the footpoints and all along the loop to disentangle viewing angle effects will reveal differences between both models in terms of the complex signatures of the transverse motions associated with the KHI and the subsequent dissipation and field-aligned flows. Such measurements are needed to detect propagation effects between loop apex and footpoints predicted by the impulsively driven model, but not by the continuously driven model. These differences are illustrated in Figs.~\ref{fig:patrick0}, \ref{fig:patrick1}, and \ref{fig:kostas}.

Figures~\ref{fig:patrick0} and \ref{fig:patrick1} show the top view (see Fig.~\ref{fig:LOSint}) of the \muse\ \fexvw\ emission for the impulsively driven model (\Cwaves). Figure~\ref{fig:patrick0} shows the long-term evolution, including the gradual onset of resonant absorption and the sudden onset of the KHI (around $t=700~s$), while Fig.~\ref{fig:patrick1} is focused on the time interval around the onset of the KHI.

In Figure~\ref{fig:patrick0} we can see the coupling of the kink mode to the azimuthal motions expected from resonant absorption (bottom row, Doppler shift along slit 19, black arrows), which leads to the wedge-like features in the time-distance plot in the bottom row. In addition, we see the sudden onset of the KHI around $t=700$~s, which manifests itself through a distinct change in the properties of the Doppler shift and broadening. This predicted behavior is qualitatively very different from the continuously driven model (bottom two rows, Fig.~\ref{fig:kostas}). 

When we focus on the time at which the KHI suddenly occurs, Figure~\ref{fig:patrick1} shows further interesting differences for the impulsively driven model. The top row shows a snapshot at t=950~s, when the KHI is fully developed across the loop. The first KHI vortices appear at the start of the time sequence shown. Steady downflows (redshifts) and upflows (blueshifts), accompanied with increased non-thermal line widths occur at the edge of the loop, due to the twisting of the field lines and reshuffling of the magnetic field from the KHI. 
The KHI-induced magnetic shear at the foopoint leads to heating and increased emission across the loop width, captured by slit 2 (3rd row, left panel). Compressive effects from the KHI vortices also lead to steady Doppler velocity and line width enhancements in the corona (captured by slit 10, shown in 4th row). 

A comparison of the temporal evolution of Doppler shift and broadening for the two selected slit positions (bottom two rows of second and third column of Figure~\ref{fig:patrick1}) shows the propagation in a short time ($\lesssim 50$~s) over a projected length of $\sim 35$\arcsec. High-cadence multi-slit \muse\ spectral observations will be able to capture these effects. These propagation effects are not predicted by the continuously driven model (\Pwaves), for which the corresponding {\em top view} (for $2/3$ of the loop) is shown in Figure~\ref{fig:kostas}.  
Unlike the \Cwaves\ model, strong upflows (blueshifts) and downflows (redshifts) can be observed across the entire loop cross-section, due to the continuous periodic driving. 
One of the results of the KHI is the continuous mixing of plasma across the loop cross-section. This process increases the loop temperature for this model, gradually reducing the emission in the \feixw\ line, as shown in Fig.~\ref{fig:kostas} for the intensity (shown here in a square-root scale) at the later stages of the time-series. The development of the KHI across the loop can be tracked through the time distance maps for the intensity, and the Doppler shift,
as shown in the two bottom rows of panels, getting stronger signatures higher up the loop.

It is clear from the above that \muse\ observations of intensity, Doppler shift, or broadening will have sufficient spatial and temporal resolution to detect the differences between the predictions of these wave-based models. Further constraints 
will come from \muse\ high-resolution images and spectra, which can detect rapid heating through tracing of loop strands as they heat through the \muse\ 171, 195, and 284\AA\ passbands (covering temperatures from 0.7 to 2.5~MK). Tracing any heating will be further facilitated by exploiting the  temperature coverage of \euvst, using rapid but small-scale rasters across loops. Similarly, quantification of the wave energy flux will be facilitated through density measurements from \euvst\ density sensitive line pairs (from large-scale rasters using a wider slit) and determination of wave amplitudes and group speed (using $\omega$-k diagrams or time-distance seismology) from \muse\ multi-slit measurements over the whole loop length \citep{Tomczyk2009}. 

\subsubsection{Wave generation}

The powerful combination of  \dkist\ (or other GBOs), \euvst, and \muse\ will also elucidate the source of the waves, whether they are propagating into the corona from the lower atmosphere or generated locally in the corona, and whether their generation is impulsive \citep{Antolin2018} or continuous \citep{Karampelas2019a,Karampelas2019b}. For example, \dkist\ spectro-polarimetric measurements can be used to determine wave modes and energy flux in the chromosphere, while \euvst\ can provide the link between the chromosphere and the large-scale coronal volume observed with \muse. This will be important to determine the dominant wave processes in the lower layers, identifying the wave modes that are transmitted into the corona as well as providing constraints on the available wave energy flux.  

The high-cadence multi-slit \muse\ observations of the propagation of oscillatory intensity, velocity and broadening signals as well as transverse displacements will reveal wave propagation along the whole loop length. \muse\ can be used to track individual wave packets \citep{Tomczyk2009} and determine the wave group speed and the direction of wave propagation. This will allow discrimination between slow-mode and other waves (based on group speed), between fast-mode, Alfvénic waves (compressibility and propagation direction), and sausage modes (phase relationships, compressibility) \citep{HinodeReview2019}. Combining \muse\ observations with estimates of the time-averaged density (from \euvst) and the time-averaged magnetic field (from Cryo-NIRSP coronal spectropolarimetry) will assist with wave-mode identification. Doppler shift and broadening measurements at high cadence along loops and in footpoints can also establish how common wave generation is in the corona, a key requirement for models of the first ionization potential (FIP) effect, the relative enrichment of low FIP elements in the corona \citep{Laming:2015cr,Laming:2017wj}. More generally, \muse\ observations at high spatial and temporal resolution of the energy flux carried by Alfv\'enic waves will provide key constraints for models of the FIP effect, which invoke the ponderomotive force associated with Alfv\'en waves (see also Section~\ref{sec:wind}).

\subsubsection{Coronal seismology}
The NGSPM approach of coordinated  \dkist, \euvst, and \muse\ observations will also lead to exciting advances in our capability to diagnose conditions in the corona using the properties of waves. Until now coronal seismology has mostly depended on coronal imaging (e.g., using AIA, \citealt{Morton2013}) or low-resolution coronal spectroscopy (using imagers like CoMP, \citealt{Tomczyk2009}). The enormous increase in spatial resolution (0.5\arcsec\ vs.\ 12\arcsec\ for CoMP) and the multiplexing advantage offered by \muse\ allows unprecedented tracking of propagation and damping of Alfv$\acute{\rm e}$nic waves in the corona \citep{DeMoortel2014,Liu2015}. First, at the spatial resolution of \muse\ such waves will have larger amplitudes of order 10-20~km~s$^{-1}$ (based on \aia\ measurements, \citealt{McIntosh:2011fk}). Secondly, \muse\ will have 600 spatial resolving elements for each of CoMP’s, and 150 for each of the upgraded UCoMP instrument that is being commissioned in 2021. \muse’s FOV is large enough to capture coronal loops and trans-equatorial loops and the roots of polar plumes in coronal holes \citep{DeMoortel2014,Liu2015}, thus complementing the global-scale observations of (U)CoMP and providing key information to link with in-situ observations with \psp\ and \solo. 

\muse’s capability of measuring POS and LOS motions as well as unresolved motions (from broadening), combined with $\omega$-k diagrams and time-distance seismology will settle many of the unresolved issues. \muse\ will be able to determine the ratio of the inward and outward propagating wave power and determine whether wave-mode coupling or dissipation play a role in the imbalance observed with CoMP \citep{Tomczyk2009}. This can be done by studying the various spatio-temporal correlations between intensities, velocities, and broadening predicted by wave dissipation models \citep{Antolin:2018fk,HinodeReview2019}.

In addition to propagating Alfv\'enic waves permeating the solar atmosphere, ubiquitous low-amplitude decay-less transverse MHD oscillations have recently been discovered \citep{Anfinogentov_2015AA...583A.136A}, which open the path for coordinated \euvst\ and \muse\ spectra to estimate the magnetic field in the corona at high cadence and resolution over a large FOV \citep{Anfinogentov_2019ApJ...884L..40A}. This is difficult to accomplish through direct methods, even with \dkist’s CryoNIRSP instrument, which is a single slit spectrograph at medium spatial resolution. It can however be achieved on a routine basis using measurements of the coronal density from relatively fast \euvst\ rasters at 0.8\arcsec\ resolution and wave group speeds from \muse’s time-distance seismology. 

Similar synergies between \euvst\ and \muse\ will provide unique coronal diagnostics by exploiting the presence of slow-mode waves. \euvst\ measurements of the temperature and density can provide insight into the coronal heating function \citep{Zavershinskii2019,Duckenfield2021}. \muse\ measurements of the damping of slow-mode waves as they propagate along coronal loops help determine the thermal conduction and polytropic index, both issues that have been the subject of controversy, but key to modeling the corona.

Combined \dkist, \euvst, and \muse\ observations of slow-mode waves can also be used to determine the magnetic field conditions above sunspots \citep{Jess2016} by measuring the wave phase speed (from \muse’s multi-slit measurements), the local sound speed (from \euvst’s temperature diagnostics), and the local density (from \euvst\ density sensitive line pairs). Such measurements provide much needed unambiguous constraints for and lead to improvements of magnetic field extrapolation methods (e.g., based on magnetograms from \dkist) that are used to track the evolution of free energy in active regions that are prone to flaring or eruptions.

\subsection{The role of magnetic flux emergence in the heating of the solar atmosphere}
\label{sec:flux_emergence}

The emergence of magnetic flux onto the photosphere and, later, higher up into the atmosphere occurs on a wide range of spatial scales from sub-granular to supergranular. When newly emerged flux reaches the coronal volume, reconnection with pre-existing magnetic field may play an important role in heating the plasma. However, the contribution of flux emergence relative to that of braiding, spicules, and waves remains unclear, and will certainly also depend on the type of region considered (e.g., quiet Sun, newly emerging or mature active region, or plage). Since emerging flux can traverse many layers in the atmosphere and its impact is felt on a wide range of scales, this topic benefits greatly from the combined NGSPM approach. 

\begin{figure*}[!ht]
    \centering
    \includegraphics[width=0.99\textwidth]{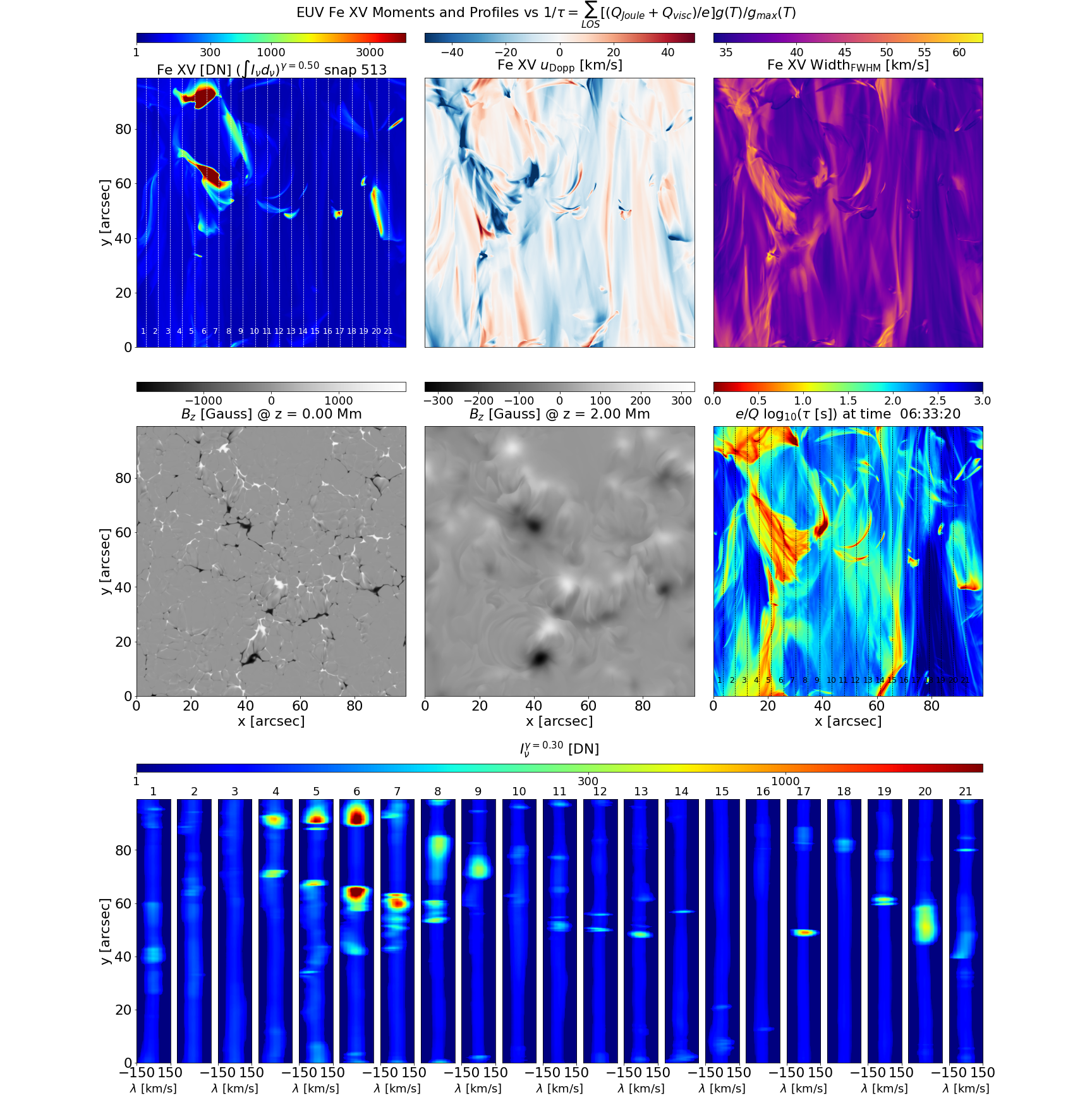}
	\caption{\label{fig:viggo2} Coronal heating resulting from flux emergence: magnetic flux is emerging in this simulation (model {\tt B\_nw072100}; see Table~\ref{table_sims} and Appendix~\ref{app:sim}), in particular in the upper left quadrant of this figure. As this flux penetrates into the corona it interacts with the pre-existing coronal field, causing increased heating rates. This causes increased intensities and line widths as well as accelerating plasma to high velocities more or less simultaneously along the ``wall'' of reconnecting field where the newly emerged and pre-existing field collide and are of equal strength. Here we show synthetic observables in the same format as for Figure~\ref{fig:viggo1},  but for the \fexvw\ line.
	The assumed exposure time is 1.5~s. Count rates are calculated as described in Appendix~\ref{app:synthesis}. Animations of this figure and the other two \muse\ spectral bands can be found online.
    }
\end{figure*}

Key measurements are the time evolution of the structure and strength of the photospheric and possibly the chromospheric fields, over an extended period of time and preferably over a spatial extent at least the size of a typical active region. These field measurements must be complemented by the intensity, velocity, and line width of coronal lines spanning a wide range of temperatures that cover a similar spatial extent in order to see how the corona reacts to the introduction of new fields.

Emergence of flux into the corona leads to a reorganization of the coronal magnetic field. This can rapidly produce highly dynamic local effects at the site of emergence, e.g., large-angle reconnection that produces strong flows and heating. However, emergence can also impact the corona at larger distances from the site of emergence, through its impact on the magnetic field connectivity and topology, and the rapid distribution of heat, flows, and waves along loops \citep{Archontis:2014yg,Hansteen2019}.

Highly sensitive measurements of the magnetic field in the photosphere (ViSP) and chromosphere (\dkist/DL-NIRSP or \sst/HeSP) can, for example, track the emergence of flux concentrations of a wide range of strengths. Simultaneously, \muse\ can observe coronal spectra along 37 slits at high cadence (of order 10-20~s)  providing the required view of both the local and non-local effects of flux emergence. 

An example of the sort of data this observation would provide is shown in Figure~\ref{fig:viggo2} where the synthetic observables from a  numerical model including flux emergence are shown: magnetic elements break through the photosphere roughly one hour before the time of Figure~\ref{fig:viggo2}, expanding and forming  bubbles of initially cool photospheric plasma that are lifted by the magnetic field into a region containing pre-existing network fields. 
At the time of the synthetic observation, the field has entered the corona and is strong enough to push the pre-existing field aside, or where field strengths are more or less equal, cause significant reconnection. This is visible as a set of 20\arcsec\ long features as observed with \muse\ in the \fexv\ 284 line, in the upper left quadrant of Fig.~\ref{fig:viggo2}. The intensity, Doppler velocity, and line width all show the impact of the emerging expanding field as it interacts with the ambient coronal field. We see high upflow velocities along a coherent structure which also is clearly visible in the increased line width there. These are signatures of reconnection and high heating rates along a current sheet that has formed as the newly emerging flux interacts and reconnects with the pre-existing field. Several sites of very broad and/or strongly shifted line profiles are visible along the current sheet.

To fully follow and disentangle the complicated set of events that emergence from the photosphere into the corona causes, such observations should be complemented with \muse\ context images (at both \heiiw\ TR and \fexiiw\ coronal temperatures). Simulations show that bright, low-lying, short-lived \heii, \feix, and \fexii\  loops will form, possibly partially obscured by EUV absorption from overlying cool plasma lifted into the coronal volume as the field rises. At sites where the angle of the reconnecting fields is large and temperatures are high we expect to see short lived \fexv\ loops as well and may also see flashes of \fexix\ as temperatures can reach 10~MK or more for a few tens of seconds at the reconnection site. 

When a significant amount of field emerges, as in a newly forming AR or ephemeral active region, the topology of the coronal field will change rapidly, driven both by reconnection and by footpoint separation. These changes will be visible in all bands except perhaps \fexix.
Much of the material carried up by the rising field will eventually drain out of newly heated coronal loops and there will be flow patterns associated with this draining concentrated near loop footpoints. This is another key measurement that distinguishes between effects from braiding and flux emergence.

Flux emergence will drive a host of highly dynamic events such as jets, surges, waves, and other brightenings that \muse\ will capture. 
\muse\ can also detect any signatures of oscillatory reconnection through the associated varying flow and heating patterns. Such reconnection is predicted by numerical simulations of reconnection around coronal null points \citep{Heggland:2009lr,Murray2009}.  

Studies of flux emergence will benefit greatly from the NGSPM approach. The magnetic field measurements with \dkist\ will provide insight into the amount of flux emerging and will constrain the field configuration. At the same time, fortuitously placed high-cadence dense \euvst\ rasters over a necessarily small FOV of order 5-10\arcsec $\times$ 140\arcsec\ could capture some of the local effects such as flows, waves, and heating, as loops emerge and traverse the chromospheric, TR, and coronal temperature regime. Coronal context from \muse\ is required to interpret these local measurements and capture the multitude of non-local effects at high spatial resolution. \muse’s multi-slit spectra over a $170\arcsec \times 170\arcsec$ FOV would thus be highly complementary to the local measurements that \euvst\ can provide for a range of temperatures spanning the chromosphere to the hot corona. \muse\ will be able to detect not only the very strong flows  ($> 100$ km~s$^{-1}$) expected from large-angle reconnection through Doppler shift and line broadening measurements \citep{Tian2018,Hansteen2019}, but also the predicted hot plasma ($\sim 10$~MK), which often does not occur right at the emergence site. In addition, \muse\ can cover the full extent of the emerging loops and make the critical distinction between local flows caused by field reorganization and evaporative flows that occur in response to local and non-local heating events. Complementary higher-cadence lower resolution rasters with \euvst\ would not capture the large-scale effects of emergence at high resolution, but would provide  low-resolution diagnostics, such as the evolution of densities and abundances as the emergence progresses. These combined measurements will thus be able to capture the multi-scale process of flux emergence and help distinguish it from other heating processes.

\subsection{Driving mechanisms at the roots of the solar wind}
\label{sec:wind}

There are many candidate processes that are thought to feed mass and energy into the roots of the solar wind. For example, jets of various sizes, from spicules to coronal jets, are ubiquitous, highly dynamic on timescales of order 10-20s, and associated with strong \alfven\ waves \citep{de-Pontieu:2007bd} that propagate into the fast solar wind along, e.g., large-scale polar plumes at speeds of hundreds to thousands of km~s$^{-1}$ \citep{Cirtain2007}. These jets are thought to originate from the interaction between magnetic fields on very small spatial scales (e.g., from flux emergence; \citealt{Nobrega-Siverio:2016qf}), but they extend over 15-80\arcsec\ at maximum length \citep{Cirtain2007}, and some appear to involve large-scale eruptions of mini-filaments into the solar wind \citep{Sterling2015}.  Similarly, the origin of some of the mass supply to the slow solar wind is thought to lie in the AR outflow regions \citep{Doschek:2008qy}, strong coronal upflows most often found at both the leading and trailing edge of active regions. While such regions persist for days, observations show a prevalence of highly dynamic events at the TR and low coronal roots of such regions on granular scales \citep{De-Pontieu:2009fk,McIntosh:2009yf,McIntosh:2009lr,Ugarte2011,Polito2020}. However, models predict that large-scale connectivity changes (including over distances of order 100\arcsec\ between the opposite edges of the active region) play a key role in their formation, through interchange reconnection \citep{Baker2009,Baker2017}.

A combined NGSPM approach is thus desired to address how these phenomena at the roots of the solar wind form and how they contribute to the mass and energy budget of the wind. \dkist\ measurements of the magnetic field in the photosphere and chromosphere (ViSP and DL-NIRSP) will determine the relative role of various mechanisms in generating jets: flux emergence, flux cancellation, and the formation (and subsequent destabilization) of mini-filaments through small-scale interactions between neighboring flux concentrations \citep{Sterling2015,Sterling2016,Moore2018}. Such highly sensitive measurements will also shed light on the likelihood of interchange reconnection, both at the roots of coronal holes and in AR outflow regions. Dense \euvst\ rasters at cadences of order 30 seconds over regions of order 5\arcsec\ $\times$ 140\arcsec\ will provide insight into the temporal evolution of the strong plasma flows (and thus mass flux, from density sensitive line pairs) and heating associated with jets. However, high cadence \muse\ rasters and coronal images over a large FOV ($170\arcsec \times 170\arcsec$) are needed to capture the further expansion of these flows and their impact on the large-scale coronal structures feeding into the solar wind. In particular, \muse’s capability of time-distance seismology \citep{Tomczyk2009}, tracing of wave packets, and determination of group speeds of waves is key to determine the properties of \alfven\ waves (including wave energy flux by combining with time-averaged densities derived from slower large-scale \euvst\ rasters), thereby constraining numerical models of the solar wind \citep{van-der-Holst:2014op, Matsumoto_Suzuki_2014MNRAS.440..971M, Shoda_2019ApJ...880L...2S}. \muse\ measurements will also help determine the role of unidirectional wave propagation \citep{Magyar_2017NatSR...714820M} or counter-propagating waves in generating turbulence and higher frequencies \citep{DeMoortel2014, Shoda_2018ApJ...859L..17S}, a key topic for solar wind studies \citep{Banerjee_2020arXiv201208802B}. Similarly, \muse\ observations will have the cadence and resolution to investigate the interplay between Alfv$\acute{\rm e}$nic waves and field-aligned motions (e.g., from spicules and other jets; \citealt{Liu2015}) at the root of the solar wind. All of these \muse\ measurements of the Alfv\'en waves and plasma properties at the roots of the solar wind will provide key constraints for in-situ measurements in the solar wind with \psp\ and \solo, including studies of the nature of switchbacks \citep{Kasper2019}, which are difficult to trace back to the corona without high-resolution spectroscopy of the corona.  

\begin{figure*}[!ht]
    \centering
    \includegraphics[width=1.0\textwidth]{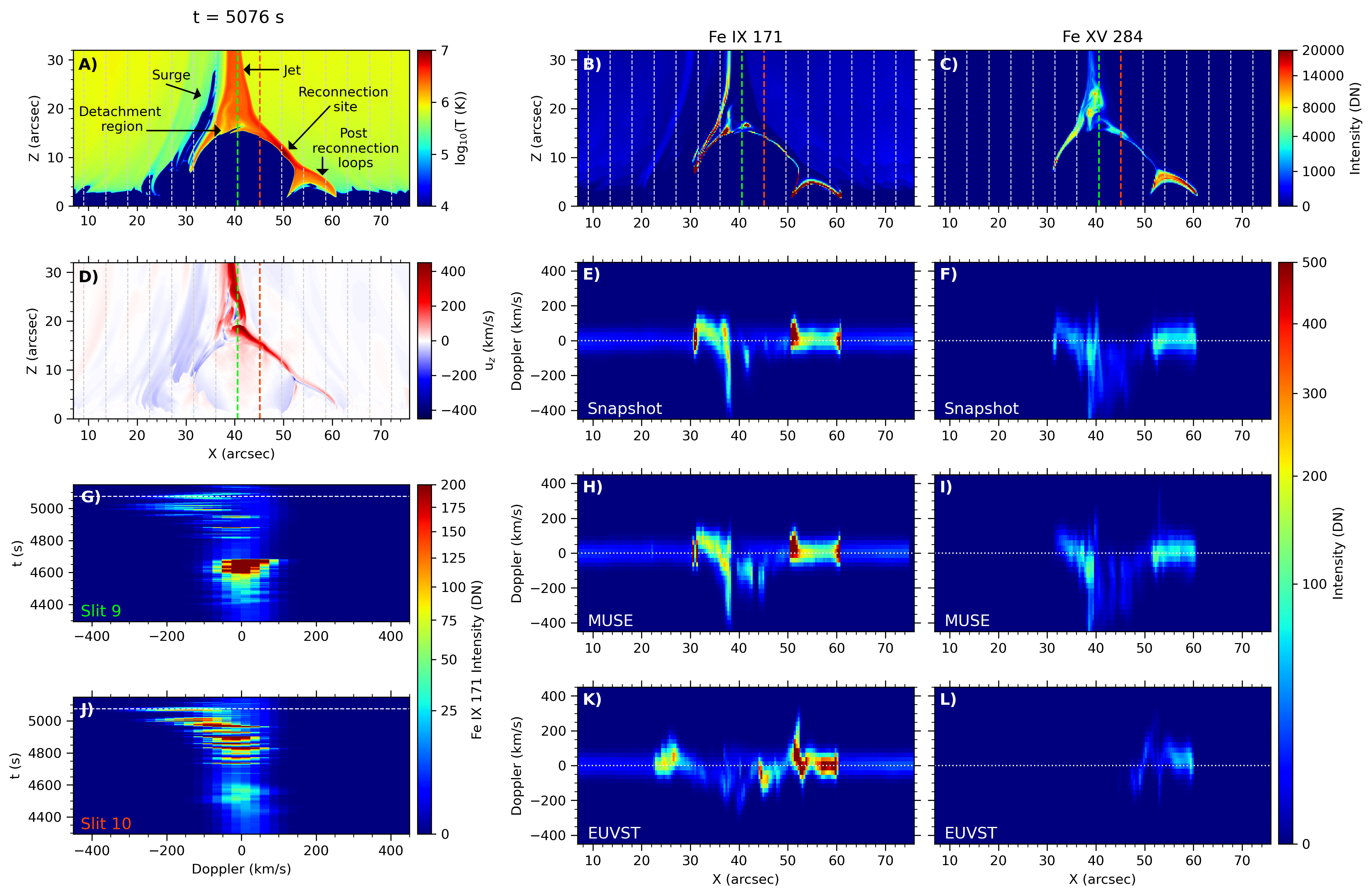}
    \caption{A coronal hot jet, from model \Bnpdns\ (see Section~\ref{sec:sim} and Appendix~\ref{app:sim})  at time $t=5076$~s .
    The limb view maps of the temperature, \feix\ and \fexv\ emission, and vertical velocity are shown in Panels, A, B, C and D respectively, with
    superimposed dashed lines indicating the location of the \muse\ slits. In Panel A, arrows mark the location of the most important regions of the simulation. Panels E-F contain the top view of spectral profiles of \feixw\ and \fexvw\ at that particular time in the simulation ($t=5076$s), assuming $2$s exposure times. Panels G and J show  \feix\ maps of wavelength versus time for the colored slits 9 and 10 (shown in panels A through D), where the horizontal line indicates the time of the maps of panels A-D. Panels H-I mimic a \muse\ dense raster while Panels K-L correspond to an \euvst\ dense (single slit) raster, with 2s slit-dwelling time and 0.4\arcsec\ steps (assuming for both instruments 0.4\arcsec\ slit width and 0.167\arcsec\ pixels along the slit). The raster duration to scan the whole numerical domain (88\arcsec) for \muse\ and \euvst\ are, respectively, 22s and 440s. The \euvst\ raster does not capture the highly dynamic evolution of this type of event, as also shown in the movie: an animation of this figure is available online.
    }
	\label{fig:daniel}
\end{figure*}

To illustrate \muse's capabilities in terms of studying the driving mechanisms of the solar wind,  Figure \ref{fig:daniel} shows a coronal hole jet in the \Bnpdns\ simulation. 
A bi-directional hot jet (Panel A) is launched from the reconnection site at high speeds (Panel D) as a by-product of magnetic flux emergence and reconnection. The coronal jet propagates upwards, even reaching the top of the numerical box at $88$\arcsec, thus potentially constituting a source of significant mass and energy input to the solar wind \citep[see the recent review by][]{Raouafi2016}. 
The strong shocks generated by the reconnection outflows produce
a detachment process in which cool material from the emerged region is peeled and launched as a surge. These shocks are key to giving the jet the canonical inverted-Y (or Eiffel tower) shape typically found in coronal hole jets \citep[see][for details about this mechanism]{Nobrega-Siverio:2016qf}. As a consequence also from the reconnection, new hot retracting loops (post reconnection loops) are created and piled up in the lower atmosphere \citep[see, e.g.,][]{Moreno-Insertis:2013}.  
The regions with enhanced intensity in \feix\ and \fexv\ seen from the limb (panels B-C, respectively)
are located at the detachment region, the hot jet, and the post reconnection loops. Plasmoids are also visible in \feix\ in the attached movie.
Spectral maps seen from the top (Panels E-F) reveal strong blue shifts (from $-100$ to $-400$ km s$^{-1}$)
associated with the upflow jet and the reconnection site as well as red shifts ($\sim 100$ km s$^{-1}$) located at the detachment region and the retracting post-reconnected loops. 

Panels G and J show \feix\ spectral-time maps in two different slits (colored ones in the context maps) that scan the reconnection site and show the rapid evolution of plasmoids \citep[see][for details about \muse's capabilities concerning plasmoids]{Cheung2021_muse_fl}. In panels H-I, we have mimicked a \muse\ dense raster with 2~s and 0.4\arcsec\ step raster, considering 0.4\arcsec\ slit width and 0.167\arcsec\ pixels along the slit. Despite the very fast plasma dynamics (see the associated movie), \muse\ is capable of spatially and temporally resolving the structure and evolution of all the regions that are critical for understanding this potential source of fast solar wind. For comparison, panels K-L contain a synthetic dense raster from the single slit of \euvst\ assuming the same exposure time as for \muse\, but a much longer raster cadence to cover the whole FOV of the numerical domain (88\arcsec). In this case we assume that \euvst\ is fortuitously placed and covers the emerging flux region and associated jet in the center of its FOV. Such an assumption is not required for \muse\ since its large FOV of $170\arcsec \times 170\arcsec$ can comfortably cover a region that is four times as large as this numerical domain. Despite this advantageous assumption for \euvst\, it cannot capture the highly dynamic evolution and key phenomena associated with the flux emergence and resulting jet, as shown in panels K-L. The high cadence rasters of \muse\, in contrast, capture the dynamic evolution fully. Consequently and as mentioned above, probably the most optimal observing program for \euvst\ is a dense raster of 5\arcsec\ $\times$ 140\arcsec\ to provide density diagnostics, chemical composition and better temperature coverage of the strong upflows. The combination of \muse\ and \euvst\ diagnostics of, respectively, Alfv\'enic wave propagation and abundance variations, will allow studies that trace the connectivity between the low corona and the solar wind \citep{Brooks2011}.

High-cadence \euvst\ rasters of single AR outflow regions will elucidate the intermittency and dynamic nature of these outflows, an important constraint for theoretical models, e.g., those based on interchange reconnection between open and closed fields. However, \muse\ observations are critical to determine the large-scale coupled dynamics, topology, and connectivity between outflow regions on opposite sides of the active region (i.e., over spatial scales of 100\arcsec\ or more), that is predicted by some interchange reconnection models \citep{Baker2009,Baker2017}. Doppler shifts from \muse\ spectra and coronal context images at high resolution will elucidate whether some of these upflows are on closed small-scale loops, resulting from reconnection around QSL as predicted by some models \citep{Baker2009,Baker2017}.

\subsection{Formation mechanisms of solar prominences}
\label{sec:promin}

Prominences play a key role in the solar atmosphere; when they become unstable they are particularly important for understanding space weather. However, a full understanding of the formation of prominences remains a major challenge, mostly because it is difficult to capture at high resolution and high cadence all aspects of these complex, multi-scale and multi-thermal phenomena that traverse all layers of the solar atmosphere. A combined NGSPM approach is thus important to address major outstanding challenges. For example, highly sensitive measurements of the magnetic field (e.g., from \dkist) are required to understand the formation of flux ropes (e.g., from cancellation; \citealt{vanBallegooijen1989,Green2011,Chintzoglou2019}). Such measurements will also elucidate the overall topology of the magnetic field (including dips and twisted flux ropes) thought to be required for accumulation of cool plasma at coronal heights \citep{Liu2012,Keppens2014}. Similarly, high-cadence rasters of \euvst\ and simultaneous \muse\ rasters and coronal images are key to constrain the evaporation-condensation model (also known as thermal non-equilibrium, TNE), the leading candidate mechanism for prominence and coronal rain formation \citep{Antolin2020}, and the recently discovered long-period intensity pulsations \citep{Froment_2015ApJ...807..158F}. In this model, sustained high-frequency heating (typically longer than the radiative cooling time of the structure) concentrated towards the footpoints of coronal structures is thought to lead to cycles of heating and cooling around an equilibrium. During one cycle, the evaporation of material caused by the heating ultimately condenses in the solar corona via thermal instability during the cooling phase \citep[e.g.,][]{Muller_2003AA...411..605M}. 

Dense \euvst\ rasters at high cadence will capture the triggering and formation of coronal rain as it cools from MK to chromospheric temperatures \citep{Antolin2015b}, while Doppler shift measurements will reveal waves and dynamic instabilities \citep[e.g., KHI][]{Okamoto2015,Antolin2015b}, and reconnection-driven nanojets \citep{Antolin2021} thought to play a role in heating prominence plasma. \muse\ TR and coronal images and multi-slit spectra will elucidate the overall mass and energy circulation at the extreme spatial and temporal scales produced by the TNE, and the signatures of dissipation of energy through various dynamic instabilities and reconnection processes. his includes impulsive flare events where coronal rain is observed and difficult to explain \citep{Reep2020}. Such data will also reveal the build-up of twist and braiding in the flux rope, key properties for understanding the (in)stability of prominences \citep{Schmieder2017,Cheng2015}. In addition, \muse\ observations will allow studies of the impact of jets on prominences and the resulting large-amplitude oscillations and counterstreaming flows in prominences \citep{Luna2021}.

\begin{figure*}[!ht]
    \centering
    \includegraphics[width=0.98\textwidth]{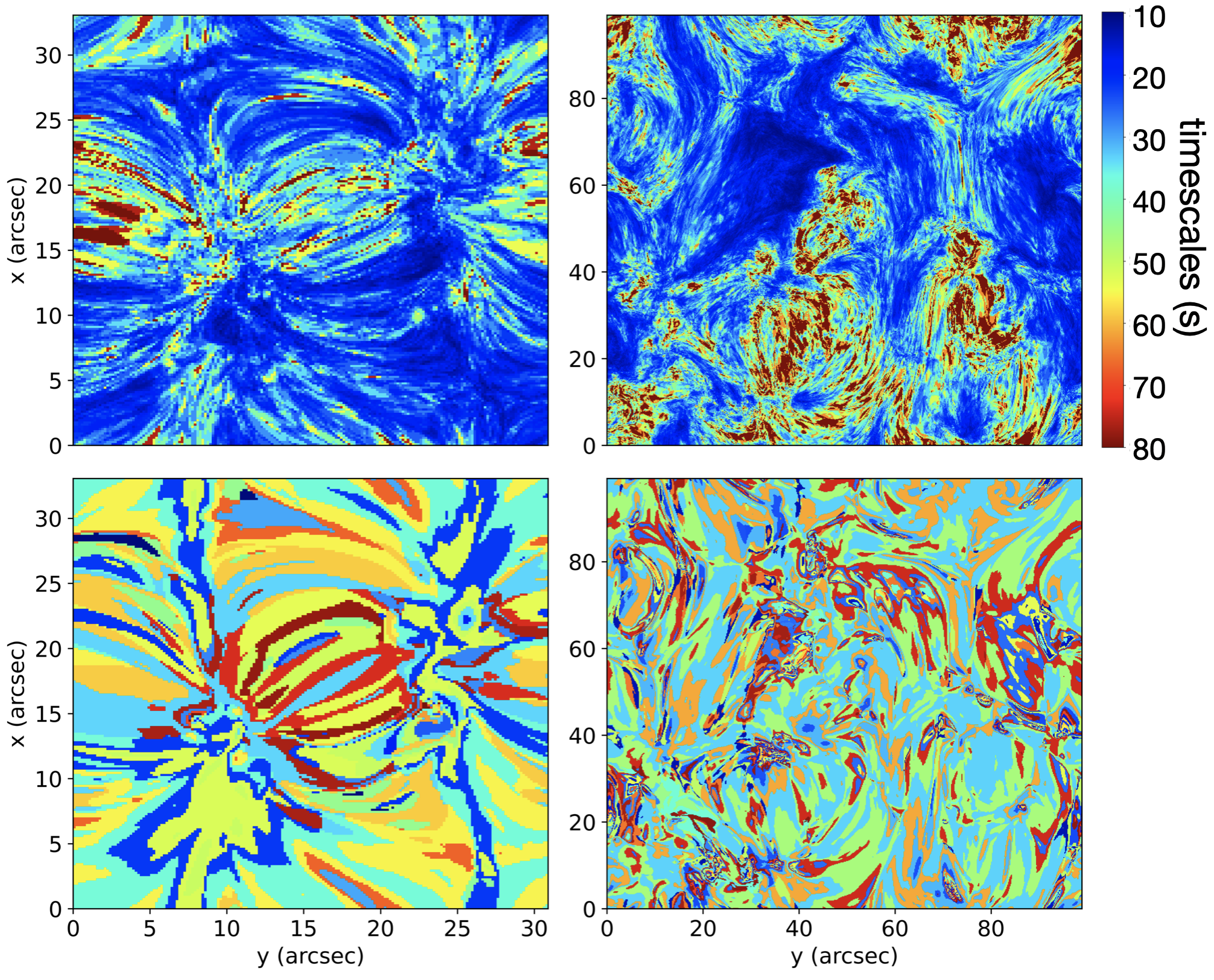}
  \caption{\label{fig:timesca} Times-scales and spatial coherence highlight the importance of scanning several tens of arcsec within a few tens of seconds to resolve the spatial and temporal properties of coronal features (and underlying physical processes). Times-scales computed as described in the Appendix~\ref{app:timescales} are shown for the \Bhion\ (top left) and \Bnw\ (top right) simulations (see Appendix~\ref{app:sim} for details). Here we show results for \feixw, but results are very similar for \fexvw. In the bottom row we show, for each simulation and a single time step, maps of the spatial distribution of clusters of locations with similar values for the first three moments of the spectral line, as derived using a supervised $k$-means algorithm (see Appendix~\ref{app:timescales} for details). The two numerical simulations represent regions with different magnetic topologies and associated spatial scales: a rather small network region (\Bhion) and a supergranular scale of a stronger network region (\Bnw).}
\end{figure*}

\section{Discussions and Conclusions}\label{sec:con}

Understanding the heating of a stellar atmosphere like the Sun's corona remains a major open issue in astrophysics \citep[e.g.,][]{Testa2015}. This is in part because many different complex physical processes are thought to play a role, which until recently have been difficult to capture in theoretical models. In addition, the spatial and temporal scales on which these processes occur have been difficult to capture with existing instrumentation, as these processes often couple plasma across widely different temperatures (from the 0.01~MK chromosphere to the 10~MK corona), and couple a wide range of spatial scales from the very small ($\sim$~0.5\arcsec) sub-granular scale to active region size scale ($\sim$~140\arcsec), all on short timescales (of order 20s). 

The past few years have seen major advances in the complexity and realism of numerical modeling of the physical processes that are thought to play a role in heating the solar atmosphere. The advances in algorithms and computing facilities now allow physics-based models, often based on state-of-the-art radiative MHD simulations, that include enough realism to allow the calculation of synthetic observables. Comparisons with observations allows direct diagnostics of the primary physical processes in the model and provide constraints on the models. The model predictions indicate that imaging observations that only capture the intensity of spectral lines are not enough: Doppler shifts and line broadening are needed to provide key diagnostics of waves, reconnection-driven flows, heating, etc. Similarly, spectroscopic observations of the corona at either low cadence over a large FOV (minutes over 140\arcsec $\times$ 140 \arcsec), at high cadence over a small FOV (20~s over $5\arcsec \times 140\arcsec)$, or at low spatial resolution ($> 1$\arcsec) are not sufficient to discriminate between the various models and determine which physical mechanisms dominate the heating of the solar corona.  

Because of the multi-scale nature of the physical processes involved, high-resolution (0.5\arcsec) coronal imaging spectroscopy at high cadence (20~s) and over a large FOV ($>$ 100 \arcsec), as well as high-resolution (0.5\arcsec), high cadence context coronal imaging are required to discriminate between these various mechanisms. As described in Section~\ref{sec:results}, the most recent models indicate that this is the case for heating associated with chromospheric jets, such as spicules (Fig.~\ref{fig:spicmaps}, \ref{fig:spicmom}); heating caused by magnetic reconnection from braiding (Figs.~\ref{fig:viggo1}, \ref{fig:viggo2_csr}, \ref{fig:fabio2}), dissipation of Alfv\'enic waves (Figs.~\ref{fig:matsufft}, \ref{fig:mah1}, \ref{fig:patrick2}, \ref{fig:patrick0}, \ref{fig:patrick1}), flux emergence (Figs.~\ref{fig:viggo2}, \ref{fig:daniel}), or cancellation. Braiding models indicate that the current sheets (and associated heating and flow signatures) caused by jostling of magnetic field show spatio-temporal coherence over spatial scales of order 40\arcsec\ and temporal scales of order 20s. Similar coherence (often on even larger scales) occurs when magnetic field reconnects as a result of new flux emerging into the atmosphere on scales between granules and supergranules. Nanojets caused by reconnection from braiding lead to coronal loop formation many tens of Mm away from the reconnection site. Furthermore, Alfv\'enic waves, whether generated in the low atmosphere or corona, propagate over tens of Mm on timescales of order 20s, dissipate over longer or similar spatial scales, and lead to tell-tale signatures that can only be identified when observing the whole loop. Finally, heating associated with spicules is predicted to occur many Mm away from the original spicular injection site into the corona. This is also illustrated nicely in Fig.~\ref{fig:timesca}, which shows that, at a spatial resolution of 0.5\arcsec, typical timescales (top row) on which the three moments of coronal lines (in this case \feixw) in the numerical models significantly vary (see Appendix~\ref{app:timescales} for a definition) are not only short (of order 20-60s), but also show spatial coherence over distances up to 60\arcsec\ (bottom row). Similar timescales have also been found for other simulations \citep{Einaudi2021}.

\begin{table*}[!htbp]
\centering%
\caption{\label{tab:observables_heating} Predicted \muse\ diagnostics for various models of heating mechanisms}
\begin{tabular}{|p{\dimexpr0.10\textwidth-1\tabcolsep-\arrayrulewidth\relax}|
                p{\dimexpr0.59\textwidth-1\tabcolsep-\arrayrulewidth\relax}|
                p{\dimexpr0.15\textwidth-1\tabcolsep-\arrayrulewidth\relax}|
                p{\dimexpr0.1\textwidth-1\tabcolsep-\arrayrulewidth\relax}|
              }
\hline
Mechanism & Predicted Diagnostic & $\lambda$\tablenotemark{*} [\AA] & Figures \\
\hline
Spicules & - type I and type II spicules & 304 & \ref{fig:spicmaps},
\ref{fig:spicmom}\\
\textcolor{white}{c} & - short-lived blue-shifted brightenings at loop footpoints associated w/ spicules & 171 & \ref{fig:spicmaps}, \ref{fig:spicmom}\\
\textcolor{white}{c} & - propagation of Alfv\'enic waves (Doppler and POS motions), triggered by spicules, along loops& 304, 171, 195, 284& \ref{fig:matsufft}a, \ref{fig:patrick2}-\ref{fig:patrick1}\\
\textcolor{white}{c} & - dissipation of Alfv\'enic waves through impulsively-driven KHI & 171, 284& \ref{fig:matsufft},  \ref{fig:patrick2}-\ref{fig:patrick1}\\
\textcolor{white}{c} & - formation of coronal loop, associated with spicules & 171, 195, 284&\ref{fig:spicmaps}, \ref{fig:spicmom}\\
\textcolor{white}{c} & - evaporative flows at loop footpoints, associated with spicules & 284 &\ref{fig:spicmaps}, \ref{fig:spicmom}\\
\hline
Braiding & - visibly braided loops & All& \ref{fig:fabio2}\\
\textcolor{white}{c} & - spatio-temporal coherence of intensity and line width along loops (20-60s, $\sim$ 5-30\arcsec)&171, 284, 108 & \ref{fig:viggo1}, \ref{fig:viggo2_csr}, \ref{fig:sanja1}\\
\textcolor{white}{c} & - short-lived ($\approx20$s single, $\approx60$s cluster) nanojets: high velocities ($< \sim 100$~km~s$^{-1}$) and line widths, transverse to guide field & 171, 284, 108 & \ref{fig:viggo1}, \ref{fig:paolo}, \ref{fig:braiding_prop}\\
\textcolor{white}{c} & - loop formation associated with nanojets & 171, 195, 284, 108 & \ref{fig:paolo}\\
\textcolor{white}{c} & - twisting and unwinding motions & 171, 195, 284 & \ref{fig:fabio2}\\
\textcolor{white}{c} & - evaporative flows in loops & 171, 284, 108 & \ref{fig:fabio}, \ref{fig:sanja1}, \ref{fig:radyn}\\
\textcolor{white}{c} & - nanoflare driven short-lived brightenings at loop footpoints, and associated short-lived hot loop emission & 171, 284, 108 & \ref{fig:sanja1}, \ref{fig:radyn}\\
\hline
Waves  & - propagating or standing oscillatory displacements of loops and jets & 304, 171, 195, 284 & \ref{fig:patrick2}-\ref{fig:kostas}\\
\textcolor{white}{c} & - oscillations in velocity, line width along loops& 171, 284& \ref{fig:ineke1}-\ref{fig:mah1},\ref{fig:patrick2}-\ref{fig:kostas} \\
\textcolor{white}{c} & - spatial dependence of FFT power spectrum along loop & 171, 284 & \ref{fig:matsufft}\\
\textcolor{white}{c} & - propagation of Doppler shift oscillations along loops& 171, 284 & \ref{fig:mah1}, \ref{fig:patrick2}-\ref{fig:patrick1} \\
\textcolor{white}{c} & - spatio-temporal coherence of velocities and line width along loops from wave propagation & 171, 284 & \ref{fig:ineke1}, \ref{fig:mah1}, \ref{fig:patrick2}-\ref{fig:patrick1}\\
\textcolor{white}{c} & - specific phase relationships between intensity, velocity, line width& 171, 284 & \ref{fig:patrick2}-\ref{fig:kostas}\\
\textcolor{white}{c} & - concentration of wave power at edge of flux tubes (KHI, RA)& 171, 195, 284& \ref{fig:patrick2}-\ref{fig:patrick1}\\
\textcolor{white}{c} & - steady downflows/upflows around edge of flux tubes (KHI, RA)& 171, 195, 284& \ref{fig:patrick2}-\ref{fig:patrick1}\\
\hline
Flux Emerg.\ & - short-lived, low-lying loops, possible EUV absorption from overlying cool plasma& 304, 171, 195, 284& \ref{fig:viggo2}\\
\textcolor{white}{c} & - flow patterns associated with draining of rising loops and topological evolution including footpoint separation & 171, 284, 195 & \ref{fig:viggo2}, \ref{fig:daniel}\\
\textcolor{white}{c} & - strong short-lived brightenings and bi-directional flows ($>$ 100~km~s$^{-1}$), large line widths (from large-angle reconnection) & 171, 284, 108& \ref{fig:viggo2}, \ref{fig:daniel}\\
\textcolor{white}{c} & - spatio-temporal coherence of highly dynamic "storms" of sudden brightenings and line width increase (10-30\arcsec, 20s) & 304, 171, 195, 284, 108& \ref{fig:viggo2}, \ref{fig:daniel}\\ 
\textcolor{white}{c} & - various types of jets, including erupting (mini-)filaments & 304, 171, 195, 284 & \ref{fig:viggo2}, \ref{fig:daniel}\\
\hline
\end{tabular}
\tablenotetext{*}{For 304 and 195 imaging is desired. For 171, 284, and 108, intensity, Doppler shift, and line broadening are typically desired. }
\end{table*}

All of these findings strongly indicate the need for an instrument like \muse\ to discriminate between these various mechanisms. Table~\ref{tab:observables_heating} attempts to capture in one table the various observable consequences for \muse\ for the different mechanisms that have been described throughout Section~\ref{sec:results}. It is not straightforward to capture all of the intricate predictions for each model into one small table, so we refer the reader to the various subsections and figures for more details. Nevertheless, this table shows that \muse\ will be able to provide critical constraints to these models on the spatio-temporal scales on which they make distinguishing and testable predictions. Detailed comparisons between \muse\ observations and these model predictions will also allow us to determine to what extent these phenomena occur on the Sun. Large-scale statistical studies of such comparisons for various solar targets will establish how common each of these phenomena are in the solar atmosphere, and establish the dominant processes in various solar targets (active regions, quiet Sun, coronal holes, etc.).

The predictions from the numerical models also indicate that coordinated observations across the whole atmosphere are desirable to address the science objectives outlined in the JAXA-NASA-ESA NGSPM study. First and foremost are measurements of the magnetic field and its dynamical evolution. Such measurements are easiest in the photosphere, but with the advent of \dkist\ (and other GBOs) are also becoming available at chromospheric and even coronal heights. Many of the science objectives we describe in Section~\ref{sec:results} benefit from coordinated observations between \muse\ and \dkist. For example, measurements of the magnetic field and plasma dynamics will greatly benefit our understanding of the driving mechanism of spicules, which can augment our understanding of the coronal impact of these ubiquitous features. Similarly, magnetic field measurements are important to help distinguish between heating caused by reconnection driven by recent flux emergence or cancellation, versus heating caused by other mechanisms. Measurements of the wave dynamics in the lower atmosphere can elucidate the source of Alfv\'en waves.

Similarly, coordinated observations of \muse\ and the recently selected single-slit \euvst\ spectrograph are desirable. While \muse\ can capture both the small and large spatial scales at high cadence, and simultaneously provide the coronal context, the large and seamless temperature coverage of \euvst\ will allow tracking of plasma as it rapidly heats and cools across a wide range of temperatures, from 10,000 K to 15 MK, while its slit-jaw imaging will provide access to the photospheric and chromospheric context. Combined \muse-\euvst\ studies will benefit our understanding of spicules, with \euvst\ rasters tracking the heating from chromospheric to coronal temperatures and \muse\ capturing the large-scale coronal loop formation. Similar synergies exist for braiding, flux emergence, and waves, as described in Section~\ref{sec:results}. For example, \euvst\ rasters will help identify the dominant wave processes in the lower solar atmosphere responsible for the observed Alfv\'enic waves in the corona, and provide stringent constraints on the transmitted wave energy flux available for coronal heating. Such measurements will be a great complement to the multi-slit rasters and imaging from MUSE that tracks the propagation and dissipation of such waves when they reach the corona.

Current observations most often do not properly constrain the models since they lack the resolution or throughput. Without new constraints at the right resolution and cadence, progress will remain limited as there is less of a driving need for improvements. All of these models, by necessity, make simplifying assumptions. It is simply not feasible from an algorithmic or computational point of view to capture all physical processes in one comprehensive model. For example, many of the 3D magneto-convective simulations have limited spatial resolution for computational reasons. While these models typically predict heating associated with reconnection from braiding, the low spatial resolution implies that the generation, propagation, and dissipation of waves is likely not properly captured. This in turns requires higher resolution models focused on wave dissipation processes that neglect self-consistent magnetoconvection. Similarly, the generation and impact of non-thermal electrons, thought to be a key component of heating resulting from reconnection, is most often not captured in 3D radiative MHD models, necessitating more simplified 1D approaches that capture the physics of the accelerated particles much better. All these types of models are highly valuable, but require both observational input to constrain the initial or boundary conditions (e.g., magnetic field distribution, input power spectrum for waves, etc.) and to allow further development in the right direction. Such development comes naturally as the models are challenged by discrepancies between predictions and novel constraints from high-resolution observations at the right scales and of the right nature (spectroscopy). This has, for example, been demonstrated by the interplay between observations with \iris\ and ground-based observations of chromospheric spicules (and their impact on the TR and corona), and successive improvements to numerical models using the Bifrost code \citep[e.g.,][]{Hansteen:2006,Martinez-Sykora:2017sci,Nobrega-Siverio:2020}, including the introduction of ion-neutral interaction effects such as ambipolar diffusion and non-equilibrium ionization. We expect that a similar approach with \muse\ will lead to a breakthrough in our understanding of the processes that heat the solar corona, and by extension, likely play a role in stellar atmospheres.

In summary, \muse, sometimes supplemented with other NGSPM instruments, will not only be able to uncover the telltale signatures of coronal heating as predicted by current state-of-the-art numerical models, but also provide key constraints that will lead to improvements of these models. This approach will lead to a breakthrough in our understanding of coronal heating.

\longacknowledgment

\appendix
\section{Numerical simulations}\label{app:sim}

In Table~\ref{table_sims} of Section~\ref{sec:sim} we summarize the numerical models we use in this paper to synthesize \muse\ observables and devise \muse\ diagnostics capable of testing these state-of-the-art models and distinguishing between different models. 
Here we provide further details on all of the models used in the paper, and listed in Table~\ref{table_sims}.

{\bf Bifrost:}
The Bifrost code \citep{Gudiksen:2011qy} aims to address the most relevant physical processes in the outer solar atmosphere, i.e., photosphere, chromosphere, TR, and lower corona.  The boundary conditions are periodic in the horizontal directions; in the vertical direction, the bottom boundary is open and sets constant entropy for the plasma entering into the domain, and the top boundary uses characteristic boundary conditions to allow waves to exit the computational domain with minimal reflection. The code includes: 1) optically thick radiative transfer including scattering, which is most important in the photosphere and lower chromosphere \citep{Skartlien2000,Hayek:2010ac}; 2) radiative losses and gains in the upper chromosphere and TR through recipes derived from detailed non-LTE calculations using RADYN \citep{Carlsson:2012uq}; 3) optically thin radiative losses in the corona based on Chianti emissivities \citep[e.g.,][]{Chianti2021}; 4) thermal conduction along the magnetic field; and it can also include: 5) ion-neutral interaction effects using the generalized Ohm's law  \citep[GOL, ][]{Martinez-Sykora:2017gol,Nobrega-Siverio:2020}; 6) and ionization balance in non-equilibrium for hydrogen and helium \citep{Leenaarts:2007sf,Golding:2016wq}; 7) non-equilibrium ionization of minority species \citep{Olluri:2013uq}. The last three physical processes are not always included due to computational costs.  

The Bifrost code is especially well suited to studying the details of interactions between the photosphere and chromosphere and the overlying corona. Here we analyze results from several Bifrost atmospheric models (see Table~\ref{table_sims}) with different properties, allowing to study a variety of solar conditions and phenomena:
\begin{itemize}
    \item Model \Bgol\ is a 2.5D MHD numerical experiment aimed at modeling the acceleration of and heating due to spicules (see Sec.~\ref{sec:spic}, and Figures~\ref{fig:spicmaps} and~\ref{fig:spicmom})  This simulation covers a spatial extent of $96\times 43$~Mm$^2$ using $6930\times 1554$ grid points on a grid with constant horizontal cell size of $14$~km, and a variable vertical grid with grid cells concentrated where gradients are large and scale heights small, i.e., in the photosphere, chromosphere and transition region. The simulation extends from $-2.5$~Mm to $40$~Mm where $z\sim0$ is the height of the photosphere. This model includes the effects of a generalized Ohm's law and is analyzed in detail in \citet{Martinez-Sykora:2017gol,Nobrega-Siverio:2020}. This model produces several spicules with properties similar to observed spicules of type~{\sc ii} \citep[e.g.,][]{Martinez-Sykora:2017sci,DePontieu:2017net}, and the processes leading to the spicule formation  also produce coronal heating in this model \citep[][see also section~\ref{sec:spic}]{De-Pontieu:2017pcd}.
    \item Model \Bnw, is a relatively large Bifrost simulation ($72  \times 72 \times 60$~Mm$^3$, extending from 8.5~Mm below the photosphere to 52~Mm above) designed to model the chromospheric network and overlying coronal regions \citep{Hansteen_AGU2020}. This spatial region is covered by a $720\times 720\times 1115$ grid, using a constant horizontal grid size, of 100~km, and a variable vertical grid size where the regions with large gradients are covered with the smallest grid size of $20$~km, the outer corona and deep convective zone have a vertical grid size of $100$~km or less. The experiment has been run for several hours solar time, initially with magnetic field strengths typical of network with an average unsigned flux of 75~Gauss. At later times a strong magnetic flux sheet ($B_y\approx 2000$~G) is injected for a period of 2.5~hours that eventually reaches the photosphere and leads to flux emergence modified by convectively driven photospheric dynamics, at which time the average unsigned flux grows to roughly 100~Gauss. This simulation shows self-consistent coronal heating through braiding (see Section~\ref{sec:nano}) as well as from the effects of flux emerging through the chromosphere into the corona (see Section~\ref{sec:flux_emergence}).
    \item The \Bhion\ is an experiment designed to study the chromosphere and corona for magnetic field conditions similar to ``enhanced network'', in which the magnetic field plays and important role dynamically and energetically mainly through braiding and reconnection. This spatial region covers a relatively small domain: $24  \times 24 \times 16$~Mm$^3$, with 2.5~Mm convection zone below the photosphere and 14~Mm of solar atmosphere. The computational box is spanned by $504\times 504\times 496$ grid zones, with a constant horizontal grid size of 48~km, and, as in the previous models, a variable vertical grid size. The model includes non-equilibrium hydrogen ionization and is described in detail in \citet{Carlsson:2016rt}.
    \item  The \Bemer\ is an experiment designed to study the chromosphere and corona for magnetic field conditions similar to ``enhanced network'', in which the magnetic field plays and important role dynamically and energetically mainly through braiding supplemented later in the simulation with flux emergence. The model covers a relatively small domain: $24  \times 24 \times 16$~Mm$^3$, with 2.5~Mm convection zone below the photosphere and 14~Mm of solar atmosphere. The computational box is spanned by $768\times 768\times 768$ grid zones, with a constant horizontal grid size of 31~km, and, as in the previous models, a variable vertical grid size. This model is unpublished but is essentially identical to that described in \citet{Hansteen2019}, differing only in the amount and location of flux injected into the bottom boundary: this model has a slightly smaller amount of flux injected than that experiment.
    \item Model  \Bnpdns\ is a 2D simulation aimed at studying coronal bright points and their conspicuous emission in the extreme-utraviolet and X-rays (see Section~\ref{sec:wind}, and Figure~\ref{fig:daniel}).  The initial condition was created imposing a potential nullpoint configuration 8~Mm above the solar surface (i.e., 10.8~Mm above the bottom boundary), in the corona over a pre-existing statistically stationary 2D snapshot mimicking a coronal hole. It encompasses a domain from the uppermost layers of the solar interior up to the corona. The physical domain is $0.0$~Mm $\leq x \leq 64.0$~Mm and $-2.8$~Mm $\leq z \leq 67.0$~Mm, where $z=0$~Mm corresponds to the solar surface. This domain is solved with $4096 \times 4096$ grid cells using a uniform numerical grid, in both the horizontal and vertical directions, with a fine grid size of $\Delta x \approx 15.6 $~km and $\Delta z \approx 17.0 $~km, respectively.
 
\end{itemize}

{\bf MURAM:}
The MURaM code is similar in concept to the Bifrost code and can cover a spatial range from deep in the convection zone to a coronal scale height ($\sim 50$~Mm) or more.  The simulations presented here are based on the coronal extension of the MURaM code as described in \cite{Rempel:2017zl}, and includes: single fluid MHD, 3D grey radiative transfer, a tabulated LTE equation of state, Spitzer heat conduction, and CHIANTI based optically thin radiative losses in the corona. As for the Bifrost experiments, the Poynting flux that heats the chromospheric and coronal parts of the simulation domain is generated through magnetoconvection in the photosphere and convection zone. 
Here we analyze model \Mplhe\ which is aimed at reproducing the plasma dynamics of solar plage, regions of moderate magnetic activity (see Section~\ref{sec:nano}, and Figure~\ref{fig:sanja1}). The simulation domain has an extent of $40 \times 40  \times 22$~Mm, with $8$~Mm protruding below the photosphere.  The resulting model is generated in phases, similarly to previous runs. The initial magnetic field of $200$~G is added to well developed non-magnetic convection simulation to form extended magnetic field concentrations at meso- to super-granular spatial scales. The computational domain was then extended to include the upper solar atmosphere and the magnetic field from the pre-existing simulation was used for potential field extrapolation into the rest of the domain. The new simulation was then run until a relaxed state is achieved. In this model, the additional bipolar flux system is advected through the bottom boundary over an ellipsoidal flux-emergence region with the major axes $(a, b) =  (3,  1) $~Mm and $B_0=8000$~G field strength \citep{Cheung2019}.  The emergence resulted in a flare after $4.6$ hours of solar time and that part of the  model is analyzed in the companion paper to this, studying flares and eruptions \citep{Cheung2021_muse_fl}.

{\bf PLUTO 3D MHD loop models heated by braiding:}
We use 3D MHD simulations of coronal loops using the PLUTO code \citep{Mignone2007a,Mignone2012a}, a modular, Godunov-type code designed for modeling astrophysical plasmas. The plasma is assumed to be fully ionized and we include optically thin radiation and thermal conduction along the magnetic field. 

\begin{itemize}
  \item Model \Ptwist: We model a loop that has been straightened into a magnetic flux tube rooted at both ends in the photosphere through two chromospheric layers at opposite sides of the box (top and bottom boundaries) \citep{Guarrasi2014a}.
  The cylindrical box is $[r, \phi, z]= [384, 256, 768]$ cells, with $-z_M < z < z_M$ along the loop axis where $z_M = 3.1 \times 10^{9} $~cm, $r_0=7 \times 10^{7} \leq r \leq r_M=3.5 \times 10^{9}$~cm across the loop, and  $0 \leq  \phi \leq 90^o$ in the azimuthal direction. To describe the transition region at sufficiently high resolution, the cell size there ($|z| \sim 2.4 \times 10^{9}$~cm) decreases to $dr \sim dz \sim 3 \times 10^6$~cm.  The resolution is uniform in the angle $\phi$, i.e., $d\phi \sim 0\degr.35$.
  The loop atmosphere consists of a corona connected to two thick and isothermal (20,000 K) chromospheric layers by thin transition regions, immersed in a magnetic field. The magnetic field is arranged to be mostly uniform in the corona and strongly tapering in the chromosphere, where the ratio of  thermal  to  magnetic pressure switches from low ($\beta < 1$) to high values ($\beta > 1$). The coronal magnetic flux tube is progressively twisted by the rotation of the plasma at the footpoints. A complete description of the model and of the results can be found in \cite{Reale2016a}.
   Heating is produced through an anomalous diffusivity that allows magnetic reconnection when gradients in the field force the current to go above a current density threshold. Thus, currents grow due to the progressive twisting of the magnetic field. The field is twisted by random rotational plasma motions at the loop footpoints, which drag the field which is line tied in the dense photospheric plasma (see Section~\ref{sec:nano}, and Figures~\ref{fig:fabio} and~\ref{fig:fabio2}).  The evolution of the plasma and magnetic field in the box is described by solving the full time-dependent MHD equations including gravity (for a curved loop), thermal conduction (including the effects of heat flux saturation), radiative losses from optically thin plasma and the anomalous magnetic diffusivity.   The model describes the 3D MHD flux tube (loop) evolution in the time range $0 < t < 2500$ s until the loop plasma reaches a maximum temperature $T \sim 4$ MK, and with a final total rotation angle is $\approx 2.7 \pi$.
  
  \item Model \Pnano: We model two neighboring loops straightened into magnetic flux tubes by letting an initial configuration with two separated concentration of magnetic flux relax. At the two ends of the straighten loops we place a high-$\beta$ layer of the atmosphere that plays the role of the chromosphere, then have the transition region, and the corona in the central part of the computational domain. The initial configuration is constructed by considering a gravitationally stratified solar atmosphere, and two flux tubes representative of coronal loops are defined by a stronger magnetic field. The numerical domain extends over 48.7~Mm in the horizontal direction with a non-uniform grid,  with the highest resolution achieved in the central 20~Mm part of the domain, where the grid cell size is 60~km. The domain is 62~Mm in the vertical direction (i.e., along the loop), covered with 128 points distributed along a stretched grid in the coronal part of the domain. Details on the numerical modeling can be found in \citep{Antolin2021}. We take as $t = 0$ the time after relaxation, when the two loops are in pressure balance and have stopped evolving. At this time the magnetic field in the corona settles to a value of approximately 15~G.
  In the driving phase, we shift the footpoints of the loop in opposite direction at either extremes of the straightened loops in a direction perpendicular to both the field-aligned direction and the direction of separation between the loops.
  In this phase, the coronal part of the loops follow the footpoint motion and a X-shape configuration of the loop is created at the centre of the domain. When the tilt is enough to trigger the magnetic reconnection, this starts leading to an increased magnetic tension that accelerates two jets sideways from the reconnection region.
  During the reconnection event,$\sim 10^{28}$~erg of magnetic energy are converted into thermal and kinetic energy, leading to maximum temperatures of $5$~MK and maximum velocities of $\sim 250$~km~s$^{-1}$. The short-lived reconnection jets disappear in $\sim 60$~s (see sec.~\ref{sec:nano}, and Figure~\ref{fig:paolo}). 
\end{itemize}

{\bf PLUTO 3D MHD loop model heated by waves:}
Model \Pwaves\ consists of a straight flux tube of radius 1~Mm and length $100$~Mm, consisting of gravitationally stratified plasma. The flux tube models half of a coronal loop, from its footpoint to its apex. The coronal background density at the footpoint is $\sim 0.84 \times 10^{-15}$~g~cm$^{-3}$, which is three times lower than the loop density at the footpoint. A straight magnetic field of $B_z = 22.8$~G is considered, while the temperature varies across the tube axis (on the $xy$-plane), from $0.9$~MK inside the tube to $2.7$~MK outside. Continuous periodic driving is applied at the footpoint of our flux tube, through a monoperiodic velocity driver, with amplitude of $8$~km~s$^{-1}$ and with a constant frequency matching that of the fundamental standing kink mode of the loop (see Section~\ref{sec:waves}, and Figures~\ref{fig:kostas} and~\ref{fig:patrick2}).
Thus, our flux tube simulates an active region coronal loop, during its cooling phase, undergoing a decay-less oscillation (e.g., \citealt{Anfinogentov2013}). The continuously driven oscillation leads to the development of the KHI and of spatially extended TWIKH (Transverse Wave Induced KH) rolls. These TWIKH rolls disrupt the initial monolithic loop profile, creating a turbulent loop cross-section \citep[][]{Karampelas2018...9K}. Inside the loop, we have energy cascade to lower scales leading to heating from energy dissipation, as well as extensive mixing of plasma with different temperatures, leading to effects of apparent heating \citep[][]{Karampelas2019a}. All calculations in the model considered here were performed in ideal 3D MHD (no explicit resistivity, viscosity, radiation nor thermal conduction) in the presence of numerical dissipation, with the use of the PLUTO code \citep{Mignone2007a,Mignone2012a}. The model has a resolution of $[40; 40; 1563]$~km and a cadence of $\sim 11$~s. For a more detailed description, the reader is referred to the analysis found in \cite{Karampelas2019b}.

{\bf CipMOCCT:} 
The CipMOCCT code \citep{Kudoh_1999_CFD.8} is here used for modelling an impulsively driven kink mode and associated dynamic instabilities in model \Cwaves. The code uses cubic-interpolated pseudoparticle/propagation scheme \citep[CIP,][]{Yabe_1991CoPhC..66..219Y} to solve the viscous-resistive MHD equations, while the method of characteristics -constrained transport \citep[MOCCT][]{Evans_1988ApJ...332..659E,Stone_1992ApJS...80..753S} is used to solve the induction equation. The CipMOCCT code is especially well suited to model instabilities thanks to its ability to maintain sharp contact surfaces \citep{Yabe_1993JPSJ...62.2537Y,Kudoh_1998ApJ...508..186K}.
In model \Cwaves\ a hot coronal loop of length 200~Mm is modelled as a straight, cylindrical flux tube of radius $R=1$~Mm in pressure equilibrium with the background. The loop is initially hotter and denser than the background, with the internal temperature $T_i=3T_e=3\times10^6~$K, and the internal total number density $n_i=3n_e=3\times10^9$~cm$^{-3}$. Correspondingly, the internal and external magnetic fields are respectively $B_i=17.87~$G and $B_i=18.63~$G. The loop's boundary layer connecting the internal and external plasma has a smoothed (hyperbolic tangent) variation over a width $0.6~R=600$~km. The numerical box is (512,512,200) grid points in size, where the plane (512,512) is uniformly distributed over the transverse plane to the loop (leading to a grid size of 15.6~km and an extent of $\approx4R$) and 200 points are uniformly distributed over the longitudinal loop axis (grid size of 2,000~km).  Further details about the geometric setup of the loop and the MHD equations solved can be found in \citep{Antolin_2019FrP.....7...85A}. 
At time $t=0$ a kink mode is impulsively excited with a velocity perturbation mimicking a fundamental mode with amplitude $v_0=16.6~$km~s$^{-1}$, leading to a nonlinear kink mode perturbation. Following the perturbation, the loop oscillates with a fundamental kink mode of period $P\approx 315$~s (see sec.~\ref{sec:waves}, and Figures~\ref{fig:patrick2}, \ref{fig:patrick0} and~\ref{fig:patrick1}). 

{\bf Lare3d:} 
The 3D simulations presented here have been obtained using the Lagrangian-Remap code, Lare3d \citep{Arber2001}. Lare3D solves the fully 3D (normalized) nonlinear MHD equations with the option to included non-ideal terms (through viscosity and resistivity) as well as additional physics such as gravity, thermal conduction, optically thin radiation and Cowling resistivity. The viscosity includes contributions from (small) shock viscosities which ensure numerical stability. Lare3D does not enforce energy conservation, i.e., numerical dissipation does not lead to an increase in the plasma temperature. 
Here we present results from 3D MHD simulations of footpoint driving in a straight magnetic field (Howson \& De Moortel 2021, in prep.).
The simulations start from a gravitationally stratified atmosphere with an initial temperature of $2\times 10^4$~K and with a very weak background heating (where without additional heating, the apex temperature would only rise to about $5 \times 10^4$~K). The lower and upper boundaries of the numerical domain are driven with either long (DC-like; model \Ldc) or short (AC-like; model \Lac) timescales (see Section~\ref{sec:waves} and Figure~\ref{fig:ineke1}); see \cite{Howson2020} for a more detailed description of the form and timescales of the boundary driver. 
The simulations are non-ideal, using a background resistivity which is non-uniform in the field-aligned direction. It is turned on in the coronal volume but does not cause heating near the driven boundaries. A grid of $256 \times 256 \times 512$ is used to cover $20 \times 20 \times 60$~Mm.
The LOS is taken to be perpendicular to the dominant magnetic field direction (represented by the $x$-coordinate in Figs~\ref{fig:ineke1}). The time interval that has been synthesised is taken from $9054-10846$ seconds after the start of the simulation, when a relatively steady state has been reached. Both the AC and DC simulations reach coronal temperatures (see Figure~\ref{fig:ineke1}). Although there are many detailed differences between the DC and AC models, we focus here on those differences enabled by the \muse\ capabilities, in particular the ability to simultaneously produce time-distance maps of the intensity and Doppler velocity (as well as Doppler width – not shown here) both along and across loop structures. Swaying motions such as those visible in slit 5 of the AC model (3$^{rd}$ row, 3$^{rd}$ column in Fig.~\ref{fig:ineke1}) have been observed previously but the combination with the time-distance map along the loop provides a clear distinction between the AC and DC models as well as the opportunity to identify the mode of oscillation. For example, the \fexvw\ Doppler velocity time-distance map along the loop for the AC model (2$^{nd}$ row, 4$^{th}$ column in Fig.~\ref{fig:ineke1}) shows an in-phase perturbation along the entire loop length, strongly suggesting the presence of the global mode. 
  
{\bf Model \Vawt:}  In the Alfv\'{e}n wave turbulence model, the photospheric footpoint motions generate transverse MHD waves that propagate upward along the magnetic field  lines. The waves reflect due to the density variations in the photosphere, chromosphere, and corona creating inward propagating waves. The counter propagating waves interact nonlinearly, resulting in turbulence. The energy is dissipated as a result of turbulence, and heats the corona to 2-3~MK.
In the version of the AWT model presented here (model \Vawt\ of Table~\ref{table_sims}), a collection of 16 photospheric flux tubes with square cross-sections is considered \citep[][]{vanBallegooijen2017AWT}. The flux tubes expand with height and merge at a height of 520~km in the low chromosphere. The merged field extends from the chromosphere at one end of the loop to the chromosphere at the other end, so the TRs are located within the merged field (the coronal loop length $L_{\rm c} = 98.4$ Mm). The imposed random footpoint motions inside each of the flux tube generates \alfven\ waves with complex transverse wave patterns. 
The transverse waves are simulated using the reduced MHD approximation \citep[e.g.,][]{Strauss1976, Strauss1997}, where the magnetic field strength $B_0 (s)$ and density $\rho (s)$ are assumed to be constant over the cross-section of the flux tubes. The dynamics of the waves are simulated for a period of 3000 s where the \alfven\ wave turbulence is developed along the field lines depositing large amount of energy and heating the corona. The numerical methods used in the present work are similar to those used in simulating \alfven\ waves in the solar wind \citep[][]{vanBallegooijen2017di}.

{\bf Model \Mac:} \cite{Matsumoto2018} use 3D MHD simulations to model a half of  a straight loop (covering a spatial extent of 3~Mm $\times$ 3~Mm $\times$ 50~Mm), and to investigate the thermal responses of the coronal plasma to the dynamic dissipation processes of MHD waves. They assume optically thick cooling (as approximated by \citealt{Anderson1989}), where the temperature is $\lesssim 4 \times 10^4$~K or the density is $\gtrsim 4.9 \times 10^{-17}$~g~cm$^{-3}$, and optically thin cooling (e.g., \citealt{Landini1990}) otherwise.
The simulation uses 64 $\times$ 64 $\times$ 1024  grid point, with a constant horizontal grid cell size of 47~km, and non-uniform vertical grid with size increasing from 25~km at the bottom to 93~km at the top. MHD waves are driven by continuous forcing at the loop footpoint, and with a total power of $2.3  \times 10^{-4}$ g$^2$ cm$^{-4}$ s$^{-4}$, and a white noise spectrum with a finite range of $\nu \in [2.5 \times 10^{-4}, 2 \times 10^{-2}]$ Hz.  
The amplitudes of the external forces is such that the root mean square of the horizontal velocity is $\sim 6$~km~s$^{-1}$ at $x = 2$~Mm, corresponding to a photospheric ($x < 0.2$~Mm) $\sim 1.6$~km~s$^{-1}$.
The resulting model atmosphere has temperature  of $0.9$~MK and low coronal density ($n_e \lesssim 10^8$~cm$^{-3}$), consistent more with open field regions than with the active region loops.

{\bf RADYN:}
To explore the diagnostic capabilities of \muse\ for uncovering the properties of nanoflare heating, including heating by non-thermal electrons (NTE) we use a set of 1D RADYN loop models, undergoing impulsive heating with various properties. The RADYN numerical code \citep{Carlsson:1992kl,Allred2005,Allred2015}  solves the 1D equations of conservation of mass, momentum, energy, and charge and the non-LTE level population rate equations for the magnetically confined plasma, and the loop's atmosphere encompasses the photosphere, chromosphere, TR, and corona. RADYN also allows to model heating by non-thermal electron (NTE) beams \citep{Allred2015}. RADYN simulates the transport of accelerated electrons through the solar atmosphere using the Fokker-Planck kinetic theory \citep{McTiernan1990,Allred2015}.
RADYN simulations  are also well suited to model larger flares, as for instance illustrated in  the companion paper on flares and eruptions \citep{Cheung2021_muse_fl}.
The models analyzed here are described in detail in \cite{Testa2020a,Polito2018}. In particular we will analyze \muse\ synthetic observables from four different simulations, all starting from an initial atmosphere characterized by a  cool ($T \sim 1$~MK) and low density ($n_e \sim 5 \times 10^8$~cm$^{-3}$) corona (see \citealt{Polito2018} for details), and undergoing an heating event characterized by 10~s duration, energy of  $1.2 \times 10^9$~erg~cm$^{-2}$~s$^{-1}$) and with the following heating properties: (1) direct heating in the corona and energy transport by thermal conduction (\Rcone\ of Table~\ref{table_sims}); (2) heating by NTE characterized by power-law distribution with energy cutoff $E_C =5$~keV (\Reone); (3) heating by NTE with $E_C =10$~keV (\Retwo); and, (4) a hybrid model with both direct heating and NTE with $E_C =10$~keV (\Rhone).
Here we map the 1D loop into a 2D domain by assuming a semi-circular shape for the coronal loop (see Figure~\ref{fig:radyn}), and assume \muse\ observes the loop from above, with its several slits sampling the whole loop length.

\section{Synthesis of \muse\ spectral observables from models} \label{app:synthesis}
Throughout the paper we have synthesized \muse\ observables from different models, using the \muse\ response functions. As described in detail in \cite{BDP:MUSE}, the \muse\ spectral response functions provide the detector response across all 1024 spectral pixels for all three channels, per unit emission measure (in $10^{27}$~cm$^{-5}$), at a specified slit (1–37), temperature, and velocity. Here in particular we focus on the main lines: \fexixw, \feixw, and \fexvw\ (and, for paper \citealt{Cheung2021_muse_fl}, which focuses on flares and eruptions, we also show predicted \fexxiw\ emission).  
The response functions are computed using the latest CHIANTI database version (10.0), and include instrumental effects (such as instrumental line broadening) and thermal broadening of the lines.
The response functions calculate the predicted spectra in units of [DN/s/pix], where the pixel is the \muse\ spectral pixel which has dimensions of 0.4\arcsec $\times$ 0.167\arcsec. 
The synthetic \muse\ SG intensities shown in the paper are calculated assuming a spectral pixel of 0.4\arcsec $\times$ 0.167\arcsec, even when the synthetic maps are displayed at a spatial sampling of 0.167\arcsec $\times$ 0.167\arcsec. In fact, we often show results with a spatial sampling of 0.167\arcsec\ in both spatial directions, to preserve the information at the highest spatial resolution achievable with \muse, given that any specific orientation of the \muse\ slits with respect to the simulated features is possible. In Fig.~\ref{fig:daniel} however the synthetic \muse\ observables are displayed with a 0.4\arcsec $\times$ 0.167\arcsec\ resolution. The synthetic \muse\ CI intensities in all figures are calculated assuming a 0.167\arcsec $\times$ 0.167\arcsec\ pixel size.

For the reduced MHD model \Vawt, the LOS velocity and non-thermal velocity are not derived from synthetic spectra, but directly obtained from the model and reduced to \muse\ resolution (0.167\arcsec $\times$ 0.167\arcsec). As the reduced-MHD model does not include temperatures, the LOS velocity and non-thermal velocity are integrated through the loop. 

\section{Definition of spectral moments} \label{sec:moments}

In the paper we show 0th ($I_0$), 1st ($I_1$) and 2nd ($I_2$), moments of the \muse\ synthetic observables from models, and they are defined, respectively, as follows:

\begin{eqnarray}
I_0 = \sum_j F_j ~[{\rm DN/pix/s}]\\
I_1 = \frac{\sum_j F_j \times v_j}{I_0}~[\text{km s}^{-1}] \\ 
I_2 = \sqrt{\frac{\sum_j F_j \times (v_j-I_1)^2}{I_0}}~[\text{km s}^{-1}]
\end{eqnarray}
\noindent where $j$ is the index of the spectral bin, $F_j$ is the intensity in spectral bin $j$ in units of DN~s$^{-1}$~pix$^{-1}$, and $v_j$ is the Doppler velocity in km~s$^{-1}$.

\section{simulated MUSE timescales}~\label{app:timescales}
In the paper we investigate the typical timescales, over which the \muse\ spectral observables change significantly, as predicted by different models (see Figure~\ref{fig:timesca}). We also study the spatial scales over which the different  models show coherence.
To compute the timescales we apply a 'supervised $k$-means' approach on the synthesized spectral data. The $k$-means clustering algorithm is a machine learning method that groups observed (in our case, synthetic) data according to their spectral properties (see e.g., \citealt{Panos2018,Bose2019} for application to solar spectral observations). Unsupervised algorithms find a certain number of clusters, each characterized by a representative spectral profile (RP), and the number can be optimized using appropriate criteria. 
Here we adopt a {\em supervised} approach where the RPs are defined a priori, and we focus on the \muse\ \feixw\ line and \fexvw\ line. In particular here we define the RPs such that we have a finely spaced grid covering the range of values for the 0th, 1st and 2nd moment with a resolution of 5~km~s$^{-1}$ for the 1st and $10$~km~s$^{-1}$ for the 2nd.
In addition, the intensity variation for the supervised representative profiles has a separation of $3\sigma$. 
Using these RPs, we then apply the $k$-means clustering analysis to the synthetic data for each model as a function of time. We then analyze, for each pixel, the times needed for the spectrum to change cluster, and define the timescale of variation in each pixel as the mean of those times.

\bibliographystyle{aasjournal}
\bibliography{collectionbib2.bib}

\end{document}